\documentstyle[preprint,aps,eqsecnum]{revtex}

\begin{document}
\baselineskip=16pt

\title{
\hfill
\parbox{4cm}{\normalsize KUNS-1317\\HE(TH)~95/02 \\hep-lat/9501032}\\
\vspace{0.3cm}
Perturbation Theory at Finite Extent of Fifth Dimension \\
for Vacuum Overlap Formula of Chiral Determinant \\
--- Continuum Limit Case ---
}
\author{ \normalsize \sc
Teruhiko Kawano\thanks{e-mail address:
kawano@gauge.scphys.kyoto-u.ac.jp}
\
\hskip 6pt and  \hskip 6pt
Yoshio Kikukawa\thanks{e-mail address:
kikukawa@gauge.scphys.kyoto-u.ac.jp}
}

\address{\normalsize \em
         Department of Physics, Kyoto University\\
         Kyoto 606-01, Japan
}
\date{\normalsize January, 1995}

\maketitle

\begin{abstract}
\baselineskip 16pt
Taking into account of the boundary condition
in the fifth direction which is derived from the lattice
Wilson fermion, we develop a theory of five-dimensional
fermion with kink-like and  homogeneous masses in finite
extent of the fifth dimension.
The boundary state wave functions are constructed explicitly and
the would-be vacuum overlap
is expanded by using the propagator of the theory.
The subtraction is performed unambiguously at the finite extent
with the help of the dimensional regularization.
Then the limit of the infinite extent is evaluated.
The consistent anomaly in four dimensional theory
is finitely obtained.
Each contribution to the vacuum polarization is vector-like.
It is the lack of the massless mode in the fermion with negative
homogeneous mass that leads to the correct chiral normalization.
Gauge noninvariant piece remains due to the breaking of
the boundary condition by the dimensional regularization.
\end{abstract}

\newcommand{\Lag}{{\cal L}}
\newcommand{\iLL}{\int^L_{-L}}
\newcommand{\iLo}{\int^L_{0}}
\newcommand{\ioL}{\int^0_{-L}}
\newcommand{\sitarel}[2]{
{\mathrel{\mathop{\kern0pt #1}\limits_{#2}}} }
\newcommand{\fs}[1]{{#1 \hskip -4.5pt/}}
\newcommand{\fsu}[1]{{#1 \hskip -6.5pt/}}
\newcommand{\ee}{\, .}
\newcommand{\ec}{\, ,}
\newcommand{\mspd}{\int\!\frac{d^Dp}{(2\pi)^D}}
\newcommand{\mskd}{\int\!\frac{d^Dk}{(2\pi)^D}}
\newcommand{\mss}{\int^{+L}_{-L}\!ds}
\newcommand{\mssh}{\int^{+L}_{0}\!ds}
\newcommand{\mst}{\int^{+L}_{-L}\!dt}
\newcommand{\bra}[1]{\left\langle #1\right\vert}
\newcommand{\ket}[1]{\left\vert #1\right\rangle}
\newcommand{\Tr}{{\rm Tr}}
\newcommand{\tr}{{\rm tr}}
\newcommand{\half}{\frac{1}{2}}
\newcommand{\ext}[1]{#1\kern-6pt\lower5pt\hbox{$\scriptstyle \sim$}}
\newcommand{\org}[1]{#1\kern-6pt\lower5pt\hbox{$\scriptstyle -$}}

\newcommand{\vns}{\sum_{n,s}}
\newcommand{\vnms}{\sum_{n,m,s}}
\newcommand{\vn}{\sum_{n}}
\newcommand{\vnm}{\sum_{nm}}
\newcommand{\vecsum}{\sum_\mu}
\newcommand{\vecsun}{\sum_\nu}
\newcommand{\vecsunn}{\sum_{\mu\nu}}

\newcommand{\mlm}{{\scriptstyle -L-1}}
\newcommand{\ml}{{\scriptstyle -L}}
\newcommand{\lp}{{\scriptstyle L}}
\newcommand{\lpp}{{\scriptstyle L^\prime}}
\newcommand{\lm}{{\scriptstyle L-1}}
\newcommand{\mlo}{{\scriptstyle -L+0}}
\newcommand{\lpo}{{\scriptstyle L-0}}
\newcommand{\bbone}{
{\mathchoice {\rm 1\mskip-4mu l} {\rm 1\mskip-4mu l}
{\rm 1\mskip-4.5mu l} {\rm 1\mskip-5mu l}}}
\def\sitarel#1#2{\mathrel{\mathop{\kern0pt #1}\limits_{#2}}}

\section{Introduction}
\label{sec:intro}

A nonperturbative regularization of chiral gauge theory,
if it would exist, could offer a consistent framework for studying
the dynamics of the standard model, especially the
dynamics of spontaneous gauge symmetry breaking.
Lattice regularization has succeeded to play such a role
for understanding the QCD dynamics.  For chiral theories,
however, it suffers from the species doubling
problem\cite{ksmit,nn,karsten,kieu}.

Recently new approaches by means of an infinite number
of Fermi fields have been proposed for such a
regularization\cite{kaplan,pvfs,cnn,olnn,latfs}.
A five-dimensional fermion has a chiral zero mode
when coupled to a domain wall\cite{callanharvey}.
Kaplan formulated such a system on a lattice with Wilson fermion
and discussed the possibility to simulate
chiral fermions\cite{kaplan,domainwall}.
On the other hand, Frolov and Slavnov
considered the possibility of
regulating chiral fermion loops
gauge-invariantly in the $SO(10)$ chiral gauge theory
with an infinite number of the
Pauli-Villars-Gupta fields\cite{pvfs,ak,genpv}.

Unified point of view on these two approaches was given
by Neuberger and Narayanan\cite{cnn}
and they have put forward the approach
to derive a lattice vacuum overlap formula
for the determinant of chiral fermion\cite{olnn}.
They first discarded the five dimensional nature
of gauge boson in the Kaplan's lattice setup.
Then the five-dimensional fermion can be seen as a collection of
infinitely many four-dimensional fermions labeled by the extra
coordinate.
They regarded the massive Dirac modes as
regulator fields for the chiral mode
(these correspond to the fermionic Pauli-Villars-Gupta fields in
the method of Frolov and Slavnov)
and gave a prescription to subtract
irrelevant bulk effects of the massive modes by
ordinary fermions with homogeneous masses
(these correspond to the bosonic Pauli-Villars-Gupta fields).
They emphasized the importance of infinite extent of the
extra space to make sure a chiral content of fermions,
which space is usually compactified on a lattice with periodic boundary
condition.
As a results of the infiniteness of the extra space,
they obtained a vacuum overlap formula for the determinant of
lattice chiral fermion, using transfer matrices in the direction of
the extra space.

By means of the infinite number of the Pauli-Villars-Gupta fields,
however,
the odd-parity part can never be regularized because
the regulator fields are the {\it Dirac} spinors\cite{fujikawa}.
Even for the even-parity part,
there exist ambiguity in the summation over
the infinite number of contributions.
In order to make the summation well-defined,
at the same time to make the number of regulator fields
finite at the first stage,
each contribution of the original or the regulator field
should be made finite by a certain subsidiary regularization.
Such subsidiary regularization necessarily breaks the gauge
invariance in the contribution of the original chiral fermion.
This may well lead to the gauge noninvariant result even in the limit of
the infinite number of the regulator fields.
The dimensional
regularization is an example
for such a subsidiary regularization\cite{Latfs}.

In the formulation of the vacuum overlap,
the problem of the odd-parity part is reflected in that
we must fix the phase of the overlap by a certain guiding
principle to reproduce the properties of the chiral determinant
by this formula, especially anomaly.
Fixing the phase of the overlap is
equivalent to the choice of the wave functions
of the boundary states.
By this choice, we must carefully place a source of
gauge noninvarinace at the boundaries to reproduce
the consistent anomaly.

Neuberger and Narayanan\cite{olnn}
proposed to fix the phase following
the Wigner-Brillouin perturbation theory,
refering to the ground state of the {\it free} Hamiltonian.
By this prescription, the continuum two-dimensional overlap
was examined and it was shown to reproduce the correct
{\it consistent} anomaly.
The lattice two- and four-dimensional overlap were also
examined numerically and correct anomaly coefficients
were observed for the Abelian background gauge field.
Its behavior under the topologically nontrivial
background gauge fields was also examined and
promising result was obtained.

They also showed the gauge invariance of the even-parity part
at nonperturbative level. Thier discussions are based on
the case of the periodic boundary condition in the fifth direction
and also on the gauge transformation property of the
ground states of the four dimensional Hamiltonians.

On the other hand, in the view point of the perturbation theory
with the five(or three)-dimensional Wilson fermion,
several authors discussed this problem.
Shamir\cite{shamir} considered the fermion in the infinite extent
of the fifth dimension but restricted first
the interaction with the gauge field into the finite region of
the fifth space.
By this restriction, the model becomes gauge noninvariant
at the boundaries of the finite region.
Then he examined how to take the limit of the infinite extent.
He claimed that the limit should be taken uniformly at
every interaction vertex with gauge boson and then
showed that this way to introduce the gauge noninvariance
leads to the correct consistent anomaly in two-dimensional model.
In the similar approach,
S. Aoki and R.B. Levien\cite{al} performed a detailed study
about the infinite extent of the extra dimension
and the subtraction procedure in the lattice two-dimensional
chiral Schwinger model.
They showed that the scheme reproduces the desired form of
the effective action: the gauge invariant real part and the
consistent anomaly from the imaginary part.

Four-dimensional perturbative study of the vacuum overlap formula
has been recently in progress.
S. Randjbar-Daemi and J. Strathdee obtained the consistent anomaly
by the four-dimensional Hamiltonian perturbation theory
in the continuum limit\cite{rs}.
Quite recently, they performed the similar analysis in the lattice
regularization\cite{lrs}.
In the view-point of the theory with infinitely many regulator fields,
it is also desirable to understand how the gauge-noninvarinace put
at the boundaries leads to the consistent anomaly
and how the gauge-invariance of the even-parity part of the
determinant is established by the subtraction.
In this article, toward this goal,
we further examine the four dimensional aspect of the vacuum overlap
formula in the continuum limit.

Our approach is as follows.
We consider the nonabelian background gauge field in general.
We start from finite extent of the fifth direction
in order to make the summation unambiguous.
We first develop the theory of the free five-dimensional fermion
with kink-like mass and positive(negative) homogeneous mass
{\it in the finite extent of fifth direction}.
As to the boundary condition in the fifth direction,
we adopt the one derived from the Wilson fermion,
by which the Dirichlet and Neumann components are determined
by the chiral projection.

{}From this free theory,
the boundary state wave functions which correspond to
the Wigner-Brillouin phase choice are explicitly constructed.
Then we formulate the perturbation expansion of
the vacuum overlap formula in terms of the propagator at
the finite extent of the fifth dimension
satisfying the boundary condition.
After performing the subtraction at the finite extent of the
fifth dimension, we examine the limit of the infinite extent.

As to a subsidiary regularization,
we adopt the dimensional regularization.
It turns out that the dimensional regularization cannot respect
the boundary condition determined by the chiral projection.
Although this fact reduces the ability of our analysis in the
continuum limit, we believe that we can make clear
in what way the vacuum overlap formula could give
the perturbative properties of the chiral determinant
in four dimensions.
We also make another technical assumption that
the dimensional regularization preserves the cluster
property.

By this perturbation theory,
we calculate the variation of the vacuum overlap
under the gauge transformation which is
induced by the boundary state wave function.
We also calculate the two-point function of the external
gauge boson.
The variation is found to be finite and does not suffer from the subtlety of
the dimensional regularization.
It reproduces the consistent anomaly in four dimensions correctly.
We also observe how the chiral normalization of the vacuum
polarization is realized in the finite extent of the fifth direction.
We, however, fail to establish its gauge invariance
due to the dimensional regularization.

This article is organized as follows.
In Sec. \ref{sec:would-be-vacuum-overlap},
we discuss
the lattice Schr\"odinger functional\cite{luscher,sint} to describe
the evolution of the boundary state during a finite ``time''
interval in the fifth direction.
It is naturally formulated from the transfer matrix given
by Neuberger and Narayanan.
The boundary condition of the fermion field is read off
from the Wilson fermion action and the boundary term is
derived.
In Sec. \ref{sec:fermion-at-finite-fifth-volume},
we formulate the free theory of
the five-dimensional fermion in the finite fifth space volume.
We solve the field equation and obtain the complete set of
solutions. The field operator is defined by the mode expansion
and the propagator is derived.
The Sommerfeld-Watson transformation is introduced, by which
we rearrange the normal modes of the fifth momentum
to be common among the fermions with the kink-like
mass and the positive(negative) homogeneous mass.
This makes it possible to do the subtraction at the finite extent
of the fifth dimension.
In Sec.
\ref{sec:perturbation-at-finite-extent-of-fifth-dimension},
the perturbation theory for the vacuum overlap is developed.
We first derive boundary state wave functions.
Then we discuss the cluster property of the contribution induced by
the boundary state wave function.
By this cluster property,
the boundary contribution turns out to be odd-parity in the
limit of the infinite extent.
In Sec. \ref{sec:anomaly-from-boundary},
we perform the calculation of the anomaly induced by
the boundary state wave function.
In Sec. \ref{sec:vacuum-polarization},
we also perform the calculation of the vacuum polarization.
Section \ref{sec:discussion} is devoted to summary and discussion.

\newpage
\section{
Would-be vacuum overlap}
\label{sec:would-be-vacuum-overlap}
In this section,
we consider
the five-dimensional Wilson fermion with kink-like mass and
its finite ``time'' evolution in the fifth direction.
In order to describe it,
we introduce the Schr\"odinger functional, which is naturally
formulated by the transfer matrix given by Neuberger and Narayanan.
Through its path-integral representation,
we read off the boundary condition imposed on the fermionic
field and also derive the boundary terms.
We also rewrite the functional in the form factorized
into the determinant of the Dirac operator
over the five-dimensional volume under the derived boundary condition
and the contribution from the boundary terms.
Next we will discuss how to prepare the boundary state wave functions which
implement the Wigner-Brillouin phase choice.
With these wave functions, we give the expression of the ``would-be''
vacuum overlap at finite extent of the fifth dimension.
We also discuss its variation under the gauge transformation.
Finally, we derive the counterpart
{\it in the continuum limit and in the Minkowski space}.
We also specify the regularization in the continuum limit analysis.

\subsection{Boundary condition in the fifth direction
and boundary terms}
\label{subsec:bc-fifth}

The action of the five-dimensional Wilson fermion with kink-like mass
is given by
\begin{eqnarray}
  \label{lattice-action}
A
&=&
\vns \left\{
\hskip 4pt
\sum_\mu \half
 \left[ \bar\psi(n,s) (1+\gamma_\mu) U_\mu(n) \psi(n+\hat\mu,s)
       +\bar\psi(n+\hat\mu,s)(1-\gamma_\mu) U^\dagger_\mu(n) \psi(n,s)
\right]
\right.
\nonumber\\
&&\hskip 38pt +
\half
\left[ \bar\psi(n,s) (1+\gamma_5) \psi(n,s+1)
+\bar\psi(n,s+1) (1-\gamma_5) \psi(n,s)     \right]
\nonumber\\
&&\hskip 38pt
\left.
+ \Big( m_0 \, sgn(s+\half) -5 \Big) \bar\psi(n,s) \psi(n,s)
\right\}
\ee
\end{eqnarray}
Here we are considering SU(N) background gauge field in general.

The transfer matrix formulation for it was first
given by Neuberger and Narayanan\cite{olnn}.
Let us assume that
the Fock space is spaned by the operators
$\hat c_{\alpha i}(n)$ and $\hat d_{\alpha i}(n)$
satisfying the following commutation relations,
\begin{equation}
  \label{commutation-relation}
  \{ \hat c_{\alpha i}(n), \hat c_{\beta}^{\dagger j}(m) \}
= \delta_{nm} \delta_{\alpha\beta} \delta_i^j
\ec \hskip 16pt
  \{ \hat d_{\alpha i}(n), \hat d_{\beta }^{\dagger j}(m) \}
= \delta_{nm} \delta_{\alpha\beta} \delta_i^j
\ec
\end{equation}
\begin{equation}
  \label{Fock-vacuum}
  \hat c_{\alpha i}(n) \ket{0}=0 \ec \hskip 16pt
  \hat d_{\alpha j}(n) \ket{0}=0 \ee
\end{equation}
Note that $\alpha,\beta$ denote the spinor index
and $i,j$ denote the index of the representation
of SU(N) gauge group.
Then the transfer matrix is given in terms of
$\hat a = ( \hat c , \hat d^\dagger )^t $ and
$\hat a^\dagger = ( \hat c^\dagger , \hat d )$ as,
\begin{equation}
  \label{Transfer-matrix}
\hat T_\pm = \exp \left( \hat a^\dagger H_\pm \hat a \right)
\ec
\end{equation}
with the matrix
\begin{eqnarray}
  \label{hamiltonian-matrix}
\exp\left( H_\pm \right)
&\equiv&
\left(\begin{array}{cc}
\frac{1}{B^\pm} & \frac{1}{B^\pm} C \\
\frac{1}{B^\pm}C^\dagger & C^\dagger \frac{1}{B^\pm} C + B^\pm
          \end{array}\right)
\ec
\end{eqnarray}
where
\begin{eqnarray}
  \label{Bfunction}
B(n,m,s)
&=&
\left(5 - m_0 \, sgn(s+\half) \right) \delta_{n,m} \delta_i^j
         -\half \vecsum \left(
                  {U_\mu(n)}_i^j         \delta_{n+\hat\mu,m}
                +{U^\dagger_\mu(m)}_i^j \delta_{n,m+\hat\mu} \right)
\ec
\\
C(n,m)
&=& \half \vecsum {\sigma_\mu}_{\alpha \beta}
       \left( {U_\mu(n)}_i^j  \delta_{n+\hat\mu,m}
             -{U^\dagger_\mu(m)}_i^j \delta_{n,m+\hat\mu}
            \right)
\equiv \vecsum {\sigma_\mu}_{\alpha \beta} \nabla_\mu(n,m)
\ec
\end{eqnarray}
and $\sigma_\mu \equiv (1,i\sigma_i)$.
$B(n,m)$ can be shown to be positive definite for $ 0 < m_0 < 1$.
We also introduce the operator
\begin{equation}
  \label{D-operator}
\hat D_\pm = \exp \left( \hat a^\dagger Q_\pm \hat a \right)
\ec
\end{equation}
with
\begin{eqnarray}
  \label{D-matrix}
\exp\left( Q_\pm \right)
&=&
\left(\begin{array}{cc}
\frac{1}{\sqrt{B^\pm}}
&\frac{1}{\sqrt{B^\pm}}C \\
0& \sqrt{B^\pm}
      \end{array}\right)
\ec
\end{eqnarray}
and we can show
\begin{equation}
  \label{hamiltonian-composition}
\exp\left( H_\pm \right)=
\exp\left( Q_\pm \right) ^\dagger \exp\left( Q_\pm \right) \ec
\hskip 24pt
\hat T_\pm = \hat D_\pm^\dagger \hat D_\pm
\ee
\end{equation}

We start from a finite ``time'' evolution in the fifth direction.
We take the symmetric region $s \in [\mlm, \lp]$. The evolution
can be described by the Schr\"odinger kernel\cite{luscher,sint}
\begin{equation}
  \label{Schrodinger-Kernel}
\bra{ c^\ast_{\mlm},d^\ast_{\mlm} }
       D_- \left(T_-\right)^\lp \left( T_+\right)^\lp D_+^\dagger
        \ket{ c^{ }_\lp,d^{ }_\lp }
\ec
\end{equation}
in the coherent state basis,
\begin{eqnarray}
  \label{coherent-state}
\ket{ c,d } &=& \exp [ -(c,{\hat c}^\dagger)-(d,{\hat d}^\dagger) ]
\ket{0} \ec\\
\bra{ c^\ast,d^\ast } &=& \bra{0} \exp [ -({\hat c},c^\ast)-({\hat d},d^\ast) ]
\ee
\end{eqnarray}
Here we are following the notation in Ref. \cite{olnn}:
$(\bar a_s,b_s) \equiv
\sum_{n,\alpha,i} \bar a_\alpha^i(n,s) b_{\alpha i}(n,s)$.

In terms of the path integral, it reads
\begin{eqnarray}
  \label{Schrodinger-Kernel-PathInt}
&&
\bra{ c^\ast_{\mlm},d^\ast_{\mlm} }
       D_- \left(T_-\right)^\lp \left( T_+\right)^\lp D_+^\dagger
        \ket{ c_\lp,d_\lp }
\prod_{0\leq s \leq \lm}(\det B_+)^{2}
\prod_{\ml \leq s \leq -1}(\det B_-)^{2}
\nonumber\\
&&
=
\int \prod_{\ml \leq s \leq \lm}
[{\cal D} \psi_s {\cal D} \hat\psi_s]
\exp\left\{ -A[\mlm,\lp] \right\}
\nonumber\\
&&
\equiv
Z[\psi_L(\mlm),\bar\psi_L(\mlm);\psi_R(\lp),\bar\psi_R(\lp)]
\ee
\end{eqnarray}
The boundary variables are given by
\begin{eqnarray}
  \label{boundary-variable}
\psi_R (n,\lp)&=& \frac{1}{\sqrt{B_+}}
\left( \begin{array}{c} c(n,\lp) \\ 0 \end{array}\right)
\ec
\hskip 16pt
\bar \psi_R(n,\lp)
=  \left( \begin{array}{cc} 0 & -b(n,\lp) \end{array}\right)
\frac{1}{\sqrt{B_+}}
\ec
\\
\psi_L(n,\mlm ) &=& \frac{1}{\sqrt{B_-}}
\left( \begin{array}{c} 0 \\ b^\ast(n,\mlm) \end{array}\right)
\ec
\hskip 16pt
\bar \psi_L(n,\mlm) =
  \left( \begin{array}{cc} c^\ast(n,\mlm) & 0 \end{array}\right)
\frac{1}{\sqrt{B_-}}
\ee
\end{eqnarray}
The action and the boundary terms are given by
\begin{equation}
  \label{lattice-action-with-boundaryterm}
  A[\mlm,\lp]= A+A^B_L+A^B_R
\ec
\end{equation}
\begin{eqnarray}
  \label{lattice-action-chiral-bc}
A
&& \equiv
\sum_{n}\sum_{s=\ml}^{\lm}
\left\{ \hskip 2pt
\vecsum
\bar\psi(n,s) \gamma_\mu \nabla_\mu \psi(n,s) -\bar\psi(n,s) B(n,m) \psi(n,s)
\right.
\nonumber\\
&&\left.
+\half
\left[ \bar\psi(n,s) (1+\gamma_5) \psi(n,s+1)
+\bar\psi(n,s+1) (1-\gamma_5) \psi(n,s)  \right]
\right\}
\Big\vert_{[-\lp,\lp]}
\ec\\
\label{left-boundary-term}
A^B_L
&&\equiv
\vn \left\{
\bar\psi(n,\ml) P_L  \psi(n,\mlm)
+\bar\psi(n,\mlm) P_R \psi(n,\ml)
\right\}
\nonumber\\
&&\hskip 16pt
+
\vn \vecsum
\bar\psi(n,\mlm) \gamma_\mu \nabla_\mu P_L \psi(n,\mlm) \ec
\\
  \label{right-boundary-term}
A^B_R
&&\equiv
\vn  \left\{
\bar\psi(n,\lp) P_L  \psi(n,\lm)
+\bar\psi(n,\lm) P_R \psi(n,\lp)
\right\}
\nonumber\\
&&\hskip 16pt
+
\vn \vecsum
\bar\psi(n,\lp) \gamma_\mu \nabla_\mu P_R \psi(n,\lp) \ec
\end{eqnarray}
where $\Big\vert_{[-\lp,\lp]}$
stands for the homogeneous boundary condition:
\begin{equation}
  \label{chiral-boundary-condition}
P_R \psi(n,\lp) = P_L \psi(n,\mlm)= 0 \ec \hskip 16pt
\bar \psi(n,\lp) P_L = \bar \psi(n,\mlm) P_R = 0  \ee
\end{equation}
We refer to this boundary condition as {\it chiral boundary condition}
here after.

Since the boundary terms
depending on $\psi_R(\lp)$, $\bar \psi_R(\lp)$ and
$\psi_L(\mlm)$, $\bar \psi_L(\mlm)$ can be regarded to be
source terms for the system with the homogeneous
boundary condition,
we obtain the following factorized form:
\begin{eqnarray}
  \label{factorized-Kernel}
Z[\psi_L(\mlm),\bar\psi_L(\mlm);\psi_R(\lp),\bar\psi_R(\lp)]
&&
= \det \left( K \right)
\Big\vert_{[-\lp,\lp]}
\exp\left\{ - \left( \bar \Psi, M \Psi \right) \right\}
\ec
\end{eqnarray}
where
$\Psi(n)=( \psi(n,\mlm), \psi(n,\lp) )^t $ and
$\bar \Psi(n)=( \bar \psi(n,\mlm), \bar \psi(n,\lp) )$, and

\begin{eqnarray}
  \label{Kinetic-term}
&&K(n,s;m,t)
\nonumber\\
&&=
\textstyle
-\delta_{s,t}
\left\{ \vecsum \gamma_\mu \nabla_\mu(n,m)-B(n,m;s)
\right\}
-
\half
\left[ (1+\gamma_5) \delta_{s+1,t}
+(1-\gamma_5) \delta_{s,t+1} \right]
\ec
\end{eqnarray}

\begin{eqnarray}
  \label{boundary-M-matrix}
&&M(n,m;[\mlm,\lp])
\nonumber\\
&&
=
 \left(  \begin{array}{cc}
      \scriptstyle
P_R S(n,\mlm;m,\mlm) P_L + P_R \vecsum \gamma_\mu \nabla_\mu(n,m)
     &\scriptstyle
P_R S(n,\mlm;m,\lp) P_R \\
\scriptstyle
     P_L S(n,\lp;m,\mlm) P_L
     & \scriptstyle
P_L S(n,\lp;m,\lp) P_R + P_L \vecsum \gamma_\mu \nabla_\mu(n,m)
         \end{array} \right)
\ec
\end{eqnarray}
and
\begin{equation}  \label{inverse-Kinetic-term}
S(n,s;m,t) = K^{-1}(n,s;m,t) \Big\vert_{[-\lp,\lp]}
\ee
\end{equation}
Note that the determinant and the inverse of $K(n,s;m,t)$
should be evaluated taking into account of
the chiral boundary condition~(\ref{chiral-boundary-condition}).

\subsection{Boundary state wave functions and Gauge non-invariance}
\label{sec:boundary wave function}

Next we consider the ``time'' evolution
of a boundary state at $s=\lp$
\begin{equation}
  \ket{b_+} \, \mbox{\rm (at $s=\lp$)} \ec \hskip 16pt
\end{equation}
and its transition to another boundary state
at $s=\mlm$,
\begin{equation}
  \ket{b_-} \, \mbox{\rm (at $s=\mlm$)} \ec
\end{equation}
The wave functions of these states, in general,
can be given in the coherent state representation as
\begin{equation}
\bra{ c^\ast,d^\ast } b_+ \rangle
=\bra{\psi_L,\bar \psi_L } b_+ \Big\rangle \ec
\hskip 16pt
\langle b_- \ket{ c^\ast,d^\ast }
=\Big\langle b_- \ket{\psi_R,\bar \psi_R } \ee
\end{equation}

To reproduce the properties of the chiral determinant
by the vacuum overlap formula, especially anomaly,
we must carefully place the source of
gauge noninvarinace at the boundaries.
This is equivalent to the choice of the wave functions
of the boundary states.
In order to implement a Wigner-Brillouin phase choice,
we regard the boundary states
as the states which evolved(would evolve)
a certain period of ``time''
by the Hamiltonian without the gauge interaction.
We take $2\lp^\prime$ period of ``time''.
Then the boundary state $\ket{b_+}$ can be written as
\begin{equation}
  \label{Wigner-Brilluin-phase-impliment}
\bra{\psi_L(\lp),\bar \psi_L(\lp) } b_\pm \Big\rangle
=
\bra{\psi_L(\lp),\bar \psi_L(\lp) } (T_+^0)^{2\lpp}
\ket{b_\pm^\prime }
\ee
\end{equation}
In the limit $L^\prime \rightarrow \infty$,
we can expect that the ground state of $\hat H_+^0$, which we denote
as $\ket{0_+}$, is projected out for any choice of the boundary state
$\ket{b_+^\prime}$.
Note that this is nothing but the way to introduce
the breaking of gauge symmetry adopted in \cite{shamir,al},
by which the extent of the fifth dimension is infinite
but the gauge field is introduced only in a finite range of the fifth
dimension.

In terms of the path integral,
Eq.~(\ref{Wigner-Brilluin-phase-impliment}) reads
\begin{eqnarray}
  \label{gauge-noninvariance-in-boundarystate}
&&
\bra{\psi_L(\lp),\bar \psi_L(\lp) } b_\pm \Big\rangle
\nonumber\\
&&
=
\int [{\cal D} \psi(\lp+2\lpp)
      {\cal D} \bar\psi(\lp+2\lpp)]
\exp\left\{ \left( \bar\psi(\lp+2\lpp), B_+^0
 \psi(\lp+2\lpp) \right) \right\}
\nonumber\\
&&
\hskip 16pt \times
Z^0_+[\psi_L(\lp),\bar \psi_L(\lp);
         \psi_R(\lp+2\lpp),\bar \psi_R(\lp+2\lpp)]
\bra{\psi_L(\lp+2\lpp),\bar \psi_L(\lp+2\lpp) }
b_\pm^\prime \Big\rangle
\nonumber\\
&&
=
\,
\int [{\cal D} \psi(\lp+2\lpp)
      {\cal D} \bar\psi(\lp+2\lpp)]
\exp\left\{ \left( \bar\psi(\lp+2\lpp), B_+^0
 \psi(\lp+2\lpp) \right) \right\}
\nonumber\\
&&
\hskip 16pt \times
\det \left( K_+^0 \right) \Big\vert_{[\lp,\lp+2\lpp]}
\exp\left\{ -\left(
        \bar\Psi^\prime, M_+^\prime \Psi^\prime
                 \right) \right\}
\bra{\psi_L(\lp+2\lp^\prime),\bar \psi_L(\lp+2\lp^\prime) }
b_\pm^\prime \Big\rangle
\ec
\end{eqnarray}
where
$\Psi^\prime(n)=( \psi(n,\lp),  \psi(n,\lp+2\lp^\prime) )^t $ and
$\bar \Psi^\prime(n)=( \bar \psi(n,\lp), \bar \psi(n,\lp+2\lpp) ) $,
\
$\Big\vert_{[\lp,\lp+2\lpp]}$ stands for the boundary condition
\begin{equation}
  \label{outside-boundary-condition}
  P_R \psi(n, \lp+2\lp^\prime)= P_L \psi(n, \lp)=0 \ec \hskip 16pt
 \bar \psi(n, \lp+2\lp^\prime) P_L =\bar \psi(n, \lp) P_R =0
\ec
\end{equation}
and
\begin{eqnarray}
  \label{outside-boundary-M-matrix}
&&M_+^\prime(n,m;\lp,\lp+2\lp^\prime)
\nonumber\\
&&
=
 \left(  \begin{array}{cc}
      \scriptstyle
 P_R S_+^0(n,\lp;m,\lp) P_L
+ P_R \vecsum \gamma_\mu \nabla_\mu^0(n,m)
     &\scriptstyle
P_R S_+^0(n,\lp;m,\lp+2\lp^\prime) P_R \\
\scriptstyle
     P_L S_+^0(n,\lp+2\lp^\prime;m,\lp) P_L
     & \scriptstyle
P_L S_+^0(n,\lp+2\lp^\prime;m,\lp+2\lp^\prime) P_R
+ P_L \vecsum \gamma_\mu \nabla_\mu^0(n,m)
         \end{array} \right)
\ec
\end{eqnarray}
\begin{equation}
  \label{inverse-Kinetic-term-rightoutside}
S^0_+(n,s;m,t) = K^{-1}(n,s;m,t) \Big\vert_{[\lp,\lp+2\lpp]}
= K^{-1}_+(n,s;m,t) \Big\vert_{[\lp,\lp+2\lpp]}
\ee
\end{equation}
The superscript $0$ stands for the quantities in which
the gauge link variables are set to unity.

At finite $\lp^\prime$, we can find the following
choice convenient:
\begin{equation}
    \label{choice-of-boundarystate}
  \ket{b_+^\prime} \, = \ket{0}  \ee
\end{equation}
In this choice,
the dependence of the wave function
in the variables $\psi_L(\lp)$, $\bar \psi_L(\lp)$
can be given explicitly as
\begin{eqnarray}
\bra{\psi_L(\lp),\bar \psi_L(\lp) }b_+ \rangle
&&
=
c_+
\exp \left\{ -\left(
\bar\psi(\lp),X_+^{\lp^\prime} \psi(\lp)
                                  \right) \right\}
\label{boundary-wave-function-left}
\ee
\end{eqnarray}
where
\begin{eqnarray}
  \label{breaking-source}
&&X_+^{\lp^\prime}(n,m)
\nonumber\\
&&
=
\scriptstyle
P_R S_+^0(n,\lp;m,\lp) P_L + P_R\vecsum\gamma_\mu \nabla^0_\mu(n,m)
\nonumber\\
&&
+
\scriptstyle
P_R S_+^0(\lp;\lp+2\lp^\prime) P_R
\left( B_+^0
  - P_L S_+^0(\lp+2\lp^\prime;\lp+2\lp^\prime) P_R
  - P_L \vecsum \gamma_\mu \nabla_\mu^0
\right)^{-1}
P_L S_+^0(\lp+2\lp^\prime;\lp) P_L (n,m)
\ec
\end{eqnarray}
and $c_+$ is a certain constant depending on $\lp$ and
$\lp+2\lp^\prime$.

We make a similar choice for the boundary state
at $s=\mlm$.
\begin{equation}
    \label{choice-of-boundarystate-left}
  \bra{b_-^\prime} \, = \bra{0}  \ec
\end{equation}
and we obtain
\begin{eqnarray}
\langle b_- \ket{\psi_R(\mlm),\bar \psi_R(\mlm) }
&&
=
c_-^\ast
\exp \left\{ -\left( \bar\psi(\mlm),X_-^{\lp^\prime}
                                \psi(\mlm) \right) \right\}
\label{boundary-wave-function-right}
\ec
\end{eqnarray}
where
\begin{eqnarray}
  \label{breaking-source-left}
&&X_-^{\lp^\prime}(n,m)
\nonumber\\
&&
=
\scriptstyle
P_L S_-^0(n,\mlm;m,\mlm) P_R + P_L\vecsum\gamma_\mu \nabla^0_\mu(n,m)
\\
&&
+
\scriptstyle
P_L S_-^0(\mlm-2\lp^\prime;\mlm) P_R
\nonumber\\
&&
\scriptstyle
\qquad \qquad
\times
\left( B_-^0
  - P_R S_-^0(\mlm-2\lp^\prime;\mlm-2\lp^\prime) P_L
  - P_R \vecsum \gamma_\mu \nabla_\mu^0
\right)^{-1}
P_R S_-^0(\mlm-2\lp^\prime;\mlm) P_R (n,m)
\ee \nonumber
\end{eqnarray}

\subsection{Would-be vacuum overlap formula
at finite extent of fifth dimension}
\label{sec:ol_at_finite_volume}

Given the explicit form of the boundary state wave functions,
the transition amplitude can be written as follows.
\begin{eqnarray}
  \label{boundary-state-amplitude}
&&
\bra{b_-}D_- \left(T_-\right)^L
             \left( T_+\right)^L D_+^\dagger \ket{b_+}
\prod_{0\leq s \leq \lp}(\det B_+)^{2}
\prod_{\mlm \leq s \leq -1}(\det B_-)^{2}
\nonumber\\
&&
=
\int \prod_{s=\mlm,\lp}
[{\cal D} \psi(s) {\cal D} \bar\psi(s)]
\exp\left\{ \sum_{s=\mlm,\lp}
\left( \bar\psi_s, B_s \psi_s \right) \right\}
\nonumber\\
&&
\times
\langle b_\pm \ket{\psi_R(\mlm),\bar \psi_R(\mlm)}
Z[\psi_L,\bar\psi_L (\mlm);\psi_R,\bar\psi_R (\lp)]
\bra{\psi_L(\lp),\bar \psi_L(\lp) }b_\pm \rangle
\nonumber\\
&&=
\det \left( K \right)
\Big\vert_{[-\lp,\lp]} \exp \Phi(b_-,b_+)
\ec
\end{eqnarray}
where the boundary contribution $\Phi(b_-,b_+)$ is given
\begin{eqnarray}
\exp \Phi(b_-,b_+)
&&
=
\int \prod_{s=\mlm,\lp}
[{\cal D} \psi_s {\cal D} \hat\psi_s]
\exp\left\{
\left( \bar \Psi, [B_{-+}-M-X_{-+}^{\lp^\prime}] \Psi \right)
\right\}
c_+ c_-^\ast
\nonumber\\
&&
\equiv
{\det}' \left( B_{-+}-M-X_{-+}^{\lp^\prime} \right)
c_+ c_-^\ast
\ec
\end{eqnarray}
with
\begin{equation}
  B_{-+}(n,m)= \left(
            \begin{array}{cc} B_-(n,m) & 0 \\ 0 & B_+(n,m)
            \end{array}
          \right)
\ec
\end{equation}
and
\begin{equation}
X_{-+}^{\lp^\prime}(n,m) =
\left(
        \begin{array}{cc} X_-^{\lp^\prime}(n,m) & 0 \\
                          0        & X_+^{\lp^\prime}(n,m)  \end{array} \right)
\ee
\end{equation}
Note that ${\det}'$ denotes the determinant over the
four-dimensional surfaces at $s=\lp$ and $\mlm$
besides over the indices of spinor and the representation of
gauge group.

It is also possible to write in a similar manner
the transition amplitudes for the five-dimensional
fermions with the positive and negative homogeneous
masses, which are needed for the subtraction scheme
proposed by Neuberger and Narayanan.
Then the subtracted transition amplitude at finite
$\lp$ and $\lp^\prime$ can be written in a factorized form as
\begin{eqnarray}
  \label{effective-action-at-finite-volume}
&&
\frac{\displaystyle
\bra{b_-}D_- \left(T_-\right)^L
             \left( T_+\right)^L D_+^\dagger \ket{b_+}}
{\displaystyle
\sqrt{ \bra{b_-}D_- \left(T_-\right)^{2L } D_-^\dagger \ket{b_-} }
\sqrt{ \bra{b_+}D_+ \left(T_+\right)^{2L}  D_+^\dagger \ket{b_+} }}
\nonumber\\
&&=\frac{\displaystyle \det \left( K \right) }
     {\sqrt{ \displaystyle
           \det \left( K_- \right) \det \left( K_+ \right)
            }}\bigg\vert_{[-\lp,\lp]}
 \exp\left\{ \Phi(b_-,b_+)
           -\half \Phi_+(b_+,b_+)
           -\half \Phi_-(b_-,b_-) \right\}
\nonumber\\
&&=\frac{\displaystyle \det \left( K \right) }
     {\sqrt{ \displaystyle
           \det \left( K_- \right) \det \left( K_+ \right)
            }}\bigg\vert_{[-\lp,\lp]}
\frac{\displaystyle {\det}' \left( B-M-X_{-+}^{\lp^\prime} \right) }
     {\sqrt{ \displaystyle
           {\det}' \left( B_--M_--X_{--}^{\lp^\prime} \right)
           {\det}' \left( B_+-M_+-X_{++}^{\lp^\prime} \right)
            }}
\ee
\end{eqnarray}
Since the phase factor
which comes from the constant $c_+$ and $c_-$,
$\frac{c_-^\ast c_+}{\vert c_- \vert \vert c_+ \vert }$,
does not depend on the gauge field and is irrelevant,
we have omitted it.
This is the expression from which we start our analysis
in the continuum limit.
Taking the limit $L^\prime \rightarrow \infty$ first
in the above formula,
we obtain the
would-be vacuum overlap formula at finite extent of the fifth dimension as
\begin{eqnarray}
  \label{effective-action-expression}
&&
\exp\left\{ -S_i[U_\mu;\lp] \right\}
\nonumber\\
&=&\frac{\displaystyle \det \left( K \right) }
     {\sqrt{ \displaystyle
           \det \left( K_- \right) \det \left( K_+ \right)
            }}\bigg\vert_{[-\lp,\lp]}
 \exp\left\{ \Phi(0_-,0_+)
           -\half \Phi_+(0_+,0_+)
           -\half \Phi_-(0_-,0_-) \right\}
\nonumber\\
&=&
\frac{\displaystyle \det \left( K \right) }
     {\sqrt{ \displaystyle
           \det \left( K_- \right) \det \left( K_+ \right)
            }}\bigg\vert_{[-\lp,\lp]}
\frac{\displaystyle {\det}' \left( B_{-+}-M-X_{-+}^\infty \right) }
     {\sqrt{ \displaystyle
           {\det}' \left( B_{--}-M_--X_{--}^\infty \right)
           {\det}' \left( B_{++}-M_+-X_{++}^\infty \right)
            }}
\ee
\nonumber\\
\end{eqnarray}

Finally we give the expression for the variation of the effective
action under the gauge transformation.
Under the gauge transformation of the link variables:
\begin{equation}
  U_\mu(n) \longrightarrow g(n) U_\mu(n) g^\dagger(n+\hat\mu)
\ec
\hskip 1cm
g(n) \in SU(N)
\ec
\end{equation}
we can easily see that
the matrices $K$, $B$ and $M$ transform covariantly.
For example,
\begin{equation}
  M(n,m) \longrightarrow  g(n) M(n,m) g^\dagger(m)
\ee
\end{equation}
On the contrary, $X^\infty$ does not transform covariantly because
it consists of $B^0$ and $M^0$ without gauge link
variables in them.
Therefore the variation of the effective action under the
infinitesimal gauge transformation with $g(n)=1+i\omega(n)$ is given by

\begin{eqnarray}
i\delta S_i[U_\mu;\lp]
&=&
{\Tr}'
\left\{
\left( B_{-+}-M-X_{-+}^\infty \right)^{-1}
\left( \omega X_{-+}^\infty
               - X_{-+}^\infty \omega \right)
\right\}
\nonumber\\
&&-\half
{\Tr}'
\left\{
\left( B_{++}-M_+-X_{++}^\infty \right)^{-1}
\left( \omega X_{++}^\infty
            - X_{++}^\infty \omega \right)
\right\}
\nonumber\\
&&-\half
{\Tr}'
\left\{
\left( B_{--}-M_--X_{--}^\infty \right)^{-1}
\left( \omega X_{--}^\infty
               - X_{--}^\infty \omega \right)
\right\}
\ec
  \label{effective-action-variation}
\end{eqnarray}
where ${\Tr}' $ denotes the trace over the
four-dimensional surfaces at $s=\lp$ and $\mlm$
besides over the indices of spinor and the representation of
gauge group.

\subsection{Continuum limit counterpart}
We start to investigate the would-be vacuum
overlap formula at finite extent of the fifth dimension
{\it in the continuum limit} and {\it in the Minkowski space}.
We first take the naive continuum limit of the action
with the boundary terms,
Eq.~(\ref{lattice-action-with-boundaryterm}),
(\ref{lattice-action-chiral-bc}),
(\ref{left-boundary-term}) and
(\ref{right-boundary-term}).

\begin{equation}
  \label{continuum-action-with-boundaryterm}
  S[\mlm,\lp]= S+S^B_L+S^B_R
\ec
\end{equation}

\begin{eqnarray}
S &=& \int^\infty_{-\infty}\!d^4x \int^L_{-L}\!ds \, \,
\bar\psi(x,s)
\big\{  i \gamma^\mu \left( \partial_\mu -igT^a A^a_\mu(x) \right)
-[\gamma_5 \partial_s  + M (s)] \big\}
\psi(x,s) \Big\vert_{[-\lp,\lp]}
\ec
\label{eqn:action-kink}
\\
S^B_L &=&  -\int^\infty_{-\infty}\!d^4x
\left\{
\bar\psi(x,\mlo) P_L  \psi(x,\ml)
+\bar\psi(x,\ml) P_R \psi(x,\mlo)
\right\}
\ec
\\
S^B_R &=&  -\int^\infty_{-\infty}\!d^4x
\left\{
\bar\psi(x,\lp) P_L  \psi(x,\lpo)
+\bar\psi(x,\lpo) P_R \psi(x,\lp)
\right\}
\ee
\end{eqnarray}
where $\Big\vert_{[-\lp,\lp]}$
stands for the chiral boundary condition
in the continuum limit:
\begin{equation}
  \label{chiral-boundary-condition-continuum-limit}
P_R \psi(x,\lp) = P_L \psi(x,\ml)= 0 \ec \hskip 16pt
\bar \psi(x,\lp) P_L = \bar \psi(x,\ml) P_R = 0  \ee
\end{equation}
Note that the kinetic parts in the boundary terms,
Eqs. (\ref{left-boundary-term}) and (\ref{right-boundary-term}),
vanish in the continuum limit because they correspond to
operators of dimension five.
The kink-like mass is defined by
\begin{eqnarray*}
M (s)
&\equiv& \left\{ \begin{array}{ll}
               +M & \quad s > 0 \\
               -M & \quad s < 0
    \end{array} \right. \nonumber\\
&=&M \, \epsilon(s) \nonumber\\
&=&M \, \int\!{d \omega \over 2\pi i }
  \biggl[\,  {1\over \omega-i 0} \,+\,  {1\over \omega+i0} \,
      \biggr]
      \, e^{i\omega s}.
\end{eqnarray*}
We assume that the Latin indices run from 0 to~4
and the Greek ones from 0 to~3.
The gamma matrices are defined as
$\{\gamma^a , \gamma^b\}= 2 \eta^{ab}$
with Minkowskian metric, $\eta^{ab}={\rm diag}.(+1,-1,-1,-1,-1)$.
We adopt the following chiral representation:
\begin{equation}
  \label{gamma-matrix}
\gamma^\mu=
    \left( \begin{array}{cc} 0 & \bar\sigma^\mu \\
                             \sigma^\mu & 0 \end{array} \right)
\hskip .5cm
\sigma^\mu = ( 1, \sigma_i) \ec
\bar\sigma^\mu = ( 1, -\sigma_i) \ec
\end{equation}
and
\begin{equation}
  \label{gamma-five}
  \gamma^{a=4} = i \gamma_5 = i^2 \gamma^0\gamma^1\gamma^2\gamma^3
=i \left( \begin{array}{cc} 1 & 0\\
                            0 & -1 \end{array} \right)
\ee
\end{equation}
$T^a$ are the hermitian generators of $SU(N)$ gauge group in a
certain representation.
We also denote the gauge potential
in the matrix form as $A_\mu(x)=-ig T^a A^a_\mu(x)$.

In the continuum limit,
$B$, $K$, its inverse, $M$, and $X$ are given formally by
\begin{eqnarray}
&&  B(x,y)=\delta^4(x-y) \ec \\
&&  K(x,s;y,t) =
\left\{
i \gamma^\mu \left( \partial_\mu +A_\mu(x) \right)
-[\gamma_5 \partial_s  + M (s)]
    \right\} \delta^4 (x-y) \delta (s-t)
\ec
\label{kinetic-term-continuum-limit}\\
&&S_{F}[A] (n,s;m,t) =
i K^{-1}(n,s;m,t) \Big\vert_{[\ml,\lp]}
\ec
\label{inverse-Kinetic-term-continuum-limit}\\
&&iM(x,y;\ml,\lp)
=
 \left(  \begin{array}{cc}
      \scriptstyle P_R S_F[A](x,\ml;y,\ml) P_L
      &\scriptstyle P_R S_F[A](x,\ml;y,\lp) P_R \\
      \scriptstyle P_L S_F[A](x,\lp;y,\ml) P_L
     & \scriptstyle P_L S_F[A](x,\lp;y,\lp) P_R
         \end{array} \right)
\ec
  \label{boundary-M-matrix-continuum-limit}
\end{eqnarray}
and
\begin{eqnarray}
iX_-^{\lp^\prime}(x,y)
&&=
\scriptstyle
P_L S_{F-}^0(x,\ml;y,\ml) P_R \\
&&
+
\scriptstyle
P_L S_{F-}^0(\ml-2\lp^\prime;\ml) P_L
\left( 1
  - P_R S_{F-}^0(\ml-2\lp^\prime;\ml-2\lp^\prime) P_L
\right)^{-1}
P_R S_{F-}^0(\ml-2\lp^\prime;\ml) P_R (x,y)
\ec
\label{breaking-source-left-continuum-limit}
\nonumber \\
iX_+^{\lp^\prime}(x,y)
&&=
\scriptstyle
P_R S_{F+}^0(x,\lp;y,\lp) P_L \\
&&
+
\scriptstyle
P_R S_{F+}^0(\lp+2\lp^\prime;\lp) P_R
\left( 1
  - P_L S_{F+}^0(\lp+2\lp^\prime;\lp+2\lp^\prime) P_R
\right)^{-1}
P_L S_{F+}^0(\lp+2\lp^\prime;\lp) P_L (x,y)
\ec
\label{breaking-source-right-continuum-limit}
\nonumber \\
X_{-+}^{\lp^\prime}(x,y)
&&=
\left(
        \begin{array}{cc} X_-^{\lp^\prime}(x,y) & 0 \\
                          0        & X_+^{\lp^\prime}(x,y)
\end{array} \right)
\label{breaking-source-total-continuum-limit}
\ee
\end{eqnarray}
$S_{F\pm}$ are the inverse of the Dirac operators of
the five dimensional fermions with positive and negative
homogeneous masses, taking into account of the
chiral boundary conditions,

\noindent
$\Big\vert_{[\lp, \lp+2\lp^\prime]}$ and
$\Big\vert_{[\ml-2\lp^\prime, \ml]}$
, respectively. The superscript $0$ stands for the quantities
in which gauge interaction is switched off.

Therefore we arrive at the formal expression of the
vacuum overlap formula at finite extent of the fifth dimension
($\lp^\prime$ is also kept finite) in the continuum limit.

\begin{equation}
  \label{overlap-formula-at-finite-fifth-volume-continuum-limit}
  \frac{\displaystyle \det \left( K \right) }
     {\sqrt{ \displaystyle
           \det \left( K_- \right) \det \left( K_+ \right)
            }}\bigg\vert_{[-\lp,\lp]}
\frac{\displaystyle {\det}' \left( 1-M-X_{-+}^{\lp^\prime} \right) }
     {\sqrt{ \displaystyle
           {\det}' \left( 1-M_--X_{--}^{\lp^\prime} \right)
           {\det}' \left( 1-M_+-X_{++}^{\lp^\prime} \right)
            }}
\ee
\end{equation}

This expression is formal because we do not yet specify the
regularization. In our continuum limit analysis, we
adopt {\it the dimensional regularization of
't Hooft and Veltman scheme}. That is,
we consider an extended space such that
the four dimensional space $(\mu=0,1,2,3)$
is extended to the D-dimensional one, but
keep the fifth space.
The gamma matrices follow the convention such that
\begin{equation}
  \label{gamma-matrix-convention-dimensional-reglarization}
\{ \gamma^\mu , \gamma^\nu \}= 2 \eta^{\mu\nu} \hskip .5cm
(\mu=0, 1, \ldots, D) \ec
\end{equation}

\begin{equation}
  \label{gamma-five-in-dimensional-regularization}
  \gamma^{a=4} = i \gamma_5 = i^2 \gamma^0\gamma^1\gamma^2\gamma^3
\ec\hskip 1cm
[ \gamma^\mu , \gamma^{a=4} ]= 0 \hskip .5cm
(\mu=4 \ldots, D) \ee
\end{equation}

Actually, in the following section,
we will find
in the evaluation of the contribution of the five-dimensional
determinant,

\begin{equation}
  \label{overlap-formula-determinant-contribution}
  \frac{\displaystyle \det \left( K \right) }
     {\sqrt{ \displaystyle
           \det \left( K_- \right) \det \left( K_+ \right)
            }}\bigg\vert_{[-\lp,\lp]}
\ec
\end{equation}
that the dimensional regularization
cannot maintain the chiral boundary condition.
Furthermore,
we also find that the subtraction by the determinants of
the fermions with the positive and negative homogeneous masses
just correspond to the subtraction by only {\it one} bosonic
Pauli-Villars-Gupta field.
Because of these facts, we obtain a gauge noninvariant result
for the vacuum polarization in four dimensions.
Since the volume contribution is expected to be gauge invariant
in the lattice regularization, our choice of the
dimensional regularization is not adequate in this case.

On the other hand, for the boundary contribution,
\begin{equation}
  \label{overlap-formula-boundary-contribution}
\frac{\displaystyle {\det}' \left( 1-M-X_{-+}^{\lp^\prime} \right) }
     {\sqrt{ \displaystyle
           {\det}' \left( 1-M_--X_{--}^{\lp^\prime} \right)
           {\det}' \left( 1-M_+-X_{++}^{\lp^\prime} \right)
            }}
\ec
\end{equation}
once we assume that
{\it the dimensional regularization preserves the cluster property},
we can obtain the following results:
The boundary contribution is purely odd-parity.
Its variation under the gauge transformation
gives the consistent form
of the gauge anomaly {\it in four dimensions},
which can be evaluated without any divergence and
is actually originated from the source of gauge noninvariance
at the boundary, but not due to the breaking of the chiral boundary
condition by the dimensional regularization.

In this sense, the continuum limit analysis with
the dimensional regularization cannot give
the whole structure of the vacuum overlap defined in
the lattice regularization.
But we think that we can see rather clearly in what way
the vacuum overlap formula could give the perturbative
properties of the chiral determinant in four dimension.

\section{
Five-dimensional fermion with kink-like mass \hskip 3cm
in a finite fifth space volume
}
\label{sec:fermion-at-finite-fifth-volume}

In this section,
we develop the theory of
free five-dimensional fermion in the finite fifth space volume.
We solve the field equation and obtain the complete set of
solutions. The field operator is defined by the mode expansion
and the propagator is derived.
The Sommerfeld-Watson transformation is introduced, by which
we rearrange the normal modes of the fifth momentum to be common
among the fermion with the kink-like mass
and the fermion with the positive(negative) homogeneous mass.

\subsection{Complete set of solutions}

We solve the free field equations of
the five dimensional fermion with kink-like mass
and positive(negative) homogeneous mass
under the chiral boundary condition,
which are derived from the actions,
\begin{eqnarray}
S_0 &=& \int^\infty_{-\infty}\!d^4x \int^L_{-L}\!ds \, \,
\bar\psi(x,s)
\left\{  i \gamma^\mu \partial_\mu
-[\gamma_5 \partial_s  + M (s)] \right\}
\psi(x,s) \Big\vert_{[-\lp,\lp]}
\ec
\\
S_\pm &=& \int^\infty_{-\infty}\!d^4x \int^L_{-L}\!ds \, \,
\bar\psi(x,s)
\left\{  i \gamma^\mu \partial_\mu
-[\gamma_5 \partial_s  \pm M \phantom{(s)}] \right\}
\psi(x,s) \Big\vert_{[-\lp,\lp]}
\ec
\end{eqnarray}
and the chiral boundary condition reads
\begin{equation}
P_R \psi(x,\lp) = P_L \psi(x,\ml)= 0 \ec \hskip 16pt
\bar \psi(x,\lp) P_L = \bar \psi(x,\ml) P_R = 0  \ee
\end{equation}
Note that, if
the parity transformation here is defined by
\begin{eqnarray}
  \label{parity-transformation}
\psi(x_0,x_i,s) \rightarrow
\psi^\prime (x^\prime_0,x^\prime_i,s^\prime)
\equiv \gamma^0 \psi(x_0,-x_i,-s)
\ec
\end{eqnarray}
both the action and the chiral boundary condition are
parity invariant for the fermion with the homogeneous
mass. For the fermion with the kink-like mass, the parity
transformation has the effect to change the signature of
the mass parameter $M$.
This is also true when the gauge field is introduced
provided that the gauge field transforms as
\begin{eqnarray}
\left(  A_0(x_0,x_i), A_i(x_0,x_i) \right) \rightarrow
\left(  A^\prime_0(x'_0,x'_i), A^\prime_i(x'_0,x'_i) \right)
\equiv
\left(  A_0(x_0,-x_i), -A_i(x_0,-x_i) \right)
\ec
\end{eqnarray}
under the parity transformation.

We treat
both $S_0$ and $S_+$ in a unified manner.
The suffix $\pm$ in the following denotes the solutions for
$S_0$ and $S_+$, respectively.
The solution for $S_-$ can be obtained by setting all the
mass $M$ to $- M$ in that for $S_+$.
We work in the momentum space for all dimensions. $\omega$ denotes
the fifth component of five momentum.

\vskip 16pt
\noindent
i) {\it Solution for $s>0$}

General solution in the region $s>0$ is given as follows:
\begin{equation}
\left( \begin{array}{l}
     \bar\sigma^\mu p_\mu \\
    -i \omega + M \\
\end{array}
\right) e^{-i\omega s}
\ec
\end{equation}
where
\begin{equation}
p^\mu p_\mu = \omega^2 + M^2 \, .
\label{dispersion-relation}
\end{equation}
We have denoted the two independent solutions in
the form of four-by-two matrix.
Then the solution satisfying the chiral boundary
condition at $s=\lp$ is given by
\begin{equation}
\left( \begin{array}{l}
     \bar \sigma^\mu p_\mu \\
    -i \omega + M \\
\end{array}
\right) e^{-i\omega (s-L)}
-
\left( \begin{array}{l}
     \bar \sigma^\mu p_\mu \\
    i \omega + M \\
\end{array}
\right) e^{i\omega (s-L)}
=
2i \left( \begin{array}{l}
     \bar \sigma^\mu p_\mu \sin \omega(L-s) \\
    - \omega \cos \omega(L-s) + M \sin \omega(L-s) \\
\end{array}
\right) \, ,
\end{equation}
where $\omega > 0 $.

\vskip 16pt
\noindent
ii) {\it Solution for $s<0$ }

Similarly, general solution in the region $s<0$ is given as follows:
\begin{equation}
\left( \begin{array}{l}
     i \omega \mp M \\
     \sigma^\mu p_\mu \\
\end{array}
\right) e^{-i\omega s}
\ee
\end{equation}
Then the solution satisfying the chiral boundary
condition at $s=\ml$ is given by
\begin{equation}
\left( \begin{array}{l}
     i \omega \mp M \\
     \sigma^\mu p_\mu \\
\end{array}
\right) e^{-i\omega (s+L)}
-
\left( \begin{array}{l}
     -i \omega \mp M \\
     \sigma^\mu p_\mu \\
\end{array}
\right) e^{i\omega (s+L)}
=
\frac{2}{i }
\left( \begin{array}{l}
    - \omega \cos \omega(L+s) \mp M \sin \omega(L+s) \\
     \sigma^\mu p_\mu \sin \omega(L+s) \\
\end{array}
\right) \, ,
\end{equation}
where $\omega > 0$.

\vskip 16pt
\noindent
iii) {\it Matching at $s=0$}

The solution should be continuous at $s=0$ and
this condition determines the normal modes of $\omega$.
The general solution satisfying the chiral boundary condition
can be written as follows:
\begin{eqnarray}
\phi(p,\omega;s) & \equiv &
C_{>0}
\left( \begin{array}{l}
     \bar \sigma^\mu p_\mu \sin \omega(L-s) \\
    - \omega \cos \omega(L-s) + M \sin \omega(L-s) \\
\end{array}
\right) \theta(s) \nonumber\\
&&\mbox +
C_{<0}
\left( \begin{array}{l}
    - \omega \cos \omega(L+s) \mp M \sin \omega(L+s) \\
     \sigma^\mu p_\mu \sin \omega(L+s) \\
\end{array}
\right)
\theta(-s) \, .
\end{eqnarray}
For both two components in the above $\phi(p,s)$ to match at $s=0$,
we should have
\begin{eqnarray}
C_{>0} &=& C \; \sigma^\mu p_\mu \sin \omega L  \, ,   \\
C_{<0} &=& C \; (-\omega \cos \omega L + M \sin \omega L ) \, ,
\end{eqnarray}
and
\begin{equation}
p^2 \sin^2 \omega L
= (\omega \cos \omega L - M \sin \omega L )
  (\omega \cos \omega L \pm M \sin \omega L ) \, .
\label{matching-condition}
\end{equation}

\vskip 16pt
\noindent
iv) {\it Spectrum of the normal modes of $\omega$}

{}From the dispersion relation Eq.~(\ref{dispersion-relation})
and the matching condition Eq.~(\ref{matching-condition}),
we can obtain the spectrum of the normal modes of $\omega$.

\begin{mathletters}
\begin{eqnarray}
(\omega^2 + M^2)   \sin^2 \omega L
&=& (\omega \cos \omega L - M \sin \omega L )
  (\omega \cos \omega L \pm M \sin \omega L ) \, ,
\label{mode-equation}
\end{eqnarray}
\end{mathletters}

For the fermion with kink-like mass term
we have
\begin{eqnarray}
\label{kink-mass-mode}
(\omega^2 + M^2)
&=&
M^2 \frac{1}{\cos 2 \omega L } \qquad ( \omega > 0) \, , \\
(-\lambda^2 + M^2)
&=&
M^2 \frac{1}{\cosh 2 \lambda L } \qquad
                               ( \lambda > 0 ; \omega=i\lambda) \, .
\label{light-mode-in-fermion-with-kink-like-mass}
\end{eqnarray}

For the fermion with ordinary positive mass
we have
\begin{eqnarray}
\label{homogeneous-mass-mode}
\omega
&=&
M \tan 2 \omega L  \qquad ( \omega > 0) \, ,
\\
\lambda
&=&
M \tanh 2 \lambda L  \qquad
( \lambda > 0, \omega =i \lambda; \mbox{only for}\ +M ) \, .
 \label{light-mode-in-fermion-with-positive-homogeneous-mass}
\end{eqnarray}
Note that both sets of solutions have the bounded modes
with the wave function which behave exponentially.

\vskip 16pt
\noindent
iv) {\it Solutions over $[-L,+L]$ }

Taking into account of the mode equation
Eq.~(\ref{mode-equation}), the solution over
the entire region can be rewritten as follows:
\begin{eqnarray}
\phi(p,\omega;s)
&=&
\left( \begin{array}{l}
    \bar \sigma^\mu p_\mu \sin \omega(L-s) \\
    - \omega \cos \omega(L-s) + M \sin \omega(L-s)
\\
\end{array}
\right) \theta(s) \nonumber\\
&&\mbox +
\frac{(-\omega \cos \omega L + M \sin \omega L )}
     {\sigma^\mu p_\mu \sin \omega L}
\left( \begin{array}{l}
    - \omega \cos \omega(L+s) \mp M \sin \omega(L+s) \\
     \sigma^\mu p_\mu \sin \omega(L+s) \\
\end{array}
\right)
\theta(-s) \\
&& =
\left( \begin{array}{l}
    \bar \sigma^\mu p_\mu \sin \omega(L-s) \\
    (-\omega \cot \omega L + M )
    \left(
    \frac{\omega \cos \omega(L-s) - M \sin \omega(L-s)}
         {\omega \cot \omega L - M} \right)  \\
\end{array}
\right) \theta(s) \nonumber\\
&&\qquad \qquad \qquad +
\left( \begin{array}{l}
 \bar \sigma^\mu p_\mu \left(
        \frac{\omega \cos \omega(L+s) \pm M \sin \omega(L+s)}
             {\omega \cot \omega L \pm M } \right) \\
     (-\omega \cot \omega L + M )\sin \omega(L+s) \\
\end{array}
\right)
\theta(-s) \\
&&=
\left( \begin{array}{l}
    \bar \sigma^\mu p_\mu \sin \omega(L-s) \\
    (-\omega \cot \omega L + M )
    \sin_- \omega(L+s) \\
\end{array}
\right) \theta(s) \nonumber\\
&& \qquad\qquad\qquad +
\left( \begin{array}{l}
 \bar \sigma^\mu p_\mu \sin_\pm \omega(L-s) \\
     (-\omega \cot \omega L + M )\sin \omega(L+s) \\
\end{array}
\right)
\theta(-s) \\
&&=
\left( \begin{array}{l}
    \bar \sigma^\mu p_\mu [\sin \omega(L-s)]_\pm \\
    (-\omega \cot \omega L + M )
    [\sin \omega(L+s)]_- \\
\end{array}
\right) \, ,
\end{eqnarray}
where we have defined
\begin{eqnarray}
\sin_\pm \omega(L-s)
&\equiv&
        \frac{\omega \cos \omega(L+s) \pm M \sin \omega(L+s)}
             {\omega \cot \omega L \pm M } \\
&=&
\left\{
\begin{array}{ll}
        \frac{\omega \cos \omega(L+s) + M \sin \omega(L+s)}
             {\omega \cot \omega L + M }
& \qquad(\omega^2+M^2=M^2 / \cos 2\omega L,\omega>0)
\\
\sin \omega(L-s) & \qquad (\omega=M \tan 2\omega L, \omega > 0) \\
\end{array} \right.
\end{eqnarray}
and we use abbreviations for the generalized  ``sin'' functions in
$[-L,+L]$ as follows,
\begin{equation}
[\sin \omega(L-s)]_\pm \equiv
\sin \omega(L-s) \theta(s) + \sin_\pm \omega(L-s) \theta(-s) \, .
\end{equation}

This generalized ``sin'' function satisfies the orthogonality:
\begin{equation}
\iLL\!ds \,
[\sin \omega(L-s)]_\pm [\sin \omega^\prime (L-s)]_\pm
= N_\pm (\omega) \delta_{\omega\omega^\prime}
\label{generalized-sin-function-orthgonarity}
\ee
\end{equation}
where the normalization factor $N_\pm(\omega)$ is given by
\begin{eqnarray}
N_\pm(\omega)
&=& \left[
         \Big(
\frac{(\omega\cot\omega L \pm M)+
              (\omega\cot\omega L - M)}
     {(\omega\cot\omega L \pm M)}
          \Big)
          \frac{1}{2}
\big( L-\frac{\sin 2\omega L}{2\omega} \big)
          +\frac{\sin^2\omega L}{(\omega\cot\omega L \pm M)}
           \right]
\nonumber\\
&\equiv& n_\pm(\omega) \frac{1}{(\omega\cot\omega L \pm M)}
\ee
\end{eqnarray}
For the fermion with kink-like mass term, it turns out to be
\begin{eqnarray}
N_+(\omega)
&=&
\frac{ \omega\cot\omega L }
     {(\omega\cot\omega L + M)}
\left[ L-
\frac{ \sin 4\omega L }
     {4 \omega \cos^2 \omega L}
           \right]
\ee
\end{eqnarray}
For the fermion with ordinary positive mass term, it reads
\begin{eqnarray}
N_-(\omega)
&=&
\left[ L-\frac{\sin4 \omega L }{4 \omega}
           \right]
\ee
\end{eqnarray}

Orthogonality of the general solutions over
the entire region, $\phi(p,\omega,s)$, is given as follows.
\begin{equation}
\iLL\!ds \phi^\dagger(p,\omega,s) \phi(p,\omega^\prime,s)
= 2 p_0 \, \bar\sigma^\mu p_\mu N(\omega) \delta_{\omega\omega^\prime}
\ee
\end{equation}
It is shown as
\begin{eqnarray*}
&& \iLL\!ds \phi^\dagger(p,\omega,s) \phi(p,\omega^\prime,s)
\nonumber\\
&=&
\bar \sigma^\mu p_\mu \bar \sigma^\mu p_\mu
\iLL \!ds \,
[\sin \omega(L-s)]_\pm [\sin \omega^\prime (L-s)]_\pm
\nonumber\\
&&
+(-\omega\cot\omega L + M)(-\omega^\prime \cot\omega^\prime L + M)
\iLL \!ds \,
[\sin \omega(L-s)]_- [\sin \omega^\prime (L-s)]_-
\nonumber\\
&=&
\left( (\bar \sigma^\mu p_\mu)^2 N_\pm(\omega)
+(-\omega\cot\omega L + M)(-\omega \cot\omega L + M) N_-(\omega)
\right)
\, \delta_{\omega\omega^\prime}
\nonumber\\
&=&
\left( (\bar \sigma^\mu p_\mu)^2
+(\omega\cot\omega L - M)(\omega \cot\omega L \pm M) \right)
\, N_\pm(\omega) \, \delta_{\omega\omega^\prime}
\nonumber\\
&=&
2 p_0 \, \bar\sigma^\mu p_\mu
N_\pm (\omega) \, \delta_{\omega\omega^\prime}
\ee
\end{eqnarray*}

Then the orthonormal positive- and negative-energy wave functions
can be obtained as
\begin{eqnarray}
u(p,\omega;s)
&=&
\frac{\left( 1+\frac{\sigma^\mu p_\mu}{-i\omega+M} \right)}
     {\sqrt{2(|p_0|+M)}}
\left( \begin{array}{l}
    \bar \sigma^\mu p_\mu [\sin \omega(L-s)]_\pm \\
    (-\omega \cot \omega L + M )
    [\sin \omega(L+s)]_- \\
\end{array}
\right)
\ec
\end{eqnarray}

\begin{eqnarray}
v(p,\omega;s)
&\equiv& u(-p,\omega;s) \nonumber\\
&=&
\frac{\left( 1-\frac{\sigma^\mu p_\mu}{-i\omega+M} \right)}
     {\sqrt{2(|p_0|+M)}}
\left( \begin{array}{l}
    -\bar \sigma^\mu p_\mu [\sin \omega(L-s)]_\pm \\
    (-\omega \cot \omega L + M )
    [\sin \omega(L+s)]_- \\
\end{array}
\right)
\ee
\end{eqnarray}
Note, for later use, that
\begin{eqnarray}
v(p_0,-\vec p,\omega;s)
&=&
\frac{\left( 1-\frac{\bar\sigma^\mu p_\mu}{-i\omega+M} \right)}
     {\sqrt{2(|p_0|+M)}}
\left( \begin{array}{l}
     - \sigma^\mu p_\mu [\sin \omega(L-s)]_\pm \\
    (-\omega \cot \omega L + M )
    [\sin \omega(L+s)]_- \\
\end{array}
\right)
\nonumber\\
&=&
\frac{\left( 1-\frac{\bar\sigma^\mu p_\mu}{-i\omega+M} \right)}
     {\sqrt{2(|p_0|+M)}}
\frac{- \sigma^\mu p_\mu}{\omega \cot \omega L \pm M}
\left( \begin{array}{l}
     (\omega \cot \omega L \pm M)[\sin \omega(L-s)]_\pm \\
     \bar \sigma^\mu p_\mu [\sin \omega(L+s)]_- \\
\end{array}
\right)
\nonumber\\
&=&
\frac{i\omega+M}{\omega \cot \omega L \pm M}
\frac{\left( 1-\frac{\sigma^\mu p_\mu}{i\omega+M} \right)}
     {\sqrt{2(|p_0|+M)}}
\left( \begin{array}{l}
     (\omega \cot \omega L \pm M)[\sin \omega(L-s)]_\pm \\
     \bar \sigma^\mu p_\mu [\sin \omega(L+s)]_- \\
\end{array}
\right)
\ee
\end{eqnarray}

\subsection{Mode expansion and Equal-time commutation relation}

We define the field operator by the mode expansion as follows.
\begin{eqnarray}
\psi(x,s) \equiv
\int\frac{d^3p}{(2\pi)^3} \sum_\omega \frac{1}{N_\pm(\omega)}
\frac{1}{\sqrt{2p_0}}
\left\{ b(\vec p,\omega) u(p,\omega;s) \, e^{-ipx}
       +d^\dagger(\vec p,\omega) v(p,\omega;s) \, e^{+ipx} \right\}
\ec
\nonumber\\
\label{mode-expansion}
\end{eqnarray}
where the canonical commutation relations are assumed as
\begin{eqnarray}
\left\{  b(\vec p,\omega) , b^\dagger(\vec q,\omega^\prime) \right\}
&=& \delta^3(\vec p-\vec q) \delta_{\omega \omega^\prime}  \, ,
\nonumber\\
\left\{  d(\vec p,\omega) , d^\dagger(\vec q,\omega^\prime) \right\}
&=& \delta^3(\vec p-\vec q) \delta_{\omega \omega^\prime}  \, ,
\end{eqnarray}
and other commutators vanish.
Then the equal-time commutation relation follows.
\begin{equation}
  \label{canonical-comulation-relation}
\left\{ \psi(x,s) , \psi^\dagger (y,t) \right\}\bigg\vert_{x^0=y^0}
=\delta^3(\vec x-\vec y)\Delta_\pm(s,t)
\ec
\end{equation}
where
$\Delta_\pm(s,t) $ is defined by

\begin{eqnarray}
\Delta_\pm(s,t)
&\equiv&
\sum_\omega\frac{1}{N_\pm(\omega)}
\frac{1}{2p_0}
\left\{
u(p_0,\vec p,\omega;s) u^\dagger(p_0, \vec p,\omega;t)
+ v(p_0,-\vec p,\omega;s) v^\dagger(p_0,-\vec p,\omega;t)
\right\}
\nonumber\\
&=&
\sum_\omega\frac{1}{n_\pm(\omega)}
\left( \begin{array}{l}
(\omega \cot \omega L \pm M)
[\sin \omega(L-s)]_\pm [\sin \omega(L-t)]_\pm \qquad\qquad 0 \\
0\qquad\qquad (\omega \cot \omega L - M)
     [\sin \omega(L+s)]_- [\sin \omega(L+t)]_- \\
\end{array}
\right)
\ee
\end{eqnarray}
We have used the following relation to obtain the above result.
\begin{eqnarray}
&& \frac{1}{2p_0}
\left\{
  u(p_0,\vec p,\omega;s) u^\dagger(p_0,\vec p,\omega;t)
+ v(p_0,-\vec p,\omega;s) v^\dagger(p_0,-\vec p,\omega;t)
\right\}
\nonumber\\
&=&
\frac{1}{2p_0 }
\left( \begin{array}{l}
    \sigma^\mu p_\mu [\sin \omega(L-s)]_\pm \\
    (-\omega \cot \omega L + M )
    [\sin \omega(L+s)]_- \\
\end{array} \right)
\left(\frac{\sigma^\mu p_\mu}{\omega^2+M^2}\right)
\nonumber\\
&&\qquad\qquad\qquad\qquad
\times
\left( \begin{array}{ll}
     \bar \sigma^\mu p_\mu [\sin \omega(L-t)]_\pm &
    (-\omega \cot \omega L + M ) [\sin \omega(L+t)]_- \\
\end{array} \right)
\nonumber\\
&&\mbox{}+
\frac{\omega \cot \omega L - M}
            {\omega \cot \omega L \pm M}
\frac{1}{2p_0}
\left( \begin{array}{l}
     (\omega \cot \omega L \pm M)[\sin \omega(L-s)]_\pm \\
     \bar\sigma^\mu p_\mu [\sin \omega(L+s)]_- \\
\end{array}
\right)
\left(\frac{\sigma^\mu p_\mu}{\omega^2+M^2}\right)
\nonumber\\
&&\qquad\qquad\qquad\qquad
\times
\left( \begin{array}{ll}
     (\omega \cot \omega L \pm M)[\sin \omega(L-t)]_\pm &
     \bar\sigma^\mu p_\mu [\sin \omega(L+t)]_- \\
\end{array}
\right)
\nonumber\\
&=&
\left( \begin{array}{ll}
[\sin \omega(L-s)]_\pm [\sin \omega(L-t)]_\pm & 0 \\
0& \frac{\omega \cot \omega L - M}
            {\omega \cot \omega L \pm M}
     [\sin \omega(L+s)]_- [\sin \omega(L+t)]_- \\
\end{array}
\right) \, .
\end{eqnarray}

\subsection{Propagator}

Once we have defined the field operator,
the two-point Green function can be obtained as follows.
We define the Green function by the time-ordered product as usual:
\begin{equation}
  \label{feynman-green-function}
S_{F\pm}(x-y;s,t) \equiv  \bra{0}T \psi(x,s) \bar\psi(y,t) \ket{0}
\ee
\end{equation}
Then we have
\begin{eqnarray}
S_{F\pm}(x-y;s,t)
&\equiv&
 \int\!\frac{d^4p}{i(2\pi)^4} \sum_\omega
\frac{1}{N_\pm(\omega)}
\frac{e^{-ip(x-y)}}{M^2+\omega^2-p^2-i \varepsilon}
s_\pm(p,\omega;s,t)
\\
&=&
 \int\!\frac{d^4p}{i(2\pi)^4} \sum_\omega
\frac{1}{n_\pm(\omega)}
\frac{e^{-ip(x-y)}}{M^2+\omega^2-p^2-i \varepsilon}
(\omega \cot \omega L \pm M) s_\pm(p,\omega;s,t)
\nonumber\\
&=&
 \int\!\frac{d^4p}{i(2\pi)^4} e^{-ip(x-y)}
\left\{
  P_R \fs p \Delta_{R\pm}(p;s,t)
+ P_L \fs p \Delta_{L-}(p;s,t) \right.
\nonumber\\
&&\qquad\qquad\qquad\qquad\qquad
\left. +P_R B_{RL\pm}(p;s,t) + P_L B_{LR\pm}(p;s,t) \right\}
\ec
\end{eqnarray}
where
\begin{eqnarray}
s_\pm(p,\omega;s,t)
&=& u(p,\omega;s) \, \bar u(p,\omega;t)
\nonumber\\
&=&
\left( \begin{array}{l}
     \bar \sigma^\mu p_\mu [\sin \omega(L-s)]_\pm \\
    (-\omega \cot \omega L + M )
    [\sin \omega(L+s)]_- \\
\end{array} \right)
\left(\frac{\sigma^\mu p_\mu}{\omega^2+M^2}\right)
\nonumber\\
&&\qquad\qquad\qquad
\times
\left( \begin{array}{ll}
    \bar  \sigma^\mu p_\mu [\sin \omega(L-t)]_\pm &
    (-\omega \cot \omega L + M ) [\sin \omega(L+t)]_- \\
\end{array} \right) \, \gamma^0
\nonumber\\
&=&
 P_R \fs p [\sin \omega(L-s)]_\pm [\sin \omega(L-t)]_\pm
\nonumber\\
&& \mbox{} + P_L \fs p [\sin \omega(L+s)]_- [\sin \omega(L+t)]_-
\left( \frac{\omega \cot \omega L - M}
            {\omega \cot \omega L \pm M} \right)
\nonumber\\
&&\mbox{}
+P_R (-\omega \cot \omega L + M)
       [\sin \omega(L-s)]_\pm [\sin \omega(L+t)]_-
\nonumber\\
&&\mbox{}
+P_L (-\omega \cot \omega L + M)
       [\sin \omega(L+s)]_- [\sin \omega(L-t)]_\pm
\ec
\end{eqnarray}
and

\begin{eqnarray}
\Delta_{R\pm}(p;s,t)
&=&
\sum_\omega
\frac{\left( \omega \cot \omega L \pm M \right)}{n_\pm(\omega)}
\frac{1}
{M^2+\omega^2-p^2-i \varepsilon}
[\sin \omega(L-s)]_\pm [\sin \omega(L-t)]_\pm
\ec
\label{propagator-right-function}
\\
\Delta_{L-}(p;s,t)
&=&
\sum_\omega
\frac{\left( \omega \cot \omega L - M \right)}{n_\pm(\omega)}
\frac{1}
{M^2+\omega^2-p^2-i \varepsilon}
[\sin \omega(L+s)]_- [\sin \omega(L+t)]_-
\ec
\label{propagator-left-function}
\\
B_{RL\pm}(p;s,t)
&=&
\sum_\omega
\frac{\left( \omega \cot \omega L \pm M \right)}{n_\pm(\omega)}
\frac{(-\omega \cot \omega L + M)}{M^2+\omega^2-p^2-i \varepsilon}
       [\sin \omega(L-s)]_\pm [\sin \omega(L+t)]_-
\ec
\label{propagator-right-to-left-function}
\\
B_{LR\pm}(p;s,t)
&=& B_{RL\pm}(p;t,s)
\nonumber\\
&=&
\sum_\omega
\frac{\left( \omega \cot \omega L \pm M \right)}{n_\pm(\omega)}
\frac{(-\omega \cot \omega L + M)}{M^2+\omega^2-p^2-i \varepsilon}
       [\sin \omega(L+s)]_- [\sin \omega(L-t)]_\pm
\label{propagator-left-to-right-function}
\ee
\end{eqnarray}

\subsection{Sommerfeld-Watson Transformation}

As we have shown in the previous subsection, the fermion
with kink-like mass has different normal modes of $\omega$
from those of
the fermion with positive (negative) homogeneous mass.
For the fermion with kink-like mass term, we have
Eq.~(\ref{kink-mass-mode}),
\begin{eqnarray*}
(\omega^2 + M^2)
&=&
M^2 \frac{1}{\cos 2 \omega L } \qquad ( \omega > 0) \, ,
\\
(-\lambda^2 + M^2)
&=&
M^2 \frac{1}{\cosh 2 \lambda L } \qquad
                               ( \lambda > 0 ; \omega=i\lambda) \, .
\end{eqnarray*}
On the other hand, for the
fermion with positive homogeneous mass, we have
Eq.~(\ref{homogeneous-mass-mode}),
\begin{eqnarray*}
\omega
&=&
M \tan 2 \omega L  \qquad ( \omega > 0) \, ,
\\
\lambda
&=&
M \tanh 2 \lambda L  \qquad
( \lambda > 0, \omega =i \lambda; \mbox{only for}\ +M ) \, .
\end{eqnarray*}
Therefore we encounter
the summations over the modes of $\omega$,
\begin{eqnarray*}
\sum_\omega \frac{1}{n_+(\omega)} F_+(\omega)
&=&
\sum_{\omega>0}
\frac{1}{n_+(\omega)} F_+(\omega)
  \biggl|_{\omega^2+M^2=M^2/\cos 2\omega L}
\; + \; \frac{1}{n_+(i\lambda)} F_+(i\lambda)
\biggl|_{\scriptstyle -\lambda^2+M^2=M^2/\cosh
2\lambda L}
\, ,
\\
\sum_\omega \frac{1}{n_-(\omega)} F_-(\omega)
&=&
\sum_{\omega>0}
\frac{1}{n_-(\omega)} F_-(\omega)
\biggl|_{\omega=M\tan 2\omega L}
\qquad\quad \; + \; \frac{1}{n_+(i\lambda)} F_-(i\lambda)
\biggl|_{\scriptstyle
\lambda=M\tan 2\lambda L}
\, ,
\end{eqnarray*}
where $F_\pm(\omega)$ are certain functions.
In order to
perform the subtraction at finite fifth space volume
$\lp$,  we need to rearrange the modes such that
the massive modes would be common.
To achieve it, we consider the Sommerfeld-Watson
transformation.

\subsubsection{General Case}

Let us consider the function
\begin{equation}
\frac{1}{\sin^2 \omega L \Delta_\pm(\omega)}\equiv
\frac{1}{
\sin^2 \omega L
\left[ (\omega^2 + M^2) -
(\omega \cot \omega L - M )
  (\omega \cot \omega L \pm M ) \right] } \, .
\end{equation}
This function has poles at the value of $\omega$
given by the mode equation Eq.~(\ref{mode-equation}),
\begin{equation}
(\omega^2 + M^2) \sin^2 \omega L
= (\omega \cos \omega L - M \sin \omega L )
  (\omega \cos \omega L \pm M \sin \omega L ) \, .
\nonumber
\end{equation}
Since we can show
\begin{eqnarray}
&&
\frac{\partial}{\partial \omega}
\left\{ \sin^2 \omega L \Delta_\pm(\omega) \right\}
\nonumber\\
&&=2 \cot\omega L \left\{ \sin^2 \omega L \Delta_\pm(\omega) \right\}
\nonumber\\
&&
+ 2 \omega
\left[
\Big(
    \frac{ (\omega\cot\omega L \pm M)
          +(\omega\cot\omega L - M) }{2} \Big)
          \big( L-\frac{\sin 2\omega L}{2\omega}  \big)
          +\sin^2\omega L     \right]
\ec
\end{eqnarray}
the residue at the pole is given as
\begin{equation}
\frac{1}{2 \omega } \frac{1}{n_\pm(\omega)} \, .
\end{equation}
For the case $\omega=i\lambda$, this expression also holds true.

Accordingly we are declined to consider the following integral.
\begin{eqnarray}
I_\pm &\equiv &
\int_C
\!\frac{d\omega}{2\pi i}
\frac{2 \omega }{\sin^2 \omega L \Delta_\pm (\omega)}
F_\pm (\omega) \, ,
\label{integration-for-sommerfeld-watson-transformation}
\end{eqnarray}
with a contour $C$ shown bellow.

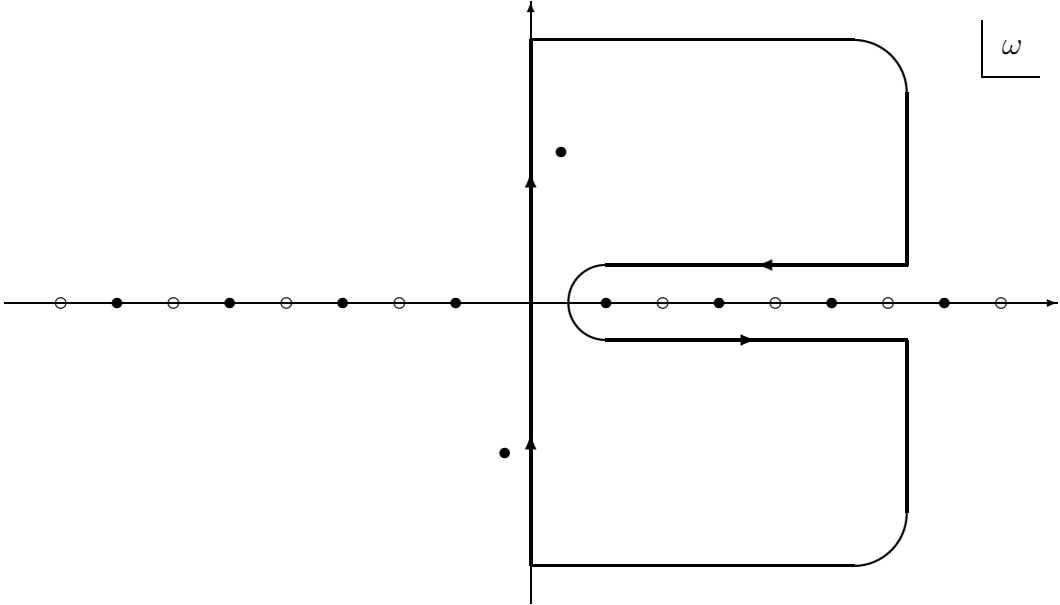
\unitlength 5mm
\begin{picture}(20,20)
\put(1,1){
\begin{picture}(101,0)
\put(12,0){\vector(0,1){16}}
\put(-2,8){\vector(1,0){28}}
\put(24,14){\line(1,0){1.5}}
\put(24,14){\line(0,1){1.5}}
\put(24.8,14.8){\makebox(0,0){$\omega$}}
\put(12.8,12){\circle*{0.3}}
\put(11.3,4){\circle*{0.3}}
\multiput(14,8)(3,0){4}{\circle*{0.3}}
\multiput(10,8)(-3,0){4}{\circle*{0.3}}
\multiput(15.5,8)(3,0){4}{\circle{0.3}}
\multiput(8.5,8)(-3,0){4}{\circle{0.3}}
\thicklines
\put(12,1){\vector(0,1){3.5}}
\put(12,4.5){\line(0,1){3.5}}
\put(12,8){\vector(0,1){3.5}}
\put(12,11.5){\line(0,1){3.5}}
\put(12,9){\oval(20,12)[rt]}
\put(12,7){\oval(20,12)[rb]}
\put(14,8){\oval(2,2)[l]}
\put(22,9){\vector(-1,0){4}}
\put(18,9){\line(-1,0){4}}
\put(14,7){\vector(1,0){4}}
\put(18,7){\line(1,0){4}}
\end{picture}
}
\end{picture}

We assume that $F_\pm$ is an even function of $\omega$,
it vanishes at the origin so that
the whole integrand does not possess any singularity at the origin,
and it also vanishes at infinity so that
the contour integral at infinity vanishes.
We allow $F_\pm(\omega)$ to have poles, for example,
on the real axis. We showed them by white-circles.

$\displaystyle \frac{2\omega}{\sin^2 \omega L \Delta_\pm (\omega)}$
has poles
on the real axis and two additional poles near the imaginary axis.
For sufficiently large $L$, we have
\begin{equation}
\omega=i\lambda \simeq iM + \varepsilon \, .
\end{equation}
Here we take into account of the Feynman boundary condition,
that is, the infinitesimal imaginary part of the mass M.
We showed these poles by black-circles. It also has a pole at
the origin, but it is assumed not to contribute the integral
due to the zero of $F_\pm$.

Then the above integral
leads to the identity
\begin{eqnarray}
\sum_\omega \frac{1}{n_\pm(\omega)} F_\pm(\omega)
= -\sum_{\omega^\prime}
\frac{2 \omega^\prime }{\sin^2 \omega^\prime L }
\frac{1}{\Delta_\pm (\omega^\prime)} \mbox{Res} F_\pm(\omega^\prime)
\biggl|_{\mbox{poles of $F_\pm$}}
\label{sommerfeld-watson-transformation-first}
\ee
\end{eqnarray}
If $F_\pm(\omega)$ possess common series of poles,
we can rearrange the different series of modes into common series of
modes. In some cases, it turns out to need
several steps of the transformations
to obtain the common series of modes.

\subsubsection{Transformation of the functions
$\Delta_{R\pm}(p;s,t)$ and $\Delta_{L-}(p;s,t)$}

\paragraph{Integrand in $\Delta_{R\pm}$}
 $\Delta_{R\pm}$ given by
Eq.~(\ref{propagator-right-function}) has the following Integrand.
\begin{eqnarray}
&&F_\pm(\omega)
\nonumber\\
&=&
\frac{1}
{M^2+\omega^2-p^2-i \varepsilon}
[\sin \omega(L-s)]_\pm [\sin \omega(L-t)]_\pm \,
      (\omega \cot \omega L \pm M)
\nonumber\\
&=&
\frac{1}
{M^2+\omega^2-p^2-i \varepsilon}
\nonumber\\
&& \times
\left\{ \,
\theta(s)\theta(t)  \,
\sin \omega(L-s) \, \sin \omega(L-t) \, (\omega \cot \omega L \pm M)
\right.
\nonumber\\
&&\quad
+\theta(-s)\theta(-t)  \,
\frac{
(\omega \cos \omega(L+s) \pm M \sin \omega(L+s) ) \,
(\omega \cos \omega(L+t) \pm M \sin \omega(L+t) ) \,
}{\omega \cot \omega L \pm M}
\nonumber\\
&&\quad
+\theta(s)\theta(-t)  \,
\sin \omega(L-s) \, (\omega \cos \omega(L+t) \pm M \sin \omega(L+t) )
\nonumber\\
&&\quad \left.
+\theta(-s)\theta(t)  \,
(\omega \cos \omega(L+s) \pm M \sin \omega(L+s) ) \,\sin \omega(L-t)
\right\}
\ee
\end{eqnarray}

\paragraph{Poles}
There are three types of poles in the above $F_\pm(\omega)$.
We summarize the poles and their residues in Table I.
\vskip 8pt

\centerline{Talbe I \
Poles and Residues in $I_\pm$ for $\Delta_{R\pm}$}
\begin{tabular}{|c|c|c|c|}
\hline
\phantom{aa} singular part in $I_\pm$ \phantom{aa}
&
\phantom{aaaaa} pole \phantom{aaaaaaaaaaa}
& \phantom{a} residue  \phantom{aaaaa}
&
${\scriptstyle
\frac{2 \omega}{\sin^2 \omega L \Delta_\pm (\omega) }} $
\phantom{aaaaaa}
\\
\hline
${\scriptstyle
\frac{2 \omega}{\sin^2 \omega L \Delta_\pm (\omega) }} $
& $\sin^2 \omega L \Delta_\pm (\omega)=0 $
& $\frac{1}{n_\pm(\omega)}$
& 1
\\
\hline\hline
  $\omega\cot\omega L=\frac{\omega\cos\omega L}{\sin\omega L}$
& $ \sin\omega L=0 $
& $ \frac{\omega}{L}$
& $-\frac{2}{\omega}$
\\
\hline
  $\frac{1}{(\omega \cot \omega L \pm M)} $
& $\omega \cot \omega L \pm M=0$
& $-\frac{\sin^2 \omega L}
         {\omega \left(L- \frac{\sin2\omega L}{2\omega}\right)}$
& $\frac{2}{\omega}$
\\
\hline
 $  \frac{1}{M^2+\omega^2-p^2-i \varepsilon} $
&$iP \equiv i\sqrt{M^2-p^2-i \varepsilon}$
&$\frac{1}{2iP}$
&$\frac{2 iP}{-\sinh^2 PL }\frac{1}{\Delta_\pm (iP)} $
\\
\hline
\end{tabular}
\vskip 16pt

\paragraph{First Stage Transformation}
By the Sommerfeld-Watson transformation given by
Eq.~(\ref{sommerfeld-watson-transformation-first}),
$\Delta_{R\pm}$ can be written as
\noindent
\baselineskip 21pt
\begin{eqnarray}
\Delta_{R\pm}(p;s,t)
&=&
\sum_\omega
\frac{\left( \omega \cot \omega L \pm M \right)}{n_\pm(\omega)}
\frac{1}
{M^2+\omega^2-p^2-i \varepsilon}
[\sin \omega(L-s)]_\pm [\sin \omega(L-t)]_\pm
\nonumber\\
&=&
\theta(s)\theta(t)  \,
\sum_{\scriptstyle \sin \omega L }
\frac{2}{L}\,
\frac{1}{M^2+\omega^2-p^2-i \varepsilon} \,
\sin \omega s \, \sin \omega t
\nonumber\\
&&+
\theta(-s)\theta(-t)  \,
\sum_{\scriptstyle \omega \cot \omega L \pm M}
\frac{2 \sin^2 \omega L}
{\omega^2 \left(L- \frac{\sin2\omega L }{2\omega} \right)} \,
\frac{1}{M^2+\omega^2-p^2-i \varepsilon} \,
\nonumber\\
&&\qquad
\times (\omega \cos \omega(L+s) \pm M \sin \omega(L+s) ) \,
(\omega \cos \omega(L+t) \pm M \sin \omega(L+t) ) \,
\nonumber\\
&&
- \frac{(P \coth P L \pm M)}{\Delta_\pm (iP)} \,
\frac{ [\sinh P(L-s)]_\pm [\sinh P(L-t)]_\pm } {\sinh^2 PL }
\nonumber\\
&=&
\theta(s)\theta(t)  \,
\sum_{\scriptstyle \sin \omega L }
\frac{2}{L}\,
\frac{1}{M^2+\omega^2-p^2-i \varepsilon} \,
\sin \omega s \, \sin \omega t
\nonumber\\
&&+
\theta(-s)\theta(-t)  \,
\sum_{\scriptstyle \omega \cot \omega L \pm M}
\frac{2}{\left(L- \frac{\sin2\omega L }{2\omega} \right)} \,
\frac{1}{M^2+\omega^2-p^2-i \varepsilon} \,
\sin \omega s \, \sin \omega t
\nonumber\\
&&
- \frac{(P \coth P L \pm M)}{\Delta_\pm (iP)} \,
\frac{ [\sinh P(L-s)]_\pm [\sinh P(L-t)]_\pm } {\sinh^2 PL }
\ec
\end{eqnarray}
\baselineskip=16pt
where we have used the relation,
\begin{equation}
\frac{\sin \omega L}{\omega}
(\omega \cos \omega(L+s) \pm M \sin \omega(L+s) )
= - \sin \omega s
\ec
\end{equation}
for $\omega$ satisfying $\omega \cot \omega L \pm M=0 $

\paragraph{Second Stage Transformation} We need further
Sommerfeld-Watson transformation for the part
in which the summation should be taken over
the different normal modes of $\omega$
given by $\omega \cot \omega L \pm M=0 $,
\begin{equation}
{\sum_\omega}\frac{2 }
{\left(L- \frac{\sin2\omega L }{2\omega} \right) } \;
\frac{1}{M^2+\omega^2-p^2-i \varepsilon} \,
\sin \omega s \sin \omega t \;
\biggl|_{\omega \cos \omega L \pm M\sin\omega L=0}
\ee
\end{equation}
For this purpose, we consider the following integral.
\begin{equation}
J_\pm \equiv
\int_C
\!\frac{d\omega}{2\pi i}\;
\frac{-2\omega }{\sin \omega L } \;
\frac{1}{\omega \cos \omega L \pm M \sin \omega L } \;
\frac{1}{M^2+\omega^2-p^2-i \varepsilon} \,
\sin \omega s \sin \omega t
\ee
\end{equation}
There are two types of poles in this case, which we summarized in
Table II.

\vskip 16pt
\centerline{Table II \
Poles and Residues in $J_\pm$ for $\Delta_{R\pm}$}
\begin{tabular}{|c|c|c|c|}
\hline
\phantom{aa} singular part in $J_\pm$ \phantom{aa}
&
\phantom{aaaaaa} pole \phantom{aaaaaaaaaaaa}
&
\phantom{a} residue \phantom{aaaaa}
&
${\scriptstyle
\frac{-2\omega }{\sin^2 \omega L } \;
\frac{1}{\omega \cot \omega L \pm M }
} $
\phantom{aaa}
\\
\hline
  ${\scriptstyle
\frac{-2\omega }{\sin \omega L } \;
\frac{1}{\omega \cot \omega L \pm M \sin \omega L } \;
} $
& $\omega \cos \omega L \pm M \sin \omega L =0 $
& ${ \scriptstyle
\frac{2}{\left(L- \frac{\sin2\omega L }{2\omega} \right) } }$
& 1
\\
\hline\hline
$\frac{-2\omega }{\sin \omega L } \;
\frac{1}{\omega \cot \omega L \pm M \sin \omega L } $
& $ \sin\omega L=0 $
& $ -\frac{2}{L}$
& 1
\\
\hline
 $  \frac{1}{M^2+\omega^2-p^2-i \varepsilon} $
&$iP \equiv i\sqrt{M^2-p^2-i \varepsilon}$
&$\frac{1}{2iP}$
&$\frac{2 iP}{\sinh^2 PL }\frac{1}{P\coth PL \pm M} $
\\
\hline
\end{tabular}
\vskip 16pt
\noindent
By the Sommerfeld-Watson transformation at the second stage,
we obtain
\begin{eqnarray}
&&{\sum_\omega}\frac{2 }
{\left(L- \frac{\sin2\omega L }{2\omega} \right) } \;
\frac{1}{M^2+\omega^2-p^2-i \varepsilon} \,
\sin \omega s \sin \omega t \,
\biggl|_{\omega \cos \omega L \pm M\sin\omega L=0}
\nonumber\\
&&={\sum_{\sin \omega L}}\frac{2}{L}
\frac{1}{M^2+\omega^2-p^2-i \varepsilon} \,
\sin \omega s \sin \omega t
+
\frac{1}{P \coth P L \pm M}
\frac{ \sinh Ps \sinh P t}{ \sinh^2 PL}
\ee
\end{eqnarray}

\paragraph{Final result} The final form of $\Delta_{R\pm}$ is
given by

\begin{eqnarray}
\Delta_{R\pm}(p;s,t)
&=&
\sum_\omega
\frac{\left( \omega \cot \omega L \pm M \right)}{n_\pm(\omega)}
\frac{1}
{M^2+\omega^2-p^2-i \varepsilon}
[\sin \omega(L-s)]_\pm [\sin \omega(L-t)]_\pm
\nonumber\\
&=&
[ \theta(s)\theta(t) + \theta(-s)\theta(-t)]  \,
\sum_{\sin\omega L}
\frac{2}{L}\,
\frac{1}{M^2+\omega^2-p^2-i \varepsilon} \,
\sin \omega s \, \sin \omega t \,
\nonumber\\
&&+
 \frac{(P \coth P L \pm M)}{-\Delta_\pm (iP)} \,
\frac{ [\sinh P(L-s)]_\pm [\sinh P(L-t)]_\pm } {\sinh^2 PL }
\nonumber\\
&&+
\theta(-s)\theta(-t)  \,
\frac{1}{P \coth P L \pm M}
\frac{ \sinh Ps \sinh P t}{ \sinh^2 PL}
\ee
\label{delta-R-sommerfeld-watson-transformed}
\end{eqnarray}

Similarly we obtain,
\begin{eqnarray}
\Delta_{L-}(p;s,t)
&=&
\sum_\omega
\frac{\left( \omega \cot \omega L - M \right)}{n_\pm(\omega)}
\frac{1}
{M^2+\omega^2-p^2-i \varepsilon}
[\sin \omega(L+s)]_- [\sin \omega(L+t)]_-
\nonumber\\
&=&
[\theta(s)\theta(t) + \theta(-s)\theta(-t)]  \,
\sum_{\sin\omega L}
\frac{2}{L}\,
\frac{1}{M^2+\omega^2-p^2-i \varepsilon} \,
\sin \omega s \, \sin \omega t \,
\nonumber\\
&&+
 \frac{(P \coth P L - M)}{-\Delta_\pm (iP)} \,
\frac{ [\sinh P(L+s)]_- [\sinh P(L+t)]_- } {\sinh^2 PL }
\nonumber\\
&&+
\theta(s)\theta(t)  \,
\frac{1}{P \coth P L -M}
\frac{ \sinh Ps \sinh P t}{ \sinh^2 PL}
\ee
\label{delta-L-sommerfeld-watson-transformed}
\end{eqnarray}

We can perform the similar transformation for
$B_\pm(p;s,t)$.
In this case, however, we need to improve the convergence
of the integration of $\omega$, using the relation
\begin{equation}
\left( \omega \cot \omega L \pm M \right)
(\omega \cot \omega L - M)
= \omega^2+M^2 = \frac{M^2}{ \cos^{(3\mp 1)/2} 2\omega L}
\ee
\label{mass-term-improve-convergence}
\end{equation}

\section{Perturbation Theory at Finite Extent of Fifth Dimension}
\label{sec:perturbation-at-finite-extent-of-fifth-dimension}

In this section,
we formulate the perturbative expansion of
the would-be vacuum overlap based on the theory
of {\it free} five-dimensional fermion
at finite extent of the fifth dimension,
which takes into account of the chiral boundary condition.
The expansion can be performed independently for
the five-dimensional volume contribution and
the four-dimensional boundary contribution.
As a subsidiary regularization,
we adopt the dimensional regularization.

As to the volume contribution,
the subtraction can be performed at {\it finite $L$},
thanks to the Sommerfeld-Watson
transformation, in each order of the expansion.
And then the limit of the infinite $L$ can be evaluated.

As to the boundary contribution,
we first derive the boundary state wave functions taking the limit
$L^\prime \rightarrow \infty$.
Since the boundary contribution is given by
the correlation between the boundaries, it is expected to be
finite in the limit $L \rightarrow \infty$
and the subtraction to be irrelevant.
As far as we do not take into account of the breaking
of the chiral boundary condition due to the dimensional regularization,
we can see that it is actually the case and the cluster property
holds:
the boundary contribution consist of the sum of the contributions
from the two boundaries and that of
the fermion with kink-like mass can be replaced by that of the fermion
with homogeneous positive(negative) mass.
Then {\it we make an assumption that this cluster property holds even under
the dimensional regularization}.
{}From the cluster property and the parity invariance of the fermion with
the homogeneous mass, we can show that
the boundary contribution is odd-parity in the limit
$L \rightarrow \infty$.

\subsection{Perturbation expansion of the determinant of $K$}

The perturbative expansion of the volume contribution,
Eq.~(\ref {overlap-formula-determinant-contribution}),
\begin{equation}
  \frac{\displaystyle \det \left( K \right) }
     {\sqrt{ \displaystyle
           \det \left( K_- \right) \det \left( K_+ \right)
            }}\bigg\vert_{[-\lp,\lp]}
\nonumber
\ec
\end{equation}
can be performed as follows.
\begin{eqnarray}
  \label{perurbative-expansion-of-detK}
\ln \det \left( K \right)\bigg\vert_{[-\lp,\lp]}
&=&
\ln \det \left(
\left\{1+i\fs A \, (K^0)^{-1}\big\vert_{[-\lp,\lp]} \right\}
K^0 \right)\bigg\vert_{[-\lp,\lp]}
\\
&=&
\sum_{n=1}^{\infty}\frac{(-i)^n }{n}
\Tr \left\{ \fs A \, (K^0)^{-1}\big\vert_{[-\lp,\lp]} \right\}^n
+ \ln \det \left( K^0 \right)\bigg\vert_{[-\lp,\lp]}
\\
&=&
\sum_{n=1}^{\infty}\frac{(-)^n}{n}
\Tr \left\{ \fs A \, S_{F+}\big\vert_{[-\lp,\lp]} \right\}^n
+\ln \det \left( K^0 \right)\bigg\vert_{[-\lp,\lp]}
\ee
\end{eqnarray}
This expansion should be evaluated with the propagator $S_{F+}$
supplemented by the dimensional regularization.
Similar expansion can be performed for the
determinant of $K_\pm$. The subtraction of them can be
performed at each order of the expansion. Thus we have
\begin{eqnarray}
  \label{perurbative-expansion-of-volume-contribution}
&& \ln \left[
\frac{\displaystyle \det \left( K \right) }
     {\sqrt{ \displaystyle
           \det \left( K_- \right) \det \left( K_+ \right)
            }}\bigg\vert_{[-\lp,\lp]}
\right]
-\ln \left[
\frac{\displaystyle \det \left( K^0 \right) }
     {\sqrt{ \displaystyle
           \det \left( K_-^0 \right) \det \left( K_+^0 \right)
            }}\bigg\vert_{[-\lp,\lp]}
\right]
\nonumber\\
&&
=
\sum_{n=1}^{\infty}\frac{(-)^n}{n}
\left[
\Tr \left\{ \fs A \, S_{F+}\big\vert_{[-\lp,\lp]} \right\}^n
-\half\Tr \left\{ \fs A \, S_{F-}(+M)\big\vert_{[-\lp,\lp]} \right\}^n
-\half\Tr \left\{ \fs A \, S_{F-}(-M)\big\vert_{[-\lp,\lp]} \right\}^n
\right]
\nonumber\\
&&
\equiv i \Gamma_K[A]
\ee
\end{eqnarray}
After the subtraction,
the limit of the infinite $L$ can be evaluated.
Explicit calculation of the vacuum polarization is given
in the later section.

\subsection{Perturbation expansion of the boundary term}

Next we consider the perturbative expansion of the
boundary contribution,
Eq.~(\ref{overlap-formula-boundary-contribution}),
\begin{equation}
\frac{\displaystyle {\det}' \left( 1-M-X_{-+}^{\lp^\prime} \right) }
     {\sqrt{ \displaystyle
           {\det}' \left( 1-M_--X_{--}^{\lp^\prime} \right)
           {\det}' \left( 1-M_+-X_{++}^{\lp^\prime} \right)
            }}
\ee
\nonumber
\end{equation}

\subsubsection{Boundary state wave function}

We first derive boundary state wave functions taking the limit
$L^\prime \rightarrow \infty$.
We have derived the propagator of the fermion with
positive and negative homogeneous masses satisfying
the chiral boundary condition $\Big\vert_{[\ml,\lp]} $
in the previous section.
Since the translational invariance hold for the
fermion with homogeneous mass, we can show
\begin{eqnarray}
&&S_{F-}[+ M](p;s+2\lpp,t+2\lpp) \bigg\vert_{[\lp,\lp+2\lpp]}
\phantom{\scriptstyle --}
=  S_{F-}[+ M](p;s,t) \bigg\vert_{[-\lpp,\lpp]}
\ec
\\
&&  S_{F-}[-M](p;s-2\lpp,t-2\lpp) \bigg\vert_{[\ml-2\lpp,\ml]}
=  S_{F-}[-M](p;s,t) \bigg\vert_{[-\lpp,\lpp]}
\ee
\end{eqnarray}
Then, using the previous results,
we can give the explicit form of the boundary state wave function
Eqs.(\ref{boundary-wave-function-left}) and
(\ref{boundary-wave-function-right}) and
the explicit form of $X_\pm^\lpp$,
Eqs.(\ref{breaking-source-left-continuum-limit}),
    (\ref{breaking-source-right-continuum-limit}) and
    (\ref{breaking-source-total-continuum-limit}).
Actually we have
\begin{eqnarray}
&& iS_{F-}[+M](p;\lpp,\lpp)
\nonumber\\
&=&
P_L \fs p \Delta_{L-}(p;\lpp,\lpp)
\nonumber\\
&=&
P_L \fs p \,
\sum_\omega
\frac{1}{n_-(\omega)}
\frac{1}{M^2+\omega^2-p^2-i \varepsilon}
\frac{\omega^2}{\left( \omega \cot \omega L^\prime - M \right)}
\nonumber\\
&=&
P_L \fs p \,
 \frac{1}{-\Delta_- (iP)} \,
\frac{P^2 }{(P \coth P L^\prime - M)}
\frac{1}{\sinh^2 PL^\prime }
+P_L \fs p \,
\frac{1}{P \coth P L' -M}
\ee
\end{eqnarray}
In the last equality, we have used the result of
the Sommerfeld-Watson transformation,
Eq.~(\ref{delta-L-sommerfeld-watson-transformed}).
Similarly, we obtain
\begin{eqnarray}
&& iS_{F-}[+M](p;-\lpp,-\lpp)
\nonumber\\
&=&
P_R \fs p \,
 \frac{1}{-\Delta_- (iP)} \,
\frac{P^2 }{(P \coth P L^\prime - M)}
\frac{1}{\sinh^2 PL^\prime }
+P_R \fs p \,
\frac{1}{P \coth P L' -M}
\ee
\end{eqnarray}

\begin{eqnarray}
 iS_{F-}[+M](p;\lpp,-\lpp)
&=&
P_L B_{LR-}(p;\lpp,-\lpp)
\nonumber\\
&=&
P_L \,
\sum_\omega
\frac{1}{n_-(\omega)}
\frac{1}{M^2+\omega^2-p^2-i \varepsilon}
(-\omega^2)
\nonumber\\
&=&
P_L \,
\sum_\omega
\frac{1}{n_-(\omega)}
\frac{1}{M^2+\omega^2-p^2-i \varepsilon}
(M^2-\frac{M^2}{ \cos^2 2\omega L'})
\nonumber\\
&=&
P_L \,
\frac{1}{-\sinh^2 PL^\prime \Delta_- (iP)} \,
(M^2-\frac{M^2}{ \cosh^2 2P L^\prime})
\label{tree-boundary-correlation-LR}
\ee
\end{eqnarray}

\begin{eqnarray}
 iS_{F-}[+M](p;-\lpp,\lpp)
&=&
P_R \,
\frac{1}{-\sinh^2 PL^\prime \Delta_- (iP)} \,
(M^2-\frac{M^2}{ \cosh^2 2P L^\prime})
\label{tree-boundary-correlation-RL}
\ee
\end{eqnarray}
The calculation of
Eqs.~(\ref{tree-boundary-correlation-LR})
and (\ref{tree-boundary-correlation-RL}) is given
in the later subsection.
Assuming that $p^2 \not =0$, we take the limit
$L^\prime \rightarrow \infty$
and have
\begin{eqnarray}
\lim_{\lpp\rightarrow \infty}
& i S_{F-}[+M](p;\lpp,\lpp)& = P_L \fs p \, \frac{1}{P -M} \ec\\
\lim_{\lpp\rightarrow \infty}
&iS_{F-}[+M](p;-\lpp,-\lpp)& = P_R \fs p \, \frac{1}{P -M} \ec\\
\lim_{\lpp\rightarrow \infty}
&iS_{F-}[+M](p;\lpp,-\lpp)& = 0 \ec\\
\lim_{\lpp\rightarrow \infty}
&iS_{F-}[+M](p;-\lpp,\lpp)& = 0 \ee
\end{eqnarray}
Therefore we obtain
\begin{eqnarray}
  \label{breaking-source-continuum-limit-explicit-form}
X_{-+}^{\infty}(p)
&=&
\left(
        \begin{array}{cc} X_-^{\infty}(p) & 0 \\
                          0        & X_+^{\infty}(p)
\end{array} \right)
=
\left(
        \begin{array}{cc} P_L \fs p \, \frac{-1}{P +M} & 0 \\
                          0        & P_R \fs p \, \frac{-1}{P -M}
\end{array} \right)
\label{breaking-source-total-continuum-limit-explicit-form}
\ee
\end{eqnarray}
In terms of the wave function, the explicit gauge symmetry
breaking term can be written as follows.
\begin{eqnarray}
&&\langle b_- \ket{\psi_R(\ml),\bar \psi_R(\ml) }
\nonumber\\
&&=
c_-^\ast
\exp \left\{
-\int d^4x d^4y \,
\bar\psi_R(x,\ml)
\int\!\frac{d^4p}{i(2\pi)^4} e^{-ip(x-y)}
\frac{ \fs p }{P +M} \,
\psi_R(y,\ml) \right\}
\label{boundary-wave-function-right-explicit-form}
\ec
\\
&&\bra{\psi_L(\lp),\bar \psi_L(\lp) }b_+ \rangle
\nonumber\\
&&=
c_+
\exp \left\{ -\int d^4x d^4y  \,
\bar\psi_L(x,\lp)
\int\!\frac{d^4p}{i(2\pi)^4} e^{-ip(x-y)}
\frac{\fs p }{P -M} \,
\psi_L(y,\lp) \right\}
\label{boundary-wave-function-left-explicit-form}
\ee
\end{eqnarray}

\subsubsection{Perturbation expansion of the boundary term}

Given the explicit form of the boundary state wave function,
we next consider the perturbative expansion of
$1-M-X_{-+}^\infty$.
$S_{F\pm}[A](x-y;s,t)$ can be expanded as
\begin{equation}
S_{F\pm}[A]
= S_{F\pm}
+ \sum_{n=1}^\infty \left\{ S_{F\pm} \cdot
(-)\fs A \cdot \right\}^n  S_{F\pm}
\ec
\end{equation}
where the following abbreviation is used,
\begin{equation}
S_{F\pm} \cdot
(-)\fs A \cdot S_{F\pm}
\equiv
\int\!d^4z \int^\lp_\ml\!du S_{F\pm}(x-z;s,u) (-) {\fs A}(z)
S_{F\pm}(z-y;u,t)
\ee
\end{equation}
Then we obtain the expansion,
\begin{eqnarray}
1-M-X_{-+}^\infty =
1-M^0-X_{-+}^\infty
+
\left( \begin{array}{c} ..P_R \\
                        . P_L   \end{array} \right)
\left[
i \sum_{n=1}^\infty \left\{ S_{F\pm} \cdot (-)\fs A \cdot \right\}^n  S_{F\pm}
\right]
\left( \begin{array}{cc} P_L..  & P_R .   \end{array} \right)
\ee
\end{eqnarray}
Here we have also introduced the abbreviation for the boundary condition:
\begin{equation}
  ..P_R \psi(x,s) = P_R \psi(x,\ml),  \hskip 16pt
\bar \psi(x,s) P_L ..= \bar \psi(x,\ml) P_L  \ee
\end{equation}
\begin{equation}
  .P_L \psi(x,s) = P_L \psi(x,\lp),  \hskip 16pt
\bar \psi(x,s) P_R .= \bar \psi(x,\lp) P_R  \ee
\end{equation}

Then we obtain the perturbative expansion of the
boundary contribution.
\begin{eqnarray}
&&\ln {\det}' \left( 1-M-X_{-+}^\infty \right)
-\ln {\det}' \left( 1-M^0-X_{-+}^\infty \right)
\nonumber\\
&&=\sum_{m=1}^\infty \frac{(-)^m}{m}
{\Tr}^\prime\left\{
D^\lp
\left( \begin{array}{c} ..P_R \\
                        . P_L   \end{array} \right)
\left[
\sum_{n=1}^\infty \left\{ S_{F+} \cdot (-)\fs A \cdot \right\}^n  S_{F+}
\right]
\left( \begin{array}{cc} P_L..  & P_R .   \end{array} \right)
\right\}^m
\ec
\end{eqnarray}
where we denote the inverse of $\left( 1-M^0-X_{-+}^\infty \right)$ as
$D^\lp$,
\begin{eqnarray}
{D^\lp}^{-1}(p)&\equiv & \left( 1-M^0-X_{-+}^\infty \right)(p)
\nonumber\\
&=&
1
+\left( \begin{array}{c} ..P_R \\
                        . P_L   \end{array} \right)
 i S_{F+}(p;s,t)
\left( \begin{array}{cc} P_L..  & P_R .   \end{array} \right)
+
\left(
        \begin{array}{cc} P_L \fs p \, \frac{1}{P +M} & 0 \\
                          0        & P_R \fs p \, \frac{1}{P -M}
\end{array} \right)
\ee
\end{eqnarray}
The contribution of the fermion with positive and negative homogeneous
mass can be expanded in a similar manner.
\begin{eqnarray}
&&\ln {\det}' \left( 1-M_\pm -X_{\pm\pm}^\infty \right)
-\ln {\det}' \left( 1-M^0_\pm -X_{\pm\pm}^\infty \right)
\nonumber\\
&&=
\sum_{m=1}^\infty \frac{(-)^m}{m}
{\Tr}^\prime\left\{
D^\lp_\pm
\left( \begin{array}{c} ..P_R \\
                        . P_L   \end{array} \right)
\left[
i \sum_{n=1}^\infty \left\{ S_{F-}[\pm M] \cdot (-)\fs A \cdot \right\}^n
S_{F-}[\pm M]
\right]
\left( \begin{array}{cc} P_L..  & P_R .   \end{array} \right)
\right\}^m
\ec
\end{eqnarray}
and
\begin{eqnarray}
{D^\lp_\pm}^{-1}(p)&\equiv & \left( 1-M^0_\pm-X_{\pm\pm}^\infty \right)
\nonumber\\
&=&
1
+\left( \begin{array}{c} ..P_R \\
                        . P_L   \end{array} \right)
 i S_{F-}[\pm M](p;s,t)
\left( \begin{array}{cc} P_L..  & P_R .   \end{array} \right)
+
\left(
        \begin{array}{cc} P_L \fs p \, \frac{1}{P \mp M} & 0 \\
                          0        & P_R \fs p \, \frac{1}{P \mp M}
\end{array} \right)
\ee
\end{eqnarray}
These expression can be also regularized by the dimensional regularization.

\subsubsection{Cluster property and parity}

Next we consider taking the limit $L \rightarrow \infty$.
Since the boundary contribution is given by
the correlations between the boundaries, it is expected to be
finite in the limit $L \rightarrow \infty$
and the subtraction to be irrelevant.
As far as we do not take into account of the breaking
of the chiral boundary condition due to the dimensional regularization,
we will see that it is actually the case and the cluster property
holds:
the correlation between the two boundaries vanishes
in the limit $L \rightarrow \infty$
and the remaining diagonal contribution from each boundary is
equal to that of the fermion with homogeneous mass
(positive or negative according to the signature of mass at that
boundary).
We will make an assumption that {\it this cluster property holds
even under the dimensional regularization}.
Then, as a result of the cluster property and the parity invariance
of the fermion with the homogeneous mass, we will show that
the boundary contribution is odd-parity in the limit
$L \rightarrow \infty$.

We first consider the leading term in the perturbative
expansion,
\begin{eqnarray}
  \left( \begin{array}{c} ..P_R \\
                        . P_L   \end{array} \right)
  S_{F\pm}(p;s,t)
\left( \begin{array}{cc} P_L..  & P_R .   \end{array} \right)
\ee
\end{eqnarray}
The diagonal components can be read off from
Eqs.~(\ref{delta-R-sommerfeld-watson-transformed}),
(\ref{delta-L-sommerfeld-watson-transformed}).
\begin{eqnarray}
 ..P_R i S_{F\pm}(p;s,t) P_L..
&=&
P_R \fs p \, \left\{
\frac{(P \coth P L - M)}{-\Delta_- (iP)} \,
\frac{ P^2}{\sinh^2 PL }
+
\frac{1}{P \coth P L \pm M}
\right\}
\nonumber\\
 .P_L i S_{F\pm}(p;s,t) P_R.
&=&
P_L \fs p \, \left\{
\frac{(P \coth P L \pm M)}{-\Delta_\pm (iP)} \,
\frac{ P^2}{\sinh^2 PL }
+
\frac{1}{P \coth P L \pm M}
\right\}
\ee
\end{eqnarray}
Off diagonal components, which give the correlation
between the boundaries at $s=-L$ and $s=L$,
are given by the following summations.
\begin{eqnarray}
 ..P_R i S_{F\pm}(p;s,t) P_R.
&=&
-P_R \,
\sum_\omega
\frac{1}{n_\pm(\omega)}
\frac{\omega^2}{M^2+\omega^2-p^2-i \varepsilon}
\nonumber\\
 .P_L i S_{F\pm}(p;s,t) P_L..
&=&
-P_L \,
\sum_\omega
\frac{-1}{n_\pm(\omega)}
\frac{\omega^2}{M^2+\omega^2-p^2-i \varepsilon}
\end{eqnarray}
These summations can be performed by the use of
the technique of the Sommerfeld-Watson transformation.
At first we need to improve the convergence
of the integration of $\omega$, using
Eq. (\ref{mass-term-improve-convergence}).
Then refering to Table III,

\vskip 16pt
\centerline{Table III \
Poles and Residues in $I_\pm$ for $ ..P_R S_{F\pm} P_L..$ }
\begin{tabular}{|c|c|c|c|}
\hline
\phantom{aa} singular part in $I_\pm$ \phantom{aa}
& \phantom{aaaaa} pole \phantom{aaaaaaaaaaa}
& \phantom{a} residue  \phantom{aaaaa}
&
${\scriptstyle
\frac{2 \omega}{\sin^2 \omega L \Delta_\pm (\omega) }} $
\\
\hline
${\scriptstyle
\frac{2 \omega}{\sin^2 \omega L \Delta_\pm (\omega) }} $
& $\sin^2 \omega L \Delta_\pm (\omega)=0 $
& $\frac{1}{n_\pm(\omega)}$
& 1
\\
\hline\hline
  ${\scriptstyle
\frac{M^2}{ \cos^{(3\mp 1)/2} 2 \omega L}}$
& $ \cos 2 \omega L =0 $
& $ \frac{M^2}{-2L \sin 2 \omega L}\frac{1\pm 1}{2} $
& $\frac{2\omega}{M^2}$
\\
\hline
 $  \frac{1}{M^2+\omega^2-p^2-i \varepsilon} $
&$iP \equiv i\sqrt{M^2-p^2-i \varepsilon}$
&$\frac{1}{2iP}$
&$\frac{2 iP}{-\sinh^2 PL }\frac{1}{\Delta_\pm (iP)} $
\\
\hline
\end{tabular}
\vskip 16pt
\noindent
we obtain
\begin{eqnarray}
&&
\sum_\omega
\frac{1}{n_\pm(\omega)}
\frac{\omega^2}{M^2+\omega^2-p^2-i \varepsilon}
\nonumber\\
&&=
\sum_\omega
\frac{1}{n_\pm(\omega)}
\frac{1}{M^2+\omega^2-p^2-i \varepsilon}
\left( \frac{M^2}{ \cos^{(3\mp 1)/2} 2\omega L}-M^2 \right)
\nonumber\\
&&=
\frac{1}{\sinh^2 PL \Delta_\pm (iP)}
\left( \frac{M^2}{ \cosh^{(3\mp 1)/2} 2 P L}-M^2 \right)
+\frac{1\pm 1}{2}
\sum_{\cos 2\omega L } \frac{1}{L} \frac{\omega}{\sin 2\omega L}
\frac{1}{M^2+\omega^2-p^2-i \varepsilon}
\nonumber\\
&&=
\frac{1}{\sinh^2 PL \Delta_\pm (iP)}
\left( \frac{M^2}{ \cosh^{(3\mp 1)/2} 2 P L}-M^2 \right)
+\frac{1\pm 1}{2}
\sum_{n=0}^\infty
\frac{(-)^n (\frac{\pi}{4}+n\frac{\pi}{2})}
{(\frac{\pi}{4}+n\frac{\pi}{2})^2+ L^2(M^2-p^2-i \varepsilon)}
\nonumber\\
&&=
\frac{1}{\sinh^2 PL \Delta_\pm (iP)}
\left( \frac{M^2}{ \cosh^{(3\mp 1)/2} 2 P L}-M^2 \right)
+\frac{1\pm 1}{2}
\frac{1}{\cosh 2PL}
\ee
\end{eqnarray}
{}From these results, we can see that
the correlations between the boundaries
have definite limit when $L\rightarrow \infty$
and we obtain
\begin{eqnarray}
\lim_{\lp\rightarrow \infty}
  \left( \begin{array}{c} ..P_R \\
                        . P_L   \end{array} \right)
  i S_{F\pm}(p;s,t)
\left( \begin{array}{cc} P_L..  & P_R .   \end{array} \right)
=
\left( \begin{array}{cc}
P_R \fs p \frac{1}{P\pm M} & 0 \\
0 & P_L \fs p \frac{1}{P-M} \end{array} \right)
\ee
\label{correlation-between-boundary-leading}
\end{eqnarray}
Thus the correlation between the two boundaries vanishes
and the remaining diagonal contribution from each boundary is
equal to that of the fermion with homogeneous mass
(positive or negative according to the signature of mass at that
boundary).

Next we consider higher order terms,
\begin{eqnarray}
  \left( \begin{array}{c} ..P_R \\
                        . P_L   \end{array} \right)
\left[
i \sum_{n=1}^\infty \left\{ S_{F \pm} \cdot (-)\fs A \cdot \right\}^n
S_{F\pm}
\right]
\left( \begin{array}{cc} P_L..  & P_R .   \end{array} \right)
\end{eqnarray}
When the dimensional regularization is not taken into account,
we can see by the explicit calculation
that the cluster property holds
(see Appendix \ref{appendix:cluster-property-of-correlation}):
\begin{eqnarray}
&&
\lim_{ \lp \rightarrow \infty}
 \left( \begin{array}{c} ..P_R \\
                        . P_L   \end{array} \right)
\left[
i \sum_{n=1}^\infty \left\{ S_{F \pm} \cdot (-)\fs A \cdot \right\}^n
S_{F\pm}
\right]
\left( \begin{array}{cc} P_L ..  & P_R .   \end{array} \right)
\nonumber\\
&& =
\lim_{\lp \rightarrow \infty}
 \left(
   \begin{array}{c}
..P_R \left[
i \sum_{n=1}^\infty \left\{ S_{F-}[\mp M]
\cdot (-)\fs A \cdot \right\}^n S_{F-}[\mp M]
\right] P_L .. \\
0   \end{array} \right.
\nonumber\\
&&
\hskip 3cm
\left.
\begin{array}{c}
0 \\
.P_L
\left[
i \sum_{n=1}^\infty \left\{ S_{F-}[+M]
\cdot (-)\fs A \cdot \right\}^n S_{F-}[+M]
\right]  P_R .
  \end{array} \right)
\ee
\label{cluster-property-in-boundary-term}
\end{eqnarray}
For $n=1$, we have
\begin{eqnarray}
&& S_{F\pm}(p+k) \cdot \gamma^\mu \cdot S_{F\pm}(p)
\nonumber\\
=&&
\left( P_R (\fs p + \fs k)\Delta_{R\pm}
+ P_L (\fs p + \fs k) \Delta_{L\pm}
+P_R B_{RL\pm}+ P_L B_{LR\pm} \right)
\nonumber\\
&&
\hskip 1cm \cdot \gamma^\mu \cdot
\left( P_R \fs p \Delta_{R\pm}
+ P_L \fs p \Delta_{L\pm}
+P_R B_{RL\pm}+ P_L B_{LR\pm} \right)
\nonumber\\
=&&
P_R
\left\{
(\fs p + \fs k) \gamma^\mu \fs p \,
                 \Delta_{R\pm} \cdot \Delta_{R\pm}
+\gamma^\mu \, B_{RL\pm} \cdot B_{LR\pm} \right\}
\nonumber\\
&&
+P_L
\left\{
(\fs p + \fs k) \gamma^\mu \fs p \,
                 \Delta_{L\pm} \cdot \Delta_{L\pm}
+ \gamma^\mu \, B_{LR\pm} \cdot B_{RL\pm} \right\}
\nonumber\\
&&
+P_R
\left\{
(\fs p + \fs k) \gamma^\mu \,
                 \Delta_{R\pm} \cdot B_{RL\pm}
+\gamma^\mu \fs p \, B_{RL\pm} \cdot \Delta_{L\pm}
\right\}
\nonumber\\
&&
+P_L
\left\{
(\fs p + \fs k) \gamma^\mu
                 \Delta_{L\pm} \cdot B_{LR\pm}
+ \gamma^\mu \fs p \, B_{LR\pm} \cdot \Delta_{R\pm}
\right\}
\ee
\end{eqnarray}
With the help of the orthogonality of the generalized ``sin'',
\begin{equation}
\iLL\!ds \,
[\sin \omega(L-s)]_\pm [\sin \omega^\prime (L-s)]_\pm
= N_\pm (\omega) \delta_{\omega\omega^\prime}
\ec
\end{equation}
we can first perform the integration over the extra coordinates
at every vertices of external gauge field.
Then we can see, for example, that
$\Delta_{R\pm} \cdot \Delta_{R\pm}$ has the similar
structure with respect to the dependence on
$\omega$ and $s,t$
to $\Delta_{R\pm}$ and that it satisfies
the same boundary condition:
\begin{eqnarray}
  \Delta_{R\pm} \cdot \Delta_{R\pm}
=
\sum_\omega
\frac{\left( \omega \cot \omega L \pm M \right)}{n_\pm(\omega)}
\frac{[\sin \omega(L-s)]_\pm [\sin \omega(L-t)]_\pm }
{[M^2+\omega^2-(p+k)^2-i \varepsilon][M^2+\omega^2-p^2-i \varepsilon]}
\end{eqnarray}
Same is true for $B_{RL\pm} \cdot B_{LR\pm}$ and it satisfies
the same boundary condition as $\Delta_{R\pm}$. Let us denote this
similarity as follows.

\begin{eqnarray}
\Delta_{R\pm} \cdot \Delta_{R\pm} \sim
B_{RL\pm} \cdot B_{LR\pm} \sim   \Delta_{R\pm}
\ee
\end{eqnarray}
Then we can write
\begin{eqnarray}
\Delta_{L-} \cdot \Delta_{L-} \sim
B_{LR\pm} \cdot B_{RL\pm} \sim   \Delta_{L-}
\ee
\end{eqnarray}
\begin{eqnarray}
\Delta_{R\pm} \cdot B_{RL\pm} \sim
B_{RL\pm} \cdot  \Delta_{L-} \sim   B_{RL\pm}
\ee
\end{eqnarray}
\begin{eqnarray}
\Delta_{L-} \cdot B_{LR\pm} \sim
B_{LR\pm} \cdot  \Delta_{R\pm} \sim   B_{LR\pm}
\ee
\end{eqnarray}
As expected from the similarity of the structure of the
Green functions,
the correlation functions between the boundaries at the order
$n=1$ can be evaluated in the same way as the leading case
by the Sommerfeld-Watson transformation.
We obtain the result that the correlation between the
two boundaries vanish in the limit of $L \rightarrow \infty$,
\begin{eqnarray}
&&
\lim_{\lp \rightarrow \infty}
\left( \begin{array}{c} ..P_R \\
                        . P_L   \end{array} \right)
\left[
i \left\{ S_{F\pm} \cdot (-) \gamma^\mu \cdot \right\} S_{F\pm}
\right]
\left( \begin{array}{cc} P_L..  & P_R .   \end{array} \right)(k+p,k)
\nonumber\\
&&
=
(i)
\left( \begin{array}{cc}
P_R V_\pm^\mu(k+p,k) P_L
& 0 \\
0 &
P_L V_-^\mu(k+p,k) P_R
\end{array} \right)
\ec
\end{eqnarray}
where
\begin{equation}
  \label{gauge-boson-vertex-one-point}
V_\pm^\mu(k+p,k) \, (P+K)  =
 \left[ (\fs k + \fs p) \gamma^\mu \fs k \right]
\frac{1}{[P\pm M][K \pm M]}
+\gamma^\mu
\ee
\end{equation}
$p$ is assumed to be the momentum incoming from the external
gauge boson attached to the vertex $\gamma^\mu$.
$P$ and $K$ are defined as $P=\sqrt{M^2-(k+p)^2-i \varepsilon}$
and $K=\sqrt{M^2-k^2-i \varepsilon}$, respectively.

For $n>1$, the correlation functions between the boundaries can
be evaluated in a similar manner.
We obtain
\begin{eqnarray}
&&
\lim_{ \lp \rightarrow \infty}
  \left( \begin{array}{c} ..P_R \\
                        . P_L   \end{array} \right)
\left[
i \prod_{i=1}^n
\left\{ S_{F\pm}(k_i) \cdot (-) \gamma^{\nu_i} \cdot \right\}
S_{F\pm}(k_{n+1})
\right]
\left( \begin{array}{cc} P_L..  & P_R .   \end{array} \right)
\nonumber\\
&& =
(i)^n \sum_{n=1}^\infty
\left( \begin{array}{cc}
P_R V_\pm^{\nu_1\nu_2\ldots\nu_{n+1}}(k_1,k_2,\ldots,k_{n+1}) P_L
& 0 \\
0 &
P_L V_-^{\nu_1\nu_2\ldots\nu_{n+1}}(k_1,k_2,\ldots,k_{n+1}) P_R
\end{array} \right)
\ec
\label{cluster-property-in-boundary-term-at-each-order}
\end{eqnarray}
where
\begin{eqnarray}
V_\pm^{\nu_1\nu_2\ldots\nu_{n+1}}(k_1,k_2,\ldots,k_{n+1})
=
\sum_{0\le 2l \le n+1} C_{2l}^n  \,
\sum_{i=1}^{n+1}\frac{(M^2-K_i^2)^l}{(K_i \pm M)}
\prod_{j\not=i} \frac{1}{K_j^2-K_i^2}
\ee
\end{eqnarray}
(See Appendix \ref{appendix:cluster-property-of-correlation} for
the definition of $C_{2l}^n$.)
This result shows that the cluster property holds in each order
of the expansion.

We will show the explicit results for $n=2$ and $n=3$ for later use.
For $n=2$,
\begin{eqnarray}
  \label{gauge-boson-vertex-two-point}
V_\pm^{\mu \nu} && (k+p,k,k-q) \, (P+K)(K+Q)(Q+P)
\nonumber\\
&&= \left[ (\fs k + \fs p) \gamma^\mu \fs k \gamma^\nu (\fs k - \fs q)\right]
\frac{P+K+Q\pm M}{[P\pm M][K \pm M][Q \pm M]}
\nonumber\\
&&\quad
+
\left[
(\fs k + \fs p) \gamma^\mu \gamma^\nu
+ \gamma^\mu \fs k \gamma^\nu
+ \gamma^\mu \gamma^\nu (\fs k - \fs q) \right]
\ee
\end{eqnarray}
$q$ is assumed to be the momentum incoming from the external
gauge boson attached to the vertex $\gamma^\nu$ and
$Q$ is defined by $Q=\sqrt{M^2-(k-q)^2-i \varepsilon}$.

For $n=3$,
\begin{eqnarray}
  \label{gauge-boson-vertex-three-point}
&&  V_\pm^{\mu \nu \lambda}(k+p,k,k-q,k-q-r)
 \, (P+K)(P+Q)(P+R)(K+Q)(K+R)(Q+R)
\nonumber\\
&&=
\left[ (\fs k + \fs p) \gamma^\mu \fs k \gamma^\nu (\fs k - \fs q)
\gamma^\lambda (\fs k - \fs q - \fs r) \right]
\frac{A_\pm (P,K,Q,R)}
{[P\pm M][K \pm M][Q \pm M][R \pm M]}
\nonumber\\
&&
+ \left[
 (\fs k + \fs p) \gamma^\mu \fs k \gamma^\nu \gamma^\lambda
+(\fs k + \fs p) \gamma^\mu \gamma^\nu (\fs k-\fs q) \gamma^\lambda
+(\fs k + \fs p) \gamma^\mu \gamma^\nu \gamma^\lambda
                                              (\fs k - \fs q - \fs r)
+  \gamma^\mu \fs k \gamma^\nu (\fs k - \fs q) \gamma^\lambda
\right. \nonumber\\
&&\qquad \left.
+  \gamma^\mu \fs k \gamma^\nu  \gamma^\lambda (\fs k - \fs q -\fs r)
+  \gamma^\mu \gamma^\nu (\fs k - \fs q) \gamma^\lambda
                                              (\fs k - \fs q -\fs r)
\right]
(P+K+Q+R)
\nonumber\\
&&
+
\left[\gamma^\mu \gamma^\nu \gamma^\lambda \right]
B (P,K,Q,R)
\ec
\end{eqnarray}
and
\begin{eqnarray}
A_\pm (P,K,Q,R) &=&
P^2(Q+R+K)+Q^2(P+R+K)+R^2(P+Q+K)+K^2(P+Q+R)
\nonumber\\
&&
+2(QRK+PRK+PQK+PQR)
\nonumber\\
&&
\pm M(P+K+Q+R)^2 + M^2(P+K+Q+R)
\ec
\\
B (P,K,Q,R) &=& (QRK+PRK+PQK+PQR) + M^2(P+K+Q+R)
\ee
\end{eqnarray}
$r$ is assumed to be the momentum incoming from the external
gauge boson attached to the vertex $\gamma^\lambda$ and
$R$ is defined by $R=\sqrt{M^2-(k-q-r)^2-i \varepsilon}$.

If we take into account of the dimensional regularization,
since $\gamma^{a=5}$ commutes with the extended part of
the gamma matrices $\gamma^\mu (\mu=5,\ldots, D)$,
the boundary condition is not respected by
the chiral structure of the gamma matrices.
There occurs the mismatch of the components on which
different boundary conditions are imposed.
For example, let us consider the component,
\begin{equation}
..P_R
\left[
i \sum_{n=1}^\infty \left\{ S_{F+} \cdot
\gamma^{\mu_i}\cdot \right\}^n  S_{F+}
\right] P_R .
\label{mismatched-correlation-R-R}
\end{equation}
Beside the regular part, we encounter the following term at $n=1$.
\begin{eqnarray}
  P_R \fs p \gamma^\mu P_L \fs q \, .. \Delta_R \cdot \Delta_L .
\end{eqnarray}
Since the boundary condition mismatches, we cannot use the
orthogonality of the generalized ``sin'' function and
cannot evaluate the correlation straightforwardly.
At $n=2$, we find the following mismatched correlation functions.
\begin{eqnarray}
.. \Delta_R \cdot \Delta_L \cdot B_{RL} .
\hskip 32pt
.. B_{RL} \cdot \Delta_L \cdot B_{RL} .  \sim
.. B_{RL} \cdot B_{RL} .  \ \ee
\label{mismatch-correlation-function-n=2}
\end{eqnarray}
At $n=3$,
\begin{eqnarray}
&& .. \Delta_R \cdot \Delta_L \cdot \Delta_R \cdot \Delta_L . \\
&&
.. B_{RL} \cdot \Delta_L \cdot \Delta_R \cdot \Delta_L .
\sim .. B_{RL} \cdot \Delta_R \cdot \Delta_L .  \\
&&
.. \Delta_R \cdot B_{LR} \cdot
\Delta_R \cdot \Delta_L .
\sim .. \Delta_R \cdot B_{LR} \cdot \Delta_L . \ \ee
\end{eqnarray}
At $n=4$,
\begin{eqnarray}
&& .. \Delta_R \cdot \Delta_L \cdot \Delta_R \cdot \Delta_L
\cdot B_{RL} . \\
&&
.. B_{RL} \cdot \Delta_L \cdot \Delta_R \cdot \Delta_L
\cdot B_{RL} .
\sim .. B_{RL} \cdot \Delta_R \cdot \Delta_L
\cdot B_{RL} .  \\
&&
.. \Delta_R \cdot \Delta_L \cdot B_{RL} \cdot
\Delta_L \cdot \cdot B_{RL} .
\sim
.. \Delta_R \cdot \Delta_L \cdot B_{RL} \cdot B_{RL} .
\\
&&
.. B_{RL}\cdot \Delta_L \cdot B_{RL} \cdot \Delta_L \cdot B_{RL} .
\sim
.. B_{RL}\cdot B_{RL} \cdot B_{RL} . \ \ee
\end{eqnarray}
Since the divergence appears in
the order $n \le 4$ by the power counting,
it is enough to consider only such orders.
Any mismatched correlation function
up to $n=4$ is ``similar'' to one of the above examples.

As to the simplest case,
Eq.~(\ref{mismatch-correlation-function-n=2}), we can see that
it actually vanishes in the limit $L \rightarrow \infty$
as follows. From
Eq.~(\ref{delta-R-sommerfeld-watson-transformed}) and
Eq.~(\ref{delta-L-sommerfeld-watson-transformed}),
the correlation between the boundary at $s=-L$ and
that at $s=L$ could emerge from the parts,
\begin{equation}
 \frac{(P \coth P L \pm M)}{-\Delta_\pm (iP)} \,
\frac{ [\sinh P(L-s)]_\pm [\sinh P(L-t)]_\pm } {\sinh^2 PL }
\ec
\end{equation}
in $\Delta_{R\pm}$ or
\begin{equation}
 \frac{(P \coth P L - M)}{-\Delta_- (iP)} \,
\frac{ [\sinh P(L+s)]_- [\sinh P(L+t)]_- } {\sinh^2 PL }
\ec
\end{equation}
in $\Delta_{L\pm}$.
At the boundaries, on the other hand,
the above correlation-mediate-parts cannot contribute
because
\begin{eqnarray}
.. \frac{[\sinh P(L-s)]_\pm}{\sinh PL }
&=& \frac{1}{\sinh PL }  \frac{P}{P \coth PL \pm M}
\sitarel{\longrightarrow}{\lp \rightarrow \infty} 0
\nonumber\\
\frac{[\sinh P(L+s)]_-}{\sinh PL } .
&=& \frac{1}{\sinh PL }  \frac{P}{P \coth PL - M}
\sitarel{\longrightarrow}{\lp \rightarrow \infty} 0
\ec
\end{eqnarray}
As a result,
\begin{equation}
  .. \Delta_R \cdot \Delta_L .
\sitarel{\longrightarrow}{\lp \rightarrow \infty} 0
\ee
\end{equation}
We can expect that other mismatched correlation functions
which appear in Eq.~(\ref{mismatched-correlation-R-R})
also vanish in the limit $L \rightarrow \infty$ in the
similar reason.
We do not enter the detail of the proof here.
{\it We rather assume that the dimensional regularization
preserves the cluster property
Eq.~(\ref {cluster-property-in-boundary-term})}.

Using this cluster property,
we obtain the formula of the perturbative expansion
of the boundary contribution
Eq.~(\ref{overlap-formula-boundary-contribution}) as follows.
\begin{eqnarray}
&&\lim_{\lp \rightarrow \infty}
\ln \left[{\scriptstyle
\frac{
     {\det}' \left( 1-M-X_{-+}^\infty \right) }
     {\sqrt{ 
           {\det}' \left( 1-M_--X_{--}^\infty \right)
           {\det}' \left( 1-M_+-X_{++}^\infty \right)
            }}} \right]
-\lim_{\lp \rightarrow \infty}
\ln \left[{\scriptstyle
\frac{
     {\det}' \left( 1-M^0-X_{-+}^\infty \right) }
     {\sqrt{ 
           {\det}' \left( 1-M^0_--X_{--}^\infty \right)
           {\det}' \left( 1-M^0_+-X_{++}^\infty \right)
            }} }\right]
\nonumber\\
&=&
\phantom{+}\half
\sum_{m=1}^\infty \frac{(-)^m}{m}
{\Tr}^\prime\left\{
d^\infty_-
\left(
\lim_{\lp \rightarrow \infty} \,
..P_R
\left[
i \sum_{n=1}^\infty
\left\{ S_{F-}[-M] \cdot (-)\fs A \cdot \right\}^n S_{F-}[-M]
\right] P_L..
\right)
\right\}^m
\nonumber\\
&&
-\half
\sum_{m=1}^\infty \frac{(-)^m}{m}
{\Tr}^\prime\left\{
d^\infty_+
\left(
\lim_{\lp \rightarrow \infty} \,
..P_R
\left[
i \sum_{n=1}^\infty
\left\{ S_{F-}[+M] \cdot (-)\fs A \cdot \right\}^n S_{F-}[+M]
\right] P_L.. \right)
\right\}^m
\nonumber\\
&&
+\half
\sum_{m=1}^\infty \frac{(-)^m}{m}
{\Tr}^\prime\left\{
d^\infty_+
\left(
\lim_{\lp \rightarrow \infty} \,
.P_L
\left[
i \sum_{n=1}^\infty
\left\{ S_{F-}[+M] \cdot (-)\fs A \cdot \right\}^n S_{F-}[+M]
\right] P_R. \right)
\right\}^m
\nonumber\\
&&
-\half
\sum_{m=1}^\infty \frac{(-)^m}{m}
{\Tr}^\prime\left\{
d^\infty_-
\left(
\lim_{\lp \rightarrow \infty} \,
.P_L
\left[
i \sum_{n=1}^\infty
\left\{ S_{F-}[-M] \cdot (-)\fs A \cdot \right\}^n S_{F-}[-M]
\right] P_R. \right)
\right\}^m
\nonumber\\
&\equiv&
 i\Gamma_X[A]
\ec
\label{perturbative-expansion-of-boundary-contribution}
\end{eqnarray}
where
\begin{eqnarray}
 d^\infty_\pm (p) = \frac{ P \mp M - \fs p }{2 P }
\ee
\end{eqnarray}
This is derived by using the result
Eq.~(\ref {correlation-between-boundary-leading}) as follows.
\begin{eqnarray}
{D^\infty}(p)^{-1}&\equiv &
\lim_{\lp \rightarrow \infty}
\left( 1-M^0-X_{-+}^\infty \right)
\nonumber\\
&=&
1
+
\left(
        \begin{array}{cc} P_R \fs p \, \frac{1}{P + M} & 0 \\
                          0        & P_L \fs p \, \frac{1}{P - M}
\end{array} \right)
+
\left(
        \begin{array}{cc} P_L \fs p \, \frac{1}{P + M} & 0 \\
                          0        & P_R \fs p \, \frac{1}{P - M}
\end{array} \right)
\nonumber\\
&=&
\left(
        \begin{array}{cc} 1+ \frac{\fs p }{P + M} & 0 \\
                          0        & 1+ \frac{\fs p }{P - M}
\end{array} \right)
\equiv
\left(
        \begin{array}{cc} {d^\infty_-(p)}^{-1} & 0 \\
                          0        & {d^\infty_+(p)}^{-1}
\end{array} \right)
\ee
\end{eqnarray}

As an important consequence of the cluster property,
we can show that the boundary contribution
Eq.~(\ref{overlap-formula-boundary-contribution}) is
purely odd-parity.
Since the fermion system with the homogeneous mass
possesses the parity invariance,
the propagator satisfies the relation,
\begin{eqnarray}
  \label{parity-invariance-relation-of-propagator}
S_{F-}[\pm M](x_0-y_0,x_i-y_i,s,t)
=
\gamma^0 S_{F-}[\pm M](x_0-y_0,-x_i+y_i,-s,-t) \gamma^0
\ee
\end{eqnarray}
We can also check explicitly that $d^\infty(x_0-y_0,x_i-y_i)$
satisfies the similar relation. From these properties,
we can easily see that
the boundary contribution given by the above trace formula
is a parity-odd functional of
the external gauge field potential.

Note also that it changes the sign when we change the sing
of $M$ to $-M$.
This means that it has
the functional form as
\begin{equation}
  \Gamma_X[A]= M F_X[A;M^2]
\ee
\end{equation}
This property is useful to reduce the superficial
degree of divergence of the loop integral in the perturbative
evaluation.

\section{Anomaly from boundary term}
\label{sec:anomaly-from-boundary}

In this section, as an application of the perturbation theory
given in the previous
section \ref{sec:perturbation-at-finite-extent-of-fifth-dimension},
we calculate the variation of the boundary term
under the gauge transformation.
We find that the consistent anomaly
is actually reproduced by the gauge
noninvariant boundary state wave functions.
This shows that
it is the correct choice
to fix the phase of the overlap
following the Wigner-Brillouin perturbation theory.

\subsection{variation under gauge transformation}

In order to examine the gauge symmetry breaking induced
by the boundary state wave function
Eq.~(\ref{boundary-wave-function-right-explicit-form}) and
Eq.~(\ref{boundary-wave-function-left-explicit-form}), we
consider the variation of the boundary contribution
$\Gamma_X[A]$
under the gauge transformation:
\begin{equation}
  \label{gauge-transformtion-infinitesimal}
  A_\mu(x) \rightarrow
  A_\mu(x)
+ \partial_\mu \omega(x) - \left[ \omega(x), A_\mu(x) \right]
\ec \hskip 16pt
\omega \in su(N) \ee
\end{equation}

First we note the Ward-Takahashi identity in which
we take into account of the breaking of the chiral boundary
condition due to the dimensional regularization,
\begin{eqnarray}
  \label{ward-takahashi-identity-under-dimensional-regularization}
&&  S_{F-}(k+p) \cdot \fs p \cdot S_{F-}(k) \nonumber\\
&&=   S_{F-}(k+p) - S_{F-}(k)
+ \Delta_1 (k+p) \cdot \fs p \cdot \Delta_2 (k)
\ec
\end{eqnarray}
where we have defined
\begin{eqnarray}
\Delta_1 (k) &\equiv&
  -\fs{ \ext{k} }
\left[ P_R \Delta_{R-}(k)-P_L \Delta_{L-}(k) \right]
\ec
\\
\Delta_2 (k) &\equiv&
  \left[ \fs k
 \left( \Delta_{R-}(k)-\Delta_{L-}(k) \right)
+\left( B_{RL-}(k)-B_{LR-}(k) \right)
      \right]
\ec
\end{eqnarray}
where $\fs{ \ext{k} }= \sum_{\mu=4}^D \gamma^\mu k_\mu$.
{}From this identity, we obtain
\begin{eqnarray}
  \label{variation-of-propagator-under-dimensional-regularization}
&& \delta_\omega
\left[  i \sum_{n=0}^\infty
\left\{ S_{F-}\cdot (-)\fs A \cdot \right\}^n S_{F-}
\right](x,y)
\nonumber\\
&=&
-\omega(x)
\left[  i \sum_{n=0}^\infty
\left\{ S_{F-}\cdot (-)\fs A \cdot \right\}^n S_{F-}
\right](x,y)
+
\left[  i \sum_{n=0}^\infty
\left\{ S_{F-}\cdot (-)\fs A \cdot \right\}^n S_{F-}
\right](x,y) \,
\omega(y)
\nonumber\\
&&
-\Delta_\omega(x,y)
\ec
\end{eqnarray}
where $\Delta_\omega(x,y)$ stands for the gauge non-invariant
correction due to the dimensional regularization,
\begin{eqnarray}
&&
\Delta_\omega(x,y)
\nonumber\\
&&=
i\,  \Delta_1 \cdot \fs \partial \omega \cdot \Delta_2 (x,y)
\nonumber\\
&&
- i
\left[
S_{F-}\cdot \fs A \cdot \Delta_1
\cdot \fs \partial \omega \cdot \Delta_2 \right]
(x,y)
- i
\left[
\Delta_1 \cdot \fs \partial \omega \cdot \Delta_2
\cdot \fs A \cdot S_{F-}
\right](x,y)
\nonumber\\
&& + \ldots
\end{eqnarray}
\noindent

Then the variation of $1-M-X_{-+}^\infty$ is given by
\begin{eqnarray}
\delta_\omega (1-M-X_{-+}^\infty )
=&&
-(\omega X_{-+}^\infty-X_{-+}^\infty \omega)
-\left( \begin{array}{c} ..P_R \\
                        . P_L   \end{array} \right)
\left[
\Delta_\omega
\right]
\left( \begin{array}{cc} P_L..  & P_R .   \end{array} \right)
\nonumber\\
&&
- \left[ \omega , (1-M-X_{-+}^\infty ) \right]
\ee
\end{eqnarray}
Therefore the variation of the determinant of
$1-M-X_{-+}^\infty$ can be written as
\begin{eqnarray}
&&\delta_\omega \ln {\det}'(1-M-X_{-+}^\infty)
\nonumber\\
&&=
{\Tr}'
\left\{
(-)
  \left(
    \omega X_{-+}^\infty- X_{-+}^\infty \omega
+
\left( \begin{array}{c} ..P_R \\
                        . P_L   \end{array} \right)
\left[
\Delta_\omega
\right]
\left( \begin{array}{cc} P_L..  & P_R .   \end{array} \right)
  \right)
\right.
\nonumber\\
&&
\left.
\hskip 32pt \times
\sum_{m=0}^\infty (-)^m
\left\{
D^\lp_+
\left( \begin{array}{c} ..P_R \\
                        . P_L   \end{array} \right)
\left[
i \sum_{n=1}^\infty \left\{ S_{F+}\cdot (-)\fs A \cdot \right\}^n
S_{F+}
\right]
\left( \begin{array}{cc} P_L..  & P_R .   \end{array} \right)
\right\}^m
D^\lp_+
\right\}
\nonumber\\
\end{eqnarray}
The variation of the determinant of
$1-M_\pm-X_{\pm\pm}^\infty$ is obtained in a similar manner.

Taking into account of the cluster property
in the limit $L \rightarrow \infty$,
the variation of $\Gamma_X$ is given by

\begin{eqnarray}
&&
i \delta_\omega \Gamma_X[A]
\nonumber\\
&&=\lim_{\lp \rightarrow \infty}
\delta_\omega
\ln \left[{\scriptstyle
\frac{
     {\det}' \left( 1-M-X_{-+}^\infty \right) }
     {\sqrt{ 
           {\det}' \left( 1-M_--X_{--}^\infty \right)
           {\det}' \left( 1-M_+-X_{++}^\infty \right)
            }}} \right]
\nonumber\\
&&=
\half
{\Tr}^\prime
\left\{
(-) \left(   \omega X^\infty_-  -  X^\infty_- \omega
+ ..P_R \Delta_\omega P_L .. \right)
\right.
\nonumber\\
&& \hskip 16pt
\left.
\times
\sum_{m=0}^\infty (-)^m
\left\{
d^\infty_-
\left(
\lim_{\lp \rightarrow \infty}
..P_R
\left[
i \sum_{n=1}^\infty
\left\{ S_{F-}[-M]\cdot (-)\fs A \cdot \right\}^n S_{F-}[-M]
\right]
P_L..  \right)
\right\}^m
d^\infty_-
\right\}
\nonumber\\
&&+
\half
{\Tr}^\prime
\left\{
(-) \left(   \omega X^\infty_+  -  X^\infty_+ \omega
+ .P_L \Delta_\omega P_R . \right)
\right.
\nonumber\\
&& \hskip 16pt
\left.
\times
\sum_{m=0}^\infty (-)^m
\left\{
d^\infty_+
\left(
\lim_{\lp \rightarrow \infty}
.P_L
\left[
i \sum_{n=1}^\infty
\left\{ S_{F-}[+M]\cdot (-)\fs A \cdot \right\}^n S_{F-}[+M]
\right]
P_R.  \right)
\right\}^m
d^\infty_+
\right\}
\nonumber\\
&&-
\half
{\Tr}^\prime
\left\{
(-) \left( \omega X^\infty_-  -  X^\infty_- \omega
+ ..P_R \Delta_\omega P_L .. \right)
\right.
\nonumber\\
&& \hskip 16pt
\left.
\times
\sum_{m=0}^\infty (-)^m
\left\{
d^\infty_-
\left(
\lim_{\lp \rightarrow \infty}
..P_R
\left[
i \sum_{n=1}^\infty
\left\{ S_{F-}[+M]\cdot (-)\fs A \cdot \right\}^n S_{F-}[+M]
\right]
P_L..  \right)
\right\}^m
d^\infty_-
\right\}
\nonumber\\
&&-
\half
{\Tr}^\prime
\left\{
(-) \left(  \omega X^\infty_+  -  X^\infty_+ \omega
+ .P_L \Delta_\omega P_R . \right)
\right.
\nonumber\\
&& \hskip 16pt
\left.
\times
\sum_{m=0}^\infty (-)^m
\left\{
d^\infty_+
\left(
\lim_{\lp \rightarrow \infty}
.P_L
\left[
i \sum_{n=1}^\infty
\left\{ S_{F-}[-M]\cdot (-)\fs A \cdot \right\}^n S_{F-}[-M]
\right]
P_R.  \right)
\right\}^m
d^\infty_+
\right\}
\ee
\label{variation-of-boundary-contribution}
\end{eqnarray}

\subsection{Consistent anomaly}

Now we perform the calculation of
Eq.~(\ref{variation-of-boundary-contribution}).
We express it in the momentum space as follows.
\begin{eqnarray}
  \label{anomaly-leading-order-term}
&& \delta_\omega \Gamma_X [A]
\nonumber\\
&&
=  \sum_{n=1}^\infty \, (i)^{n+1}
\int\! \frac{d^4 l}{(2\pi)^4}\prod_{i=1}^n \frac{d^4 p_i}{(2\pi)^4}
(2\pi)^4\delta\left( l+\sum_i^n p_i \right)
\Gamma_X^{\nu_1 \nu_2 \ldots \nu_n}
(p_1,\ldots, p_n) \,
\tr\{ \omega(l) \prod_{i=1}^n A_{\nu_i}(p_i) \}
\ee
\nonumber\\
\end{eqnarray}

\subsubsection{Finiteness of $\Gamma_X$}
Since $\Gamma_X [A]$ is a parity-odd functional of gauge field
potential, it must involve
the $\epsilon$-tenser in four-dimensions
$\epsilon_{\nu_1\nu_2\nu_3 \nu_4} \, (\nu_i=0,1,2,3)$.
It is also proportional to $M$.
The Lorentz indices $\nu_1,\nu_2,\nu_3,\nu_4$ should be
contracted with those of the gauge field potential $A^\nu$ and
the momentum $p^\mu$'s.
For $n=1$, we can easily see that such structure cannot appear.
For $n=2$, we can have the form
\begin{eqnarray}
M \, \epsilon_{\mu \nu \nu_1 \nu_2 } \,
p_1^{\nu_1} p_2^{\nu_2} F(p_1,p_2, M^2 )
\ee
\end{eqnarray}
In this case, the superficial degree of divergence of
$\Gamma_X^{\mu\nu}$ is two by the power counting rule
of the five-dimensional theory.
Since the above structure reduces it by three and
we have minus one. This means that
there does not appear ultraviolet divergence in this term.
This also means that the additional term including $\Delta_\omega$
due to the dimensional regularization does not contribute.
For $n=3$, we can have the term
\begin{eqnarray}
M \, \epsilon_{\mu \nu \rho \nu_i } \,
p_i^{\nu_i} \, F(p_1,p_2,p_3,M^2)
\ee
\end{eqnarray}
In this case, the superficial degree of divergence is also reduced to
minus one and
the additional term due to the dimensional regularization
does not contribute.
For $n=4$, we can have the term
\begin{equation}
M \, \epsilon_{\mu \nu \rho \sigma }
\ee
\end{equation}
The superficial degree of divergence is zero and it is reduced by one
because of the coefficient $M$. It is also finite.
For $n \ge 5$, they are finite by the power counting rule.
Therefore, in every orders of the expansion, the boundary contribution
is finite and
the additional term due to the dimensional regularization
does not contribute.
Therefore, in the following calculation, we can omit the terms
due to the dimensional regularization.

\subsubsection{First order}
The first order term $(n=1)$ is given by the expression
\begin{eqnarray}
&&\Gamma_X^{\nu}(p) \nonumber\\
&&=\half \int\! \frac{d^D k}{(2\pi)^D}
\tr \left\{
P_L \left(
\frac{\fs k + \fs p}{P+M} - \frac{\fs k}{K+M}
\right) P_R
\right.
\nonumber\\
&& \qquad \qquad \qquad \qquad
\left.
\cdot
\frac{P+M - (\fs k + \fs p)}{2P}
P_R
V_+^\nu(k+p,k)
P_L
\frac{K+M - \fs k}{2K}
\right\}
\nonumber\\
&&+ \half \int\! \frac{d^D k}{(2\pi)^D}
\tr \left\{
i \Delta_1(k+p) \fs p \Delta_2(k)
\right.
\nonumber\\
&& \qquad \qquad \qquad \qquad \qquad
\left.
\cdot
\frac{P+M - (\fs k + \fs p)}{2P}
P_R
V_+^\nu(k+p,k)
P_L
\frac{K+M - \fs k}{2K}
\right\}
\nonumber\\
&&
+ \ldots
\nonumber\\
&&=-\half \int\! \frac{d^D k}{(2\pi)^D}
\tr \left\{
\gamma_5 \left(
\frac{\fs k + \fs p}{P+M} - \frac{\fs k}{K+M}
         \right)
\frac{P+M}{2P}
V_+^\nu(k+p,k)
\frac{K+M}{2K}
\right\}
\nonumber\\
&& \
+\half \int\! \frac{d^D k}{(2\pi)^D}
\tr \left\{
\gamma_5 \left(
\frac{\fs k + \fs p}{P-M} - \frac{\fs k}{K-M}
\right)
\frac{P-M}{2P}
V_-^\nu(k+p,k)
\frac{K-M}{2K}
\right\}
\ec
\end{eqnarray}
where $\ldots$ stands for the contributions from
the second, third and fourth terms
in Eq.~(\ref{variation-of-boundary-contribution}).
$V_+^\nu(k+p,k)$ is defined by
Eq.~(\ref{gauge-boson-vertex-one-point}).
This contribution
actually vanishes because of the trace over gamma matrices.

\subsubsection{Second order}

The second order term $(n=2)$ is given by the following expression.
\begin{eqnarray}
&&\Gamma_X^{\mu\nu}(p,q) \nonumber\\
&=&\half \int\! \frac{d^D k}{(2\pi)^D}
\tr \left\{
P_L \left(
   \frac{\fs k + \fs p}{P+M} - \frac{\fs k - \fs q }{Q+M}
    \right) P_R
\right.
\nonumber\\
&& \qquad
\left.
\cdot
\frac{P+M - (\fs k + \fs p )}{2P}
P_R V_+^{\mu\nu}(k+p,k,k-q) P_L
\frac{Q+M - \fs k - \fs q }{2Q}
\right\}
\nonumber\\
&&-\half \int\! \frac{d^D k}{(2\pi)^D}
\tr \left\{
P_L \left(
   \frac{\fs k + \fs p}{P+M} - \frac{\fs k - \fs q }{Q+M}
    \right) P_R
\right.
\nonumber\\
&& \qquad \quad\quad
\cdot
\frac{P+M - (\fs k  + \fs p)}{2P}
P_R V_+^{\mu}(k+p,k) P_L
\nonumber\\
&& \qquad \quad\quad\quad\quad
\left.
\cdot
\frac{K+M - \fs k}{2K}
P_R  V_+^{\nu}(k,k-q) P_L
\frac{Q+M - \fs k - \fs q }{2Q}
\right\}
\nonumber\\
&&
+ \ldots
\label{boundary-contribution-second-order}
\\
&=&
-\half \int\! \frac{d^D k}{(2\pi)^D}
\tr \left\{
\gamma_5 \left(
   \frac{\fs k + \fs p }{P+M} - \frac{\fs k - \fs q }{Q+M}  \right)
\frac{P+M }{2P}
V_+^{\mu\nu}(k+p,k,k-q)
\frac{Q+M }{2Q}
\right\}
\label{boundary-contribution-second-order-first-line}
\\
&&
+\half \int\! \frac{d^D k}{(2\pi)^D}
\tr \left\{
\gamma_5 \left(
   \frac{\fs k + \fs p }{P-M} - \frac{\fs k - \fs q }{Q-M}  \right)
\frac{P-M }{2P}
V_-^{\mu\nu}(k+p,k,k-q)
\frac{Q-M }{2Q}
\right\}
\label{boundary-contribution-second-order-second-line}
\\
&&+\half \int\! \frac{d^D k}{(2\pi)^D}
\tr \left\{
\gamma_5 \left(
   \frac{\fs k + \fs p}{P+M} - \frac{\fs k -\fs q }{Q+M}  \right)
\frac{K+M }{2K}
 V_+^{\mu}(k+p,k)
\right.
\nonumber\\
&& \qquad \qquad \qquad \qquad \qquad \qquad \qquad \qquad
\qquad \qquad
\left.
\cdot
\frac{- \fs k }{2K}
 V_+^{\nu}(k,k-q)
\frac{Q+M }{2Q}
\right\}
\label{boundary-contribution-second-order-third-line}
\\
&&-\half \int\! \frac{d^D k}{(2\pi)^D}
\tr \left\{
\gamma_5 \left(
   \frac{\fs k + \fs p }{P-M} - \frac{\fs k -\fs q }{Q-M}  \right)
\frac{K-M }{2K}
 V_-^{\mu}(k+p,k)
\right.
\nonumber\\
&& \qquad \qquad \qquad \qquad \qquad \qquad \qquad \qquad
\qquad \qquad
\left.
\cdot
\frac{- \fs k}{2K}
 V_-^{\nu}(k,k-q)
\frac{Q-M }{2Q}
\right\}
\label{boundary-contribution-second-order-forth-line}
\ec
\end{eqnarray}
where
$V_+^{\mu\nu}(k+q+p,k+q,k)$ is defined by
Eq.~(\ref{gauge-boson-vertex-two-point}).

The first line (\ref{boundary-contribution-second-order-first-line})
in the second equality can be evaluated as

\begin{eqnarray}
&&  -\half \int\! \frac{d^D k}{(2\pi)^D}
\frac{1}{4PQ(P+K)(K+Q)(Q+P)}
\nonumber\\
&&\qquad
\tr \gamma_5  \left\{
(k+p)^2 \gamma^\mu \fs k \gamma^\nu (\fs k -\fs q)
\frac{P+K+Q+M}{(P+M)(K+M)}
\right.
\nonumber\\
&&\qquad
+ \left[
(k+p)^2 \gamma^\mu \gamma^\nu
+(\fs k + \fs p) \gamma^\mu \fs k \gamma^\nu
+(\fs k + \fs p) \gamma^\mu \gamma^\nu (\fs k -\fs q)
\right](Q+M)
\nonumber\\
&&\qquad
+ (k-q)^2 (\fs k + \fs p) \gamma^\mu \fs k \gamma^\nu
\frac{P+K+Q+M}{(K+M)(Q+M)}
\nonumber\\
&&\left.\qquad
+ \left[
(k-q)^2 \gamma^\mu \gamma^\nu
+(\fs k + \fs p) \gamma^\mu \gamma^\nu (\fs k -\fs q)
+ \gamma^\mu \fs k \gamma^\nu (\fs k -\fs q)
\right](P+M)
\right\}
\ee\end{eqnarray}
Subtracting the contribution with the mass of opposite signature
(\ref{boundary-contribution-second-order-second-line}), we obtain

\begin{eqnarray}
&&  -\half \int\! \frac{d^D k}{(2\pi)^D}
\frac{2M}{4PQ(P+K)(K+Q)(Q+P)}
\nonumber\\
&&\qquad
\tr \gamma_5  \left\{
\gamma^\mu \fs k \gamma^\nu (\fs k -\fs q)
\frac{(P+K+Q)(P+K)-(PK+M^2)}{(K^2-M^2)}
\right.
\nonumber\\
&&\qquad
+ \left[
(k+p)^2 \gamma^\mu \gamma^\nu
+(\fs k + \fs p) \gamma^\mu \fs k \gamma^\nu
+(\fs k + \fs p) \gamma^\mu \gamma^\nu (\fs k -\fs q)
\right]
\nonumber\\
&&\qquad
+ (\fs k + \fs p) \gamma^\mu \fs k \gamma^\nu
\frac{(P+K+Q)(K+Q)-(KQ+M^2)}{(K^2-M^2)}
\nonumber\\
&&\left.\qquad
+ \left[
(k-q)^2 \gamma^\mu \gamma^\nu
+(\fs k + \fs p) \gamma^\mu \gamma^\nu (\fs k -\fs q)
+ \gamma^\mu \fs k \gamma^\nu (\fs k -\fs q)
\right]
\right\}
\ee\end{eqnarray}

Assuming that $p,q \ll M$, we make expansion with respect to
$p,q$. Taking into account of the property of the trace
of gamma matrix, we can calculate it as follows.

\begin{eqnarray}
&&
-M \int\! \frac{d^D k}{(2\pi)^D}
\left\{
-\tr(\gamma_5 \gamma^\mu \fs k \gamma^\nu \fs q)
\frac{1}{8K^2(K^2-M^2)}
\frac{(P+2K)(P+K)-(PK+M^2)}{P(P+K)^2}
\right.
\nonumber\\
&&\qquad\qquad\qquad
+\tr(\gamma_5 \fs p \gamma^\mu \fs k \gamma^\nu )
\frac{1}{8K^2(K^2-M^2)}
\frac{(Q+2K)(Q+K)-(QK+M^2)}{Q(Q+K)^2}
\nonumber\\
&&\qquad\qquad\qquad
-2 \tr(\gamma_5 \fs p \gamma^\mu \gamma^\nu \fs q)
\frac{1}{32K^5}
\nonumber\\
&&\left.\qquad\qquad\qquad
-\tr(\gamma_5 \fs k \gamma^\mu \gamma^\nu \fs q)
\frac{1}{8K^2P(P+K)^2}
+\tr(\gamma_5 \fs p \gamma^\mu \gamma^\nu \fs k)
\frac{1}{8K^2Q(Q+K)^2}
\right\}
+ {\cal O}\left( \frac{1}{M} \right)
\nonumber\\
&&
=
-M \int\! \frac{d^D k}{(2\pi)^D}
\left\{
-\tr(\gamma_5 \gamma^\mu \fs k \gamma^\nu \fs q)
\frac{(k \cdot p)}{8K^2(K^2-M^2)}
\left(
\frac{1}{4K^3}+ 2\frac{3}{4K^3}-K \frac{1}{4K^4}-M^2 \frac{1}{2K^5}
\right)
\right.
\nonumber\\
&&\qquad\qquad\qquad\quad
-\tr(\gamma_5 \fs p \gamma^\mu \fs k \gamma^\nu )
\frac{(k \cdot q)}{8K^2(K^2-M^2)}
\left(
\frac{1}{4K^3}+ 2\frac{3}{4K^3}-K \frac{1}{4K^4}-M^2 \frac{1}{2K^5}
\right)
\nonumber\\
&&\qquad\qquad\qquad\quad
- \tr(\gamma_5 \fs p \gamma^\mu \gamma^\nu \fs q)
\frac{1}{16K^5}
\nonumber\\
&&\left. \qquad\qquad\qquad\quad
-\tr(\gamma_5 \fs k \gamma^\mu \gamma^\nu \fs q)
\frac{(k \cdot p)}{8K^2}\frac{1}{2K^5}
-\tr(\gamma_5 \fs p \gamma^\mu \gamma^\nu \fs k)
\frac{(k \cdot q)}{8K^2}\frac{1}{2K^5}
\right\}
+ {\cal O}\left( \frac{1}{M} \right)
\nonumber\\
&&
= M \tr(\gamma_5 \gamma^\mu \gamma^\nu \fs p \fs q)
\int\! \frac{d^D k}{(2\pi)^D} \frac{1}{8K^5}
+ {\cal O}\left( \frac{1}{M} \right)
\nonumber\\
&&
=
M  \tr(\gamma_5 \gamma^\mu \gamma^\nu \fs p \fs q)
\, \frac{-i}{16\pi^2}\frac{4}{3}\frac{1}{|M|}
+ {\cal O}\left( \frac{1}{M} \right)
\nonumber\\
&&
=
- \frac{1}{24\pi^2}
\epsilon^{\mu\nu\rho\sigma} p_\rho q_\sigma
+ {\cal O}\left( \frac{1}{M} \right)
\end{eqnarray}

On the other hand, the third line
(\ref{boundary-contribution-second-order-third-line}) can be
evaluated as

\begin{eqnarray}
&&  -\half \int\! \frac{d^D k}{(2\pi)^D}
\frac{1}{8PKQ(P+K)(K+Q)}
\nonumber\\
&&\qquad
\tr \gamma_5  \left\{
\left(  \gamma^\mu \frac{(k+p)^2 k^2}{(P+M)(K+M)}
+(\fs k + \fs p) \gamma^\mu \fs k
\right)
\left( \fs k \gamma^\nu (\fs k -\fs q) \frac{1}{(K+M)}
      +\gamma^\nu(Q+M)  \right)
\right.
\nonumber\\
&&\left.\qquad\qquad
+\left( (\fs k + \fs p) \gamma^\mu \fs k \frac{1}{(K+M)}
      +\gamma^\nu (P+M)  \right)
\left(  \gamma^\nu \frac{k^2 (k-q)^2}{(K+M)(Q+M)}
+\fs k \gamma^\mu (\fs k - \fs q)
\right)
\right\}
\nonumber\\
&&=
-\half \int\! \frac{d^D k}{(2\pi)^D}
\frac{1}{8PKQ(P+K)(K+Q)}
\nonumber\\
&&
\tr \gamma_5  \left\{
 \gamma^\mu \fs k \gamma^\nu (\fs k - \fs q)
                \frac{(k+p)^2 k^2}{(P+M)(K+M)^2}
+ (\fs k + \fs p) \gamma^\mu \gamma^\nu (\fs k - \fs q)
                \frac{k^2}{(K+M)}
\right.
\nonumber\\
&& \qquad
+ (\fs k + \fs p) \gamma^\mu \fs k \gamma^\nu (Q+M)
+  \gamma^\mu \fs k \gamma^\nu (\fs k - \fs q)(P+M)
\nonumber\\
&&\left.\qquad
+(\fs k + \fs p)  \gamma^\mu \fs k \gamma^\nu
                \frac{k^2 (k-q)^2}{(K+M)^2(Q+M)}
+ (\fs k + \fs p) \gamma^\mu \gamma^\nu (\fs k - \fs q)
                \frac{k^2}{(K+M)}
\right\}
\ee
\nonumber\\
\end{eqnarray}
Subtracting the contribution with the mass of opposite signature
(\ref{boundary-contribution-second-order-forth-line}), we obtain

\begin{eqnarray}
&&
 -\half \int\! \frac{d^D k}{(2\pi)^D}
\frac{2M}{8PKQ(P+K)(K+Q)}
\nonumber\\
&&
\tr \gamma_5  \left\{
-\gamma^\mu \fs k \gamma^\nu (\fs k - \fs q)
                \frac{2PK+(K^2+M^2)}{(K^2-M^2)}
+ (\fs k + \fs p) \gamma^\mu \gamma^\nu (\fs k - \fs q)
+ (\fs k + \fs p) \gamma^\mu \fs k \gamma^\nu
\right.
\nonumber\\
&&\left.\qquad\quad
-(\fs k + \fs p)  \gamma^\mu \fs k \gamma^\nu
                \frac{2KQ+(K^2+M^2)}{(K^2-M^2)}
+ (\fs k + \fs p) \gamma^\mu \gamma^\nu (\fs k - \fs q)
+  \gamma^\mu \fs k \gamma^\nu (\fs k - \fs q)
\right\}
\ee
\nonumber\\
\end{eqnarray}
The expansion with respect to $p,q$ leads to the result,
\begin{eqnarray}
&&
 -M \int\! \frac{d^D k}{(2\pi)^D}
\left\{
 \tr(\gamma_5 \gamma^\mu \fs k \gamma^\nu \fs q)
\frac{1}{16K^3(K^2-M^2)}\frac{2PK+(K^2+M^2)}{P(P+K)}
\right.
\nonumber\\
&&
\phantom{ -M \int\! \frac{d^D k}{(2\pi)^D} }
-\tr(\gamma_5 \fs p \gamma^\mu \fs k \gamma^\nu )
\frac{1}{16K^3(K^2-M^2)}\frac{2QK+(K^2+M^2)}{Q(Q+K)}
\nonumber\\
&&
\phantom{ -M \int\! \frac{d^D k}{(2\pi)^D} }
-\tr(\gamma_5 \fs p \gamma^\mu \gamma^\nu \fs q)
\frac{1}{16K^5}
\nonumber\\
&& \left.
\phantom{ -M \int\! \frac{d^D k}{(2\pi)^D} }
-\tr(\gamma_5 \fs k \gamma^\mu \gamma^\nu \fs q)
\frac{1}{16K^3}\frac{1}{P(P+K)}
+\tr(\gamma_5 \fs p \gamma^\mu \gamma^\nu \fs k)
\frac{1}{16K^3}\frac{1}{Q(Q+K)}
\right\} + {\cal O}\left( \frac{1}{M} \right)
\nonumber\\
&&
= -M \int\! \frac{d^D k}{(2\pi)^D}
\left\{
 \tr(\gamma_5 \gamma^\mu \fs k \gamma^\nu \fs q)
\frac{(k\cdot p)}{16K^3(K^2-M^2)}
\left( 2K \frac{1}{4K^3} +(K^2+M^2) \frac{3}{4K^4} \right)
\right.
\nonumber\\
&&
\phantom{ = -M \int\! \frac{d^D k}{(2\pi)^D} }
+\tr(\gamma_5 \fs p \gamma^\mu \fs k \gamma^\nu )
\frac{(k\cdot q)}{16K^3(K^2-M^2)}
\left( 2K \frac{1}{4K^3} +(K^2+M^2) \frac{3}{4K^4} \right)
\nonumber\\
&&
\phantom{ = -M \int\! \frac{d^D k}{(2\pi)^D} }
-\tr(\gamma_5 \fs p \gamma^\mu \gamma^\nu \fs q)
\frac{1}{16K^5}
\nonumber\\
&& \left.
\phantom{ = -M \int\! \frac{d^D k}{(2\pi)^D} }
-\tr(\gamma_5 \fs k \gamma^\mu \gamma^\nu \fs q)
\frac{(k\cdot p)}{16K^3}\frac{3}{4K^4}
-\tr(\gamma_5 \fs p \gamma^\mu \gamma^\nu \fs k)
\frac{(k\cdot q)}{16K^3}\frac{3}{4K^4}
\right\} + {\cal O}\left( \frac{1}{M} \right)
\nonumber\\
&&
= M \tr(\gamma_5 \gamma^\mu \gamma^\nu \fs p \fs q)
\int\! \frac{d^D k}{(2\pi)^D}
\left\{
\frac{1}{4 \cdot 16 K^5}
\left( \frac{5}{2} + \frac{3}{2} \frac{M^2}{K^2}
- \frac{3}{2} \frac{k^2}{K^2} -4  \right)
\right\}+ {\cal O}\left( \frac{1}{M} \right)
\nonumber\\
&&
=0 + {\cal O}\left( \frac{1}{M} \right)
\ee
\end{eqnarray}
Therefore, we obtain
\begin{eqnarray}
\Gamma_X^{\mu\nu}(p,q)
= - \frac{1}{24\pi^2}
\epsilon^{\mu\nu\rho\sigma} p_\rho q_\sigma
+ {\cal O}\left( \frac{1}{M} \right)
\ee
\end{eqnarray}

\subsubsection{Third order}
The third order term is given by the following expression.
\begin{eqnarray}
&&\Gamma_X^{\mu\nu\lambda}(p,q,r) \nonumber\\
&=&\half \int\! \frac{d^D k}{(2\pi)^D}
\tr \left\{
P_L \left(
   \frac{\fs k + \fs p}{P+M} - \frac{\fs k - \fs q - \fs r}{R+M}
    \right) P_R
\right.
\nonumber\\
&& \qquad
\left.
\cdot
\frac{P+M - (\fs k + \fs p )}{2P}
P_R V_+^{\mu\nu\lambda}(k+p,k,k-q,k-q-r) P_L
\frac{R+M - (\fs k - \fs q - \fs r)}{2R}
\right\}
\nonumber\\
&&-\half \int\! \frac{d^D k}{(2\pi)^D}
\tr \left\{
P_L \left(
   \frac{\fs k + \fs p}{P+M} - \frac{\fs k - \fs q - \fs r}{R+M}
    \right) P_R
\right.
\nonumber\\
&& \qquad \quad\quad
\cdot
\frac{P+M - (\fs k  + \fs p)}{2P}
P_R V_+^{\mu\nu}(k+p,k,k-q) P_L
\nonumber\\
&& \qquad \quad\quad\quad\quad
\left.
\cdot
\frac{Q+M - (\fs k - \fs q)}{2Q}
P_R  V_+^{\lambda}(k-q,k-q-r) P_L
\frac{R+M - (\fs k - \fs q - \fs r)}{2R}
\right\}
\nonumber\\
\nonumber\\
&&-\half \int\! \frac{d^D k}{(2\pi)^D}
\tr \left\{
P_L \left(
   \frac{\fs k + \fs p}{P+M} - \frac{\fs k - \fs q - \fs r}{R+M}
    \right) P_R
\right.
\nonumber\\
&& \qquad \quad\quad
\cdot
\frac{P+M - (\fs k  + \fs p)}{2P}
P_R V_+^{\mu}(k+p,k) P_L
\nonumber\\
&& \qquad \quad\quad\quad\quad
\left.
\cdot
\frac{K+M - \fs k }{2K}
P_R  V_+^{\nu\lambda}(k,k-q,k-q-r) P_L
\frac{R+M - (\fs k - \fs q - \fs r)}{2R}
\right\}
\nonumber\\
&&+\half \int\! \frac{d^D k}{(2\pi)^D}
\tr \left\{
P_L \left(
   \frac{\fs k + \fs p}{P+M} - \frac{\fs k - \fs q - \fs r}{R+M}
    \right) P_R
\right.
\nonumber\\
&& \qquad \quad\quad
\cdot
\frac{P+M - (\fs k  + \fs p)}{2P}
P_R V_+^{\mu}(k+p,k) P_L
\nonumber\\
&& \qquad \quad\quad\quad
\cdot
\frac{K+M - \fs k}{2P}
P_R V_+^{\nu}(k,k-q) P_L
\nonumber\\
&& \qquad \quad\quad\quad\quad
\left.
\cdot
\frac{Q+M - (\fs k - \fs q)}{2K}
P_R  V_+^{\lambda}(k-q,k-q-r) P_L
\frac{R+M - (\fs k - \fs q - \fs r)}{2R}
\right\}
\nonumber\\
&&
+ \ldots
\label{boundary-contribution-third-order}
\\
&=&
-\half \int\! \frac{d^D k}{(2\pi)^D}
\tr \left\{
\gamma_5 \left(
   \frac{\fs k + \fs p }{P+M} - \frac{\fs k - \fs q - \fs r}{R+M}
\right)
\right.
\nonumber\\
&& \qquad \qquad \qquad \qquad \qquad \qquad
\left.
\cdot
\frac{P+M }{2P}
V_+^{\mu\nu\lambda}(k+p,k,k-q,k-q-r)
\frac{R+M}{2R}
\right\}
\label{boundary-contribution-third-order-term-one}
\\
&&+\half \int\! \frac{d^D k}{(2\pi)^D}
\tr \left\{
\gamma_5 \left(
   \frac{\fs k + \fs p}{P+M} - \frac{\fs k -\fs q -\fs r}{R+M}
\right)
\frac{K+M }{2K}
 V_+^{\mu\nu}(k+p,k,k-q)
\right.
\nonumber\\
&& \qquad \qquad \qquad \qquad \qquad \qquad
\left.
\cdot
\frac{- (\fs k -\fs q)}{2Q}
 V_+^{\lambda}(k-q,k-q-r)
\frac{R+M }{2R}
\right\}
\label{boundary-contribution-third-order-term-two-a}
\\
&&+\half \int\! \frac{d^D k}{(2\pi)^D}
\tr \left\{
\gamma_5 \left(
   \frac{\fs k + \fs p}{P+M} - \frac{\fs k -\fs q -\fs r}{R+M}
\right)
\frac{K+M }{2K}
 V_+^{\mu}(k+p,k)
\right.
\nonumber\\
&& \qquad \qquad \qquad \qquad \qquad \qquad
\left.
\cdot
\frac{- \fs k }{2K}
 V_+^{\nu\lambda}(k,k-q,k-q-r)
\frac{R+M }{2R}
\right\}
\label{boundary-contribution-third-order-term-two-b}
\\
&&-\half \int\! \frac{d^D k}{(2\pi)^D}
\tr \left\{
\gamma_5 \left(
   \frac{\fs k + \fs p }{P+M} - \frac{\fs k -\fs q -\fs r}{R+M}
\right)
\frac{K+M }{2K}
 V_+^{\mu}(k+p,k)
\right.
\nonumber\\
&& \qquad \qquad \qquad \qquad
\left.
\cdot
\frac{- \fs k}{2K}
 V_+^{\nu}(k,k-q)
\frac{-(\fs k - \fs q)}{2Q}
 V_+^{\lambda}(k-q,k-q-r)
\frac{R+M }{2R}
\right\}
\label{boundary-contribution-third-order-term-three}
\\
&&
- \left( M \leftrightarrow -M \right)
\label{boundary-contribution-third-order-terms-opposite-mass}
\ee
\end{eqnarray}

The expansion with respect to $p$, $q$ and $r$ in
(\ref{boundary-contribution-third-order-term-one}),
(\ref{boundary-contribution-third-order-term-two-a}),
(\ref{boundary-contribution-third-order-term-two-b}) and
(\ref{boundary-contribution-third-order-term-three})
leads to the result,
\begin{eqnarray}
-\half \int\! &&\frac{d^D k}{(2\pi)^D}
\frac{1}{2^7 K^7} \,
\nonumber\\
&& \tr \gamma_5  \left\{
-\gamma^\mu \fs k \gamma^\nu \fs k \gamma^\lambda
(\fs p + \fs q + \fs r)
\left[ 10M \right]
\right.
\nonumber\\
&& \phantom{\tr \gamma_5}
+\gamma^\mu \gamma^\nu \gamma^\lambda \fs k
\left[ 4k\cdot(p+q+r) M \right]
\nonumber\\
&& \phantom{\tr \gamma_5}
-\gamma^\mu \gamma^\nu \gamma^\lambda (\fs p + \fs q +\fs r)
\left[ -2MK^2 + 10M^3 \right]
\nonumber\\
&& \phantom{\tr \gamma_5}
+\gamma^\mu \fs k \gamma^\nu \gamma^\lambda
\left[
  -4k\cdot (p+q+r) K
  +4k\cdot(p+q+r) \frac{-k^2}{(K+M)}
\right]
\nonumber\\
&& \left. \phantom{\tr \gamma_5}
-\fs k \gamma^\mu \fs k \gamma^\nu \gamma^\lambda
(\fs q +\fs r)
\left[ 10M \right]
+ \fs p \gamma^\mu \gamma^\nu \fs k \gamma^\lambda \fs k
\left[ 10M \right]
\right\}
+ {\cal O}\left( \frac{1}{M}\right)
\ee
\end{eqnarray}
Subtracting the contribution with the mass of opposite signature
(\ref{boundary-contribution-third-order-terms-opposite-mass}),
we finally obtain
\begin{eqnarray}
&&  \Gamma_X^{\mu\nu\lambda}(p,q,r) \nonumber\\
&&
=\half
\tr \gamma_5 \gamma^\mu \gamma^\nu \gamma^\lambda
(\fs p + \fs q +\fs r)
\nonumber\\
&& \qquad\qquad\qquad
\times
\int\! \frac{d^D k}{(2\pi)^D}
\frac{1}{2^7 K^7}
\left[ -10 M k^2 - 2M k^2 -4MK^2 + 20M^3  + 2M k^2 -10Mk^2 \right]
\nonumber\\
&&
=\half
\tr \gamma_5 \gamma^\mu \gamma^\nu \gamma^\lambda
(\fs p + \fs q +\fs r)
\int\! \frac{d^D k}{(2\pi)^D}
\frac{M}{2^3 K^5}
\nonumber\\
&&
=-\half
\frac{1}{24\pi^2}
\epsilon^{\mu\nu\rho\sigma}(p+q+r)_\sigma
+ {\cal O}\left( \frac{1}{M} \right)
\ee
\end{eqnarray}

\subsubsection{Forth order and Higher orders}
As to the forth order term $(n=4)$, we find that
the leading term vanishes in the expansion with respect to the external
momentum. Then we obtain
\begin{equation}
\Gamma_X^{\mu\nu\lambda\rho}(p,q,r,s)
= 0 + {\cal O}\left( \frac{1}{M} \right)
\ee
\end{equation}

The higher order terms $(n \ge 5)$ have negative mass dimensions and
they are expected to be suppressed by the factor
$\frac{1}{M^{n-4}}$.

\subsubsection{Final result}
{}From these results,
we obtain
the variation of the boundary term under the gauge transformation
as follows.

\begin{eqnarray}
&& \delta_\omega \Gamma_X [A]
\nonumber\\
&&
=
\int\! \frac{d^4 p}{(2\pi)^4}\frac{d^4 q}{(2\pi)^4}
\left\{ -\frac{i^3 }{24\pi^2}
\epsilon^{\mu\nu\rho\sigma} p_\rho q_\sigma
\right\}
\tr\{ \omega(-p-q) A_{\mu}(p)A_{\nu}(q) \}
\nonumber\\
&&
+
\int\! \frac{d^4 p}{(2\pi)^4}\frac{d^4 q}{(2\pi)^4}\frac{d^4 r}{(2\pi)^4}
\left\{ - \frac{i^4}{24\pi^2}
\epsilon^{\mu\nu\lambda\sigma} \frac{(p+q+r)_\sigma}{2}
\right\}
\tr\{ \omega(-p-q) A_{\mu}(p)A_{\nu}(q)A_{\lambda}(r) \}
+ {\cal O}\left( \frac{1}{M} \right)
\nonumber\\
&&
=
\frac{i}{24\pi^2}
\int\! dx^4
\epsilon^{\mu\nu\rho\sigma}
\tr\left\{ \omega(x)
\left[
\partial_\mu A_{\nu}(x) \partial_\rho A_{\sigma}(x)
+\half
\partial_\mu \left(A_{\nu}(x)A_{\rho}(x)A_{\sigma}(x) \right)
\right]
\right\}
+ {\cal O}\left( \frac{1}{M} \right)
\ee
\nonumber\\
\end{eqnarray}
We can see that
the consistent anomaly is correctly reproduced by the
Wigner-Brillouin phase fixing procedure.

\section{Vacuum Polarization}
\label{sec:vacuum-polarization}

In this section, as another application of the perturbation theory,
we perform the calculation of the two-point function
(vacuum polarization function) in the expansion of the
five-dimensional determinant contribution,
Eq.~(\ref{perurbative-expansion-of-volume-contribution}).
In the momentum space, it is written as follows.
\begin{eqnarray}
i\Gamma_K [A]
=  \sum_{n=1}^\infty \, (i)^n
\int\! \prod_{i=1}^n \frac{d^4 p_i}{(2\pi)^4}
(2\pi)^4\delta\left(\sum_i^n p_i \right)
\Gamma_K^{\nu_1 \nu_2 \ldots \nu_n}
(p_1,\ldots, p_n) \,
\tr\{ \prod_{i=1}^n A_{\nu_i}(p_i) \}
\ec
\end{eqnarray}
where
\begin{eqnarray}
\Gamma_K^{\nu_1 \nu_2 \ldots \nu_n}(p_1,\ldots, p_n;L)
&&=\Pi_+^{\nu_1 \nu_2 \ldots \nu_n}(p_1,\ldots, p_n;L)
\nonumber\\
&&
-\half \Pi_-^{\nu_1 \nu_2 \ldots \nu_n}(p_1,\ldots, p_n;L)[+M]
-\half \Pi_-^{\nu_1 \nu_2 \ldots \nu_n}(p_1,\ldots, p_n;L)[-M]
\ec
\end{eqnarray}
\begin{eqnarray}
  \Pi_\pm^{\nu_1 \nu_2 \ldots \nu_n}(p_1,\ldots, p_n)
=
\int\!\frac{d^Dk}{i (2\pi)^D}
\int^{+L}_{-L}\prod_{i=1}^n ds_i \,
\Tr
\left\{
\prod_{i=1}^n \left[ \gamma^{\nu_i}
iS_{F\pm}(k+\sum_{j>i}p_j;s_i,s_{i+1})
\right]
       \right\}
\ee
\end{eqnarray}
Note that $s_{n+1}=s_1$.

We will find that each contribution from the fermion with the kink-like
mass or the fermion with the homogeneous mass is never chiral.
The fermion with
the {\it positive} homogeneous mass contains the light mode
(massless mode in the limit $L \rightarrow \infty$)
just as well as the fermion with the kink-like mass.
On the contrary, the fermion with the {\it negative} homogeneous
mass does not contain such light mode.
Therefore, by the subtraction, the normalization of the vacuum
polarization becomes correctly a half of that of the massless
Dirac fermion.

Unfortunately, it turns out that the dimensional regularization is
not adequate for the calculation of this volume contribution.
It leads to the gauge noninvariant term proportional to $M^2$.
Since in the lattice regularization
the volume contribution is expected to be gauge invariant,
this fact means our bad choice of the subsidiary regularization.

\subsection{Expression of Vacuum Polarization}

The two point function is given explicitly as follows.
\begin{eqnarray}
\Pi_\pm^{\mu\nu}(p;L)
&=&
\int\!\frac{d^Dk}{i (2\pi)^D}
\int^{+L}_{-L}\!\!dsdt \,
\Tr
\left\{
   \gamma^\mu iS_\pm(k+p;s,t)\gamma^\nu iS_\pm(k;t,s)
       \right\}
\\
&=&
\int\!\frac{d^Dk}{i (2\pi)^D}
\left\{
\Tr\left(\gamma_\mu P_R (\fs k+ \fs p) \gamma_\nu P_R \fs k \right)
\int^{+L}_{-L}\!\!dsdt \,
\Delta_{R\pm}(k+p;s,t) \, \Delta_{R\pm}(k;t,s) \,
\right.
\nonumber\\
&&\qquad\qquad\quad
+\Tr\left(\gamma_\mu P_L (\fs k+ \fs p) \gamma_\nu P_L \fs k \right)
\int^{+L}_{-L}\!\!dsdt \,
\Delta_{L-}(k+p;s,t) \, \Delta_{L-}(k;t,s) \,
\nonumber\\
&&\qquad\qquad\quad +
\Tr\left(\gamma_\mu P_R \gamma_\nu P_L \right)
\int^{+L}_{-L}\!\!dsdt \,
B_{R\pm}(k+p;s,t) \, B_{L\pm}(k;t,s) \,
\nonumber\\
&&\qquad\qquad\quad
+\Tr\left(\gamma_\mu P_L \gamma_\nu P_R \right)
\int^{+L}_{-L}\!\!dsdt \,
B_{L\pm}(k+p;s,t) \, B_{R\pm}(k;t,s) \,
\nonumber\\
&&\qquad\qquad\quad
+\Tr\left(\gamma_\mu P_R (\fs k+ \fs p) \gamma_\nu P_L \fs k \right)
\int^{+L}_{-L}\!\!dsdt \,
\Delta_{R\pm}(k+p;s,t) \, \Delta_{L-}(k;t,s) \,
\nonumber\\
&&\qquad\qquad\quad
\left.
+\Tr\left(\gamma_\mu P_L (\fs k+ \fs p) \gamma_\nu P_R \fs k \right)
\int^{+L}_{-L}\!\!dsdt \,
\Delta_{L-}(k+p;s,t) \, \Delta_{R\pm}(k;t,s) \,
\right\}
\ee
\end{eqnarray}
Using the orthogonality of the generalized ``sin'' function
Eq.~(\ref{generalized-sin-function-orthgonarity}), we perform
the integration over $s,t$ and obtain,

\begin{eqnarray*}
\phantom{\Pi_\pm^{\mu\nu}(p;L) }
&=&
\int\!\frac{d^Dk}{i (2\pi)^D}
\sum_{\omega(\pm)} \,
\frac{
\left\{
\Tr\left(\gamma_\mu P_R (\fs k+ \fs p) \gamma_\nu P_R \fs k \right)
+\Tr\left(\gamma_\mu P_L (\fs k+ \fs p) \gamma_\nu P_L \fs k \right)
\right\}
      }
{[M^2+\omega^2-(k+p)^2-i \varepsilon] \,
         [M^2+\omega^2-k^2-i \varepsilon]}
\\
&+&
\int\!\frac{d^Dk}{i (2\pi)^D}
\sum_{\omega(\pm)} \,
\frac{(\omega^2+M^2)\,  \Tr\left(\gamma_\mu \gamma_\nu \right) }
{[M^2+\omega^2-(k+p)^2-i \varepsilon] \,
         [M^2+\omega^2-k^2-i \varepsilon]}
\\
&+&
\int\!\frac{d^Dk}{i (2\pi)^D}
\left\{
\Tr\left(\gamma_\mu P_R (\fs k+ \fs p) \gamma_\nu P_L \fs k \right)
\int^{+L}_{-L}\!\!dsdt \,
\Delta_{R\pm}(k+p;s,t) \, \Delta_{L-}(k;t,s) \,
\right.
\\
&&\qquad\qquad\quad
\left.
+\Tr\left(\gamma_\mu P_L (\fs k+ \fs p) \gamma_\nu P_R \fs k \right)
\int^{+L}_{-L}\!\!dsdt \,
\Delta_{L-}(k+p;s,t) \, \Delta_{R\pm}(k;t,s) \,
\right\}
\\
&=&
\int\!\frac{d^Dk}{i (2\pi)^D}
\sum_{\omega(\pm)} \,
\frac{\Tr\left(\gamma_\mu (\fs k+ \fs p) \gamma_\nu \fs k \right) }
{[M^2+\omega^2-(k+p)^2-i \varepsilon] \,
         [M^2+\omega^2-k^2-i \varepsilon]}
\\
&+&
\int\!\frac{d^Dk}{i (2\pi)^D}
\sum_{\omega(\pm)} \,
\frac{(\omega^2+M^2) \, \Tr\left(\gamma_\mu \gamma_\nu \right) }
{[M^2+\omega^2-(k+p)^2-i \varepsilon] \,
         [M^2+\omega^2-k^2-i \varepsilon]}
\\
&+&
\int\!\frac{d^Dk}{i (2\pi)^D}
\left\{
\Tr\left(\gamma_\mu P_R (\fs k+ \fs p) \gamma_\nu P_L \fs k \right)
\right.
\nonumber\\
&& \left. \qquad \qquad \qquad \qquad
\cdot \int^{+L}_{-L}\!\!dsdt \,
\Delta_{R\pm}(k+p;s,t) \,
( \Delta_{L-}(k;t,s) -\Delta_{R\pm}(k;t,s) )\,
\right.
\\
&&\qquad\qquad\quad
+\Tr\left(\gamma_\mu P_L (\fs k+ \fs p) \gamma_\nu P_R \fs k \right)
\nonumber\\
&& \left. \qquad \qquad \qquad \qquad
\cdot \int^{+L}_{-L}\!\!dsdt \,
\Delta_{L-}(k+p;s,t) \,
( \Delta_{R\pm}(k;t,s)-\Delta_{L-}(k;t,s)) \,
\right\}
\ee
\end{eqnarray*}
Therefore we can write the two-point function as follows.
\begin{eqnarray}
\Pi_\pm^{\mu\nu}(p;L)
&=&
\int\!\frac{d^Dk}{i (2\pi)^D} \,
\left\{
\Tr\left(\gamma_\mu (\fs k+ \fs p) \gamma_\nu \fs k \right)
\right\}
{\cal K}_\pm (k+p,k)
\\
&+&
\int\!\frac{d^Dk}{i (2\pi)^D} \,
\left\{ \Tr\left(\gamma_\mu \gamma_\nu \right) \right\}
{\cal B}_\pm(k+p,k)
\\
&-& \half
\int\!\frac{d^Dk}{i (2\pi)^D}
\left\{
\Tr\left(\gamma_\mu \fs{\ext{k}} \gamma_\nu \fs{\ext{k}} \right)
\right\}
{\cal M}_{\pm}(k+p,k)
\ec
\end{eqnarray}
where

\begin{eqnarray}
{\cal K}_\pm (k+p,k)
&=&
\int^{+L}_{-L}\!\!dsdt \,
\Delta_{R\pm(L-)}(k+p;s,t) \, \Delta_{R\pm(L-)}(k;t,s)
\ec
\\
&=&
\sum_{\omega(\pm)} \,
\frac{1}{[M^2+\omega^2-(k+p)^2-i \varepsilon] \,
         [M^2+\omega^2-k^2-i \varepsilon]}
\\
{\cal B}_\pm(k+p,k)
&=&
\int^{+L}_{-L}\!\!dsdt \,
B_{\pm}(k+p;s,t) \, B_{\pm}(k;s,t)
\\
&=&
\sum_{\omega(\pm)} \,
\frac{(\omega^2+M^2)}{[M^2+\omega^2-(k+p)^2-i \varepsilon] \,
         [M^2+\omega^2-k^2-i \varepsilon]}
\ec
\\
{\cal M}_{\pm}(k+p,k)
&=&
\int^{+L}_{-L}\!\!dsdt \,
\left( \Delta_{R\pm}(k+p;s,t)- \Delta_{L-}(k+p;s,t) \right) \,
\left( \Delta_{R\pm}(k;t,s)  - \Delta_{L-}(k;t,s)   \right)
\ee
\nonumber\\
\end{eqnarray}

{}From this expression, we find that each term $\Pi_\pm^{\mu\nu}$
is vector-like, is even-parity and does not have any chiral structure.
(Even for the extra term due to the dimensional regularization.)
However both $\Pi_-^{\mu\nu}$ and $\Pi_+^{\mu\nu}[+M]$
contain the contribution of the light mode
(massless mode in the limit $L \rightarrow \infty$)
given by
Eq.~(\ref{light-mode-in-fermion-with-kink-like-mass}) and
Eq.~(\ref{light-mode-in-fermion-with-positive-homogeneous-mass}).
On the contrary, $\Pi_+^{\mu\nu}[-M]$ does not contain such a contribution
of the light mode.
By this fact, the subtracted two-point function shows the
correct {\it chiral normalization}: one half of the vacuum polarization
of the massless Dirac fermion.
We can see it explicitly in the following calculation.

\subsection{Evaluation of Vacuum Polarization}

By the Sommerfeld-Watson transformation,
it is possible to express ${\cal K}_\pm (k+p,k)$
and ${\cal B}_\pm (k+p,k)$ by the common normal modes.
Then we can perform the subtraction explicitly
at the finite extent of the fifth dimension.

${\cal K}_\pm (k+p,k;s,t)$ has the following Integrand.
\begin{eqnarray}
F_\pm(\omega)
&=&
\frac{n_\pm(\omega)}
{[ M^2+\omega^2-k^2-i \varepsilon]
 [ M^2+\omega^2-(k+p)^2-i \varepsilon]}
\end{eqnarray}
where
\begin{equation}
n_\pm(\omega) =
\left[
\Big(
    \frac{ (\omega\cot\omega L \pm M)
          +(\omega\cot\omega L - M) }{2} \Big)
          \big( L-\frac{\sin 2\omega L}{2\omega}  \big)
          +\sin^2\omega L     \right]
\ee
\end{equation}
The poles and its residues are summarized in
Table IV.
\vskip 16pt
\centerline{Table IV \
Poles and Residues in $I_\pm$ for ${\cal K}_\pm$}
\begin{tabular}{|c|c|c|c|}
\hline
\phantom{aa} singular part in ${\cal K}_\pm$ \phantom{aa}
& \phantom{aaaaa} pole \phantom{aaaaaaaaaaa}
& \phantom{a} residue \phantom{aaaaa}
&
${\scriptstyle
\frac{2 \omega}{\sin^2 \omega L \Delta_\pm (\omega) }} $
\\
\hline
  ${\scriptstyle
\frac{2 \omega}{\sin^2 \omega L \Delta_\pm (\omega) }} $
& $ \sin^2 \omega L \Delta_\pm (\omega)=0 $
& $ \frac{1}{n_\pm(\omega)}$
& 1
\\
\hline\hline
  $\omega\cot\omega L=\frac{\omega\cos\omega L}{\sin\omega L}$
& $ \sin\omega L=0 $
& $ \frac{\omega}{L}$
& $-\frac{2}{\omega}$
\\
\hline
 $  \frac{1}{M^2+\omega^2-p^2-i \varepsilon} $
&$iP \equiv i\sqrt{M^2-p^2-i \varepsilon}$
&$\frac{1}{2iP}$
&$\frac{2 iP}{-\sinh^2 PL }\frac{1}{\Delta_\pm (iP)} $
\\
\hline
\end{tabular}
\vskip 16pt
\noindent
By the Sommerfeld-Watson transformation,
${\cal K}_\pm (k+p,k)$ can be rewritten as
\begin{eqnarray}
{\cal K}_\pm (k+p,k)
&=&
\sum_{\sin\omega L=0} \,
\frac{2}{[M^2+\omega^2-(k+p)^2-i \varepsilon] \,
         [M^2+\omega^2-k^2-i \varepsilon]}
\nonumber\\
&&
+
\frac{n_\pm(iP)}{\sinh^2 PL \, \Delta_\pm (iP)}
\frac{1}{K^2-P^2}
+
\frac{n_\pm(iK)}{\sinh^2 KL \, \Delta_\pm (iK)}
\frac{1}{P^2-K^2}
\ee
\end{eqnarray}
Similarly we obtain,

\begin{eqnarray}
{\cal B}_\pm (k+p,k)
&=&
\sum_{\sin\omega L=0} \,
\frac{2(\omega^2+M^2)}{[M^2+\omega^2-(k+p)^2-i \varepsilon] \,
         [M^2+\omega^2-k^2-i \varepsilon]}
\nonumber\\
&&+
\frac{n_\pm(iP)}{\sinh^2 PL \, \Delta_\pm (iP)}
\frac{M^2-P^2}{K^2-P^2}
+
\frac{n_\pm(iK)}{\sinh^2 KL \, \Delta_\pm (iK)}
\frac{M^2-K^2}{P^2-K^2}
\ee
\end{eqnarray}
{}From these results, we see that
the subtraction can be performed rather simply.

In order to evaluate the remaining terms in the limit
$L \rightarrow \infty$,  we need to know the limit
of $\Delta_\pm(iP)$ and $\frac{n_\pm(iP)}{\sinh^2 PL}$.
It is given by
\begin{eqnarray}
\Delta_\pm(iP) &=& M^2-P^2-(P\coth PL \pm M)(P\coth PL -M) \\
               &\sitarel{=}{L\rightarrow \infty}&
               \left\{ \begin{array}{l}
               2(M-P)(M+P) \\
               2(M-P)P
    \end{array} \right.
\ec
\end{eqnarray}

\begin{eqnarray}
  \frac{n_\pm(iP)}{\sinh^2 PL}
&=&
\frac{
\left[
\Big(
    \frac{ (P \coth PL \pm M)
          +(P \coth PL - M) }{2} \Big)
          \big( L-\frac{\sinh 2PL}{2P}  \big)
          -\sinh^2 PL     \right]
}{\sinh^2 PL}
\\
&\sitarel{=}{L\rightarrow \infty}&
 \left\{ \begin{array}{l}
               -2 \\
               \frac{(-2P+M)}{P}
    \end{array} \right.
\ee
\end{eqnarray}
Then we have
\begin{eqnarray}
\bar{\cal K} (k+p,k)
&\equiv&
{\cal K}_+ (k+p,k)
-\half \left\{ {\cal K}_- (k+p,k)
            + {\cal K}_- (k+p,k)[M\rightarrow -M] \right\}
\\
&\sitarel{=}{L\rightarrow \infty}&
\half \left\{
\frac{1}{[P^2-M^2][K^2-M^2]} -\frac{1}{P^2K^2} \right\}
\ee
\end{eqnarray}

\begin{eqnarray}
\bar{\cal B} (k+p,k)
&\equiv&
{\cal B}_+ (k+p,k)
-\half \left\{ {\cal B}_- (k+p,k)
            + {\cal B}_- (k+p,k)[M\rightarrow -M] \right\}
\\
&\sitarel{=}{L\rightarrow \infty}&
\half \left\{
-\frac{M^2}{P^2K^2} \right\}
\ee
\end{eqnarray}

Using these results, the two-point function is written as
\begin{eqnarray}
\lim_{\lp \rightarrow \infty} \Pi^\pm_{\mu\nu}(p;L)
&=&
\half
\int\!\frac{d^Dk}{i (2\pi)^D} \,
\left\{
\frac{\Tr\left(\gamma_\mu (\fs k+ \fs p) \gamma_\nu \fs k \right)}
{[-(k+p)^2][-k^2]}
-\frac{\Tr\left(\gamma_\mu (\fs k+ \fs p) \gamma_\nu \fs k \right)
+\Tr\left(\gamma_\mu \gamma_\nu \right) M^2}{[M^2-(k+p)^2][M^2-k^2]}
\right\}
\nonumber\\
&-& \lim_{\lp \rightarrow \infty}
\half
\int\!\frac{d^Dk}{i (2\pi)^D}
\left\{
\Tr\left(\gamma_\mu \fs{\ext{k}} \gamma_\nu \fs{\ext{k}} \right)
\right\}
{\cal M}_{\pm}(k+p,k)
\ee
\end{eqnarray}
The first term in the r.h.s. is nothing but the contribution of massless
Dirac fermion
subtracted by {\it one Pauli-Villars-Gupta bosonic spinor field
with mass $M$}, except for the factor one half before it.
It is gauge invariant (even under the dimensional regularization).
This factor gives the correct normalization of the vacuum
polarization derived from the chiral determinant.

The remaining term is due to the dimensional regularization.
In four dimensions, we can show that it gives a finite term
proportional to $M^2$, which means quadratic divergence in the limit
$M \rightarrow \infty$.
It breaks gauge invariance.

The determinant of $K$, however, is expected to be
gauge invariant in the lattice regularization as we can see from
Eq. (\ref{effective-action-variation}).
Of course, this fact is not yet established at the perturbative level.
We need careful investigation of the two-point function
{\it in the continuum limit of the lattice theory}.

As far as the continuum limit theory is concerned,
the above result tells that
our choice of the dimensional regularization
is not suitable for the calculation of the part of the
determinant of $K$.
Besides this failure due to the dimensional regularization,
we think that our continuum limit analysis so far
have clarified the structure of the vacuum overlap formula
by taking the limit from at finite extent of the fifth dimension.

\section{Summary and Discussion}
\label{sec:discussion}

We have formulated the perturbation theory of
the vacuum overlap formula, based on the theory of
the fermion (with kink-like and  homogeneous masses)
in the finite extent of the fifth dimension.
The chiral projection entered the boundary condition
of the fermion field in the fifth direction.
Different series of discrete normal modes of the fifth momentum
occured and they were rearranged by the Sommerfeld-Watson transformation.
We have assumed that
the dimensional regularization preserves the cluster property.

The gauge non-invariance introduced by the boundary state
wave function actually led to the consistent anomaly.
The normalization of the vacuum polarization is a half of the
massless Dirac fermion.
This is because both the fermion with kink-like mass and
the fermion with positive homogeneous mass involve the light
(massless) modes but the fermion with negative homogeneous mass
does not.

We find that the dimensional regularization is not suitable as a
subsidiary regularization. It cannot respect the chiral
boundary condition and it induced a gauge non-invariant
piece in the vacuum polarization of the four dimensional theory.
The determinant of $K$, however, is expected to be
gauge invariant in the lattice regularization as we can see from
Eq. (\ref{effective-action-variation}).
Of course, this important point
is not yet established at the perturbative level.
We need careful investigation of the two-point function
{\it in the continuum limit of the lattice theory}.

Finally we make a comment about the case of the two-dimensional theory.
In this case, the subtle point due to the
dimensional regularization does not cause any difficulty.
The
{\it once-subtraction by the Pauli-Villars-Gupta bosonic spinor field}
is enough to make the two-point function finite.
We find that
the two-point function from the volume contribution
is gauge invariant and has the correct chiral normalization.
We also find that the boundary term reproduces the consistent anomaly.

Therefore the next desired step is to examine the perturbative
aspect of the vacuum overlap in the lattice regularization
in four dimensions.
We hope that the technique developed in the continuum limit analysis
given here may be also useful in the lattice case.

\acknowledgments

The authors would like to thank T.~Kugo for enlightening discussions,
especially about the Sommerfeld-Watson transformation.
The authors would like to express sincere thanks to S.~Randjbar-Daemi
and J. Strathdee for informing us about thier recent work.

\appendix
\section{Orthogonality of generalized ``sin'' function}

The orthogonality of the generalized ``sin'' function
\begin{equation}
\iLL\!ds \,
[\sin \omega(L-s)]_\pm [\sin \omega^\prime (L-s)]_\pm
= N_\pm (\omega) \delta_{\omega\omega^\prime}
\label{generalized-sin-function-orthgonarity-appendix}
\ec
\end{equation}
can be shown as follows.
\begin{eqnarray}
&& \iLL\!ds \,
[\sin \omega(L-s)]_\pm [\sin \omega^\prime (L-s)]_\pm
\nonumber\\
&=& \iLo \!ds \, \sin\omega(L-s)\sin\omega^\prime(L-s)
  + \iLo \!ds \, \sin_\pm\omega(L-s) \sin_\pm\omega^\prime(L-s)
\nonumber\\
&=&
\left[ \frac{\sin (\omega-\omega^\prime)L }{2(\omega-\omega^\prime)}
      -\frac{\sin (\omega+\omega^\prime)L }{2(\omega+\omega^\prime)}
\right]
\nonumber\\
&+&
\frac{1}{(\omega\cot\omega L \pm M)(\omega^\prime\cot\omega^\prime L \pm M)}
    \left(
     (\omega \omega^\prime + M^2)
        \left[ \frac{\sin (\omega-\omega^\prime)L }{2(\omega-\omega^\prime)}
      -\frac{\sin (\omega+\omega^\prime)L }{2(\omega+\omega^\prime)} \right]
\right.
\nonumber\\
&& \qquad\qquad\qquad\qquad\qquad\qquad \qquad\quad
+\omega \omega^\prime
\left[ \frac{\sin (\omega+\omega^\prime)L }{(\omega+\omega^\prime)} \right]
\nonumber\\
&& \qquad\qquad\qquad\qquad \qquad\qquad\qquad\quad
\pm M\omega^\prime
\left[-\frac{\cos (\omega-\omega^\prime)L-1 }{2(\omega-\omega^\prime)}
      -\frac{\cos (\omega+\omega^\prime)L-1 }{2(\omega+\omega^\prime)} \right]
\nonumber\\
&&\left. \qquad\qquad\qquad\qquad\qquad\qquad\qquad\quad
\pm M\omega
\left[+ \frac{\cos (\omega-\omega^\prime)L-1 }{2(\omega-\omega^\prime)}
      -\frac{\cos (\omega+\omega^\prime)L-1 }{2(\omega+\omega^\prime)} \right]
\right)
\nonumber\\
&=&
\frac{1}{(\omega^2-\omega^{\prime 2})
          (\omega\cot\omega L \pm M)(\omega^\prime\cot\omega^\prime L \pm M)}
\nonumber\\
&& \times \left(
(\omega\cot\omega L \pm M)(\omega^\prime\cot\omega^\prime L \pm M)
\right.
\nonumber\\
&&\qquad\qquad\quad
\times
\left[
\sin \omega L
(\omega^\prime \cos \omega^\prime L - M \sin \omega^\prime L)
- (\omega \cos\omega L - M \sin\omega L) \sin \omega^\prime L
\right]
\nonumber\\
&&\quad
+(\omega \omega^\prime + M^2)
\nonumber\\
&&\quad\qquad\qquad
\times
\left[
\sin\omega L
(\omega^\prime \cos\omega^\prime L \pm M \sin\omega^\prime L)
- (\omega \cos\omega L \pm M \sin\omega L) \sin \omega^\prime L
\right]
\nonumber\\
&&\quad
+\omega \omega^\prime (\omega-\omega^\prime)
\left[ \sin\omega L \cos\omega^\prime L
      +\cos\omega L \sin\omega^\prime L \right]
\nonumber\\
&&\left. \quad
\pm M (\omega^2-\omega^{\prime2}) \sin\omega L \sin\omega^\prime L
\right)
\end{eqnarray}
using Eq.~(\ref{mode-equation})

\begin{eqnarray}
&=&
\frac{1}{(\omega^2-\omega^{\prime 2})
          (\omega\cot\omega L \pm M)(\omega^\prime\cot\omega^\prime L \pm M)}
\nonumber\\
&&
\times
\left(
(\omega^{\prime 2} + M^2)
\left[
(\omega \cos\omega L \pm M \sin\omega L) \sin \omega^\prime L
\right]
\right.
\nonumber\\
&&\quad
-(\omega^2 + M^2)
\left[
\sin \omega L
(\omega^\prime \cos \omega^\prime L \pm M \sin \omega^\prime L)
\right]
\nonumber\\
&&\quad
+(\omega \omega^\prime + M^2)
\nonumber\\
&&\quad\qquad\qquad
\times
\left[
\sin\omega L
(\omega^\prime \cos\omega^\prime L \pm M \sin\omega^\prime L)
- (\omega \cos\omega L \pm M \sin\omega L) \sin \omega^\prime L
\right]  \phantom{aaaa}
\nonumber\\
&&\quad
+\omega \omega^\prime (\omega-\omega^\prime)
\left[ \sin\omega L \cos\omega^\prime L
      +\cos\omega L \sin\omega^\prime L \right]
\nonumber\\
&&\left. \quad
\pm M (\omega^2-\omega^{\prime2}) \sin\omega L \sin\omega^\prime L
\right)
\nonumber\\
&=&
N_\pm (\omega) \, \delta_{\omega\omega^\prime}
\ee
\end{eqnarray}
The normalization factor $N_\pm(\omega)$ is evaluated by also
using Eq.~(\ref{mode-equation}) as
\begin{eqnarray}
N_\pm(\omega)
&=& \left\{ 1
           +\frac{(\omega^2+M^2)}{(\omega\cot\omega L \pm M)^2}
         \right\}
    \frac{1}{2}
\big( L-\frac{\sin 2\omega L}{2\omega} \big)
\nonumber\\
&+& \frac{1 }{(\omega\cot\omega L \pm M)^2}
    \left\{
      \frac{\omega}{2}\sin2\omega L \pm \frac{M}{2}(1-\cos2\omega L)
      \right\}
\nonumber\\
&=& \left[
         \Big(
\frac{(\omega\cot\omega L \pm M)+
              (\omega\cot\omega L - M)}
     {(\omega\cot\omega L \pm M)}
          \Big)
          \frac{1}{2}
\big( L-\frac{\sin 2\omega L}{2\omega} \big)
          +\frac{\sin^2\omega L}{(\omega\cot\omega L \pm M)}
           \right]
\nonumber\\
&\equiv& n_\pm(\omega) \frac{1}{(\omega\cot\omega L \pm M)}
\ee
\end{eqnarray}

\section{
correlation functions between boundaries}
\label{appendix:cluster-property-of-correlation}

In this appendix, we perform the calculation of the
correlation functions between boundaries of
the order $n \ge 1$ in the perturbative expansion given
in Sec. \ref{sec:perturbation-at-finite-extent-of-fifth-dimension}.
We do not take into account of the breaking of the
chiral boundary condition due to the dimensional regularization.
In the momentum space, it reads

\begin{eqnarray}
&&
  \left( \begin{array}{c} ..P_R \\
                        . P_L   \end{array} \right)
\left[
i \sum_{n=1}^\infty \left\{ S_{F \pm} \cdot (-)\fs A \cdot \right\}^n
S_{F\pm}
\right]
\left( \begin{array}{cc} P_L..  & P_R .   \end{array} \right)
\nonumber\\
&&=
 \sum_{n=1}^\infty
\int\! \frac{d^4 k}{(2\pi)^4}\prod_{i=1}^n \frac{d^4 p_i}{(2\pi)^4}
\exp\{-ik_1x+ik_{n+1}y\}
\nonumber\\
&&\qquad \qquad
  \left( \begin{array}{c} ..P_R \\
                        . P_L   \end{array} \right)
\left[
i \prod_{i=1}^n
\left\{ S_{F\pm}(k_i) \cdot (-) \gamma^{\nu_i} \cdot \right\} S_{F\pm}(k)
\right]
\left( \begin{array}{cc} P_L..  & P_R .   \end{array} \right)
\prod_{i=1}^n A_{\nu_i}(p_i)
\ec
\end{eqnarray}
where
$p_i$ is assumed to be the momentum incoming from the external
gauge boson attached to the vertex $\gamma^i$ and
$k_i = k+\sum_{j \ge i}p_j$.

We first perform the integration over the extra coordinates at every
vertices of external gauge field, with the help of
the orthogonality of the generalized ``sin'',
\begin{equation}
\iLL\!ds \,
[\sin \omega(L-s)]_\pm [\sin \omega^\prime (L-s)]_\pm
= N_\pm (\omega) \delta_{\omega\omega^\prime}
\ee
\end{equation}
Then we obtain
\begin{eqnarray}
..P_R
\left[
i \prod_{i=1}^n
\left\{ S_{F\pm}(k_i) \cdot (-) \gamma^{\nu_i} \cdot \right\}
S_{F\pm}(k_{n+1})
\right]
P_L ..
&=&
(i)^n \sum_{0\le 2l \le n+1} P_R C_{2l}^n  \,
\Delta_{R\pm}^{(n+1,2l)} 
\ec
\\
.P_L
\left[
i \prod_{i=1}^n
\left\{ S_{F\pm}(k_i) \cdot (-) \gamma^{\nu_i} \cdot \right\}
S_{F\pm} (k_{n+1})
\right]
P_R .
&=&
(i)^n \sum_{0 \le 2l \le n+1} P_L C_{2l}^n  \,
\Delta_{L-}^{(n+1,2l)} 
\ec
\\
..P_R
\left[
i \prod_{i=1}^n
\left\{ S_{F\pm}(k_i) \cdot (-) \gamma^{\nu_i} \cdot \right\}
S_{F\pm}(k_{n+1})
\right]
P_R .
&=&
(i)^n \sum_{1 \le 2l+1 \le n+1} P_R C_{2l+1}^n \,
B_{\pm}^{(n+1,2l)} 
\ec
\\
.P_L
\left[
i \prod_{i=1}^n
\left\{ S_{F\pm}(k_i) \cdot (-) \gamma^{\nu_i} \cdot \right\}
S_{F\pm}(k_{n+1})
\right]
P_L ..
&=&
(i)^n \sum_{1 \le 2l+1 \le n+1} P_L C_{2l+1}^n \,
B_{\pm}^{(n+1,2l)} 
\ec
\end{eqnarray}
and

\begin{eqnarray}
\Delta_{R\pm}^{(n+1,2l)}
&=&
\sum_\omega
\frac{1}{n_\pm(\omega)}
\frac{\omega^2 }
     {\left(\omega \cot \omega L \pm M \right)}
(\omega^2+M^2)^l
\prod_{i=1}^{n+1} \frac{1}{[M^2+\omega^2-k_i^2-i \varepsilon]}
\ec
\\
\Delta_{L-}^{(n+1,2l)}
&=&
\sum_\omega
\frac{1}{n_\pm(\omega)}
\frac{\omega^2 }
     {\left(\omega \cot \omega L - M \right)}
(\omega^2+M^2)^l
\prod_{i=1}^{n+1} \frac{1}{[M^2+\omega^2-k_i^2-i \varepsilon]}
\ec
\\
B_{\pm}^{(n+1,2l)}
&=&
\sum_\omega
\frac{1}{n_\pm(\omega)}
\left(-\omega^2 \right)
(\omega^2+M^2)^l
\prod_{i=1}^{n+1} \frac{1}{[M^2+\omega^2-k_i^2-i \varepsilon]}
\ee
\end{eqnarray}
$C_0^n$ is the product of the gamma matrices defined by
\begin{eqnarray}
C_0^n&=&
\fs k_1 \gamma^{\nu_1}
\fs k_2 \gamma^{\nu_2}
\ldots
\fs k_{n-2} \gamma^{\nu_{n-2}}
\fs k_{n-1} \gamma^{\nu_{n-1}}
\fs k_n \gamma^{\nu_n}
\fs k_{n+1}
\ec
\end{eqnarray}
and $C_m^n$ is defined as the summation over the possible products
of the gamma matrices which can be obtained from $C_0^n$ by replacing
$m$-number of $\fs k_i$ by the unit matrix.

The summations over the normal modes of $\omega$
in $\Delta_{R\pm}^{(n+1,2l)}$,
$\Delta_{L-}^{(n+1,2l)}$ and
$B_{\pm}^{(n+1,2l)}$
can be performed by the use of
the technique of the Sommerfeld-Watson transformation.
Refering to the Tables
I and II, we obtain

\begin{eqnarray}
\Delta_{R\pm}^{(n+1,2l)}
&=&
 \sum_\omega
\frac{1}{n_\pm(\omega)}
\frac{\omega^2 }
     {\left(\omega \cot \omega L \pm M \right)}
(\omega^2+M^2)^l
\prod_{i=1}^{n+1} \frac{1}{[M^2+\omega^2-k_i^2-i \varepsilon]}
\nonumber\\
&=&
-
\sum_{i=1}^{n+1}
\frac{1}{\sinh^2 K_i L \Delta_\pm (iK_i)} \,
\frac{K_i^2(M^2-K_i^2)^l }{(K_i \coth K_i L \pm M)}
\prod_{j\not=i} \frac{1}{K_j^2-K_i^2}
\nonumber\\
&&
+\sum_{\scriptstyle \omega \cot \omega L \pm M}
\frac{2 \sin^2 \omega L}
{\omega^2 \left(L- \frac{\sin2\omega L }{2\omega} \right)} \,
\omega^2 (\omega^2+M^2)^l
\prod_{i=1}^{n+1} \frac{1}{[M^2+\omega^2-k_i^2-i \varepsilon]}
\nonumber\\
&=&
-
\sum_{i=1}^{n+1}
\frac{1}{\sinh^2 K_i L \Delta_\pm (iK_i)} \,
\frac{K_i^2 (M^2-K_i^2)^l }{(K_i \coth K_i L \pm M)}
\prod_{j\not=i} \frac{1}{K_j^2-K_i^2}
\nonumber\\
&&
+
\sum_{i=1}^{n+1}
\frac{(M^2-K_i^2)^l}{(K_i \coth K_i L \pm M)}
\prod_{j\not=i} \frac{1}{K_j^2-K_i^2}
\ec
\\
  B_{\pm}^{(n+1,2l)}
&=&
\sum_\omega
\frac{1}{n_\pm(\omega)}
\left(-\omega^2 \right)
(\omega^2+M^2)^l
\prod_{i=1}^{n+1} \frac{1}{[M^2+\omega^2-k_i^2-i \varepsilon]}
\nonumber\\
&=&
\sum_{i=1}^{n+1}
\frac{1}{\sinh^2 K_i L \Delta_\pm (iK_i)} \,
 K_i^2 (M^2-K_i^2)^l
\prod_{j\not=i} \frac{1}{K_j^2-K_i^2}
\ec
\end{eqnarray}

In the limit $L\rightarrow \infty$, we have
\begin{eqnarray}
\lim_{L\rightarrow \infty}
\Delta_{R\pm}^{(n+1,2l)}
&=&
\sum_{i=1}^{n+1}\frac{(M^2-K_i^2)^l}{(K_i \pm M)}
\prod_{j\not=i} \frac{1}{K_j^2-K_i^2}
\ec
\\
\lim_{L\rightarrow \infty}
\Delta_{L-}^{(n+1,2l)}
&=&
\sum_{i=1}^{n+1}\frac{(M^2-K_i^2)^l}{(K_i - M)}
\prod_{j\not=i} \frac{1}{K_j^2-K_i^2}
\ec
\\
\lim_{L\rightarrow \infty}
B_{\pm}^{(n+1,2l)}
&=& 0
\ec
\end{eqnarray}
Therefore we obtain
\begin{eqnarray}
&&
\lim_{ \lp \rightarrow \infty}
  \left( \begin{array}{c} ..P_R \\
                        . P_L   \end{array} \right)
\left[
i \prod_{i=1}^n
\left\{ S_{F\pm}(k_i) \cdot (-) \gamma^{\nu_i} \cdot \right\}
S_{F\pm}(k_{n+1})
\right]
\left( \begin{array}{cc} P_L..  & P_R .   \end{array} \right)
\nonumber\\
&& =
(i)^n \sum_{n=1}^\infty
\left( \begin{array}{cc}
P_R V_\pm^{\nu_1\nu_2\ldots\nu_{n+1}}(k_1,k_2,\ldots,k_{n+1}) P_L
& 0 \\
0 &
P_L V_-^{\nu_1\nu_2\ldots\nu_{n+1}}(k_1,k_2,\ldots,k_{n+1}) P_R
\end{array} \right)
\ec
\label{cluster-property-in-boundary-term-at-each-order-appendix}
\end{eqnarray}
where
\begin{eqnarray}
V_\pm^{\nu_1\nu_2\ldots\nu_{n+1}}(k_1,k_2,\ldots,k_{n+1})
=
\sum_{0\le 2l \le n+1} C_{2l}^n  \,
\sum_{i=1}^{n+1}\frac{(M^2-K_i^2)^l}{(K_i \pm M)}
\prod_{j\not=i} \frac{1}{K_j^2-K_i^2}
\ee
\end{eqnarray}
This result shows that the cluster property holds:

\begin{eqnarray}
&&
\lim_{ \lp \rightarrow \infty}
 \left( \begin{array}{c} ..P_R \\
                        . P_L   \end{array} \right)
\left[
i \sum_{n=1}^\infty \left\{ S_{F \pm} \cdot (-)\fs A \cdot \right\}^n
S_{F\pm}
\right]
\left( \begin{array}{cc} P_L ..  & P_R .   \end{array} \right)
\nonumber\\
&& =
\lim_{\lp \rightarrow \infty}
 \left(
   \begin{array}{c}
..P_R \left[
i \sum_{n=1}^\infty \left\{ S_{F-}[\mp M]
\cdot (-)\fs A \cdot \right\}^n S_{F-}[\mp M]
\right] P_L .. \\
0   \end{array} \right.
\nonumber\\
&&
\hskip 3cm
\left.
\begin{array}{c}
0 \\
.P_L
\left[
i \sum_{n=1}^\infty \left\{ S_{F-}[+M]
\cdot (-)\fs A \cdot \right\}^n S_{F-}[+M]
\right]  P_R .
  \end{array} \right)
\ee
\label{cluster-property-in-boundary-term-appendix}
\end{eqnarray}

For $n=1,2,3$,
the explicit form of $V_\pm^{\nu_1\nu_2\ldots\nu_{n+1}}$
is given as follows.
For $n=1$,
\begin{equation}
  \label{gauge-boson-vertex-one-point-appendix}
V_\pm^\mu(k+p,k) \, (P+K)  =
 \left[ (\fs k + \fs p) \gamma^\mu \fs k \right]
\frac{1}{[P\pm M][K \pm M]}
+\gamma^\mu
\ee
\end{equation}
$p$ is assumed to be the momentum incoming from the external
gauge boson attached to the vertex $\gamma^\mu$.
$P$ and $K$ is defined as $P=\sqrt{M^2-(k+p)^2-i \varepsilon}$
and $K=\sqrt{M^2-k^2-i \varepsilon}$.

For $n=2$,
\begin{eqnarray}
  \label{gauge-boson-vertex-two-point-appendix}
V_\pm^{\mu \nu} && (k+p,k,k-q) \, (P+K)(K+Q)(Q+P)
\nonumber\\
&&= \left[ (\fs k + \fs p) \gamma^\mu \fs k \gamma^\nu (\fs k - \fs q)\right]
\frac{P+K+Q\pm M}{[P\pm M][K \pm M][Q \pm M]}
\nonumber\\
&&\quad
+
\left[
(\fs k + \fs p) \gamma^\mu \gamma^\nu
+ \gamma^\mu \fs k \gamma^\nu
+ \gamma^\mu \gamma^\nu (\fs k - \fs q) \right]
\ee
\end{eqnarray}
$q$ is assumed to be the momentum incoming from the external
gauge boson attached to the vertex $\gamma^\nu$ and
$Q$ is defined by $Q=\sqrt{M^2-(k-q)^2-i \varepsilon}$.

For $n=3$,
\begin{eqnarray}
  \label{gauge-boson-vertex-three-point-appendix}
&&  V_\pm^{\mu \nu \lambda}(k+p,k,k-q,k-q-r)
 \, (P+K)(P+Q)(P+R)(K+Q)(K+R)(Q+R)
\nonumber\\
&&=
\left[ (\fs k + \fs p) \gamma^\mu \fs k \gamma^\nu (\fs k - \fs q)
\gamma^\lambda (\fs k - \fs q - \fs r) \right]
\frac{A_\pm (P,K,Q,R)}
{[P\pm M][K \pm M][Q \pm M][R \pm M]}
\nonumber\\
&&
+ \left[
 (\fs k + \fs p) \gamma^\mu \fs k \gamma^\nu \gamma^\lambda
+(\fs k + \fs p) \gamma^\mu \gamma^\nu (\fs k-\fs q) \gamma^\lambda
+(\fs k + \fs p) \gamma^\mu \gamma^\nu \gamma^\lambda
                                              (\fs k - \fs q - \fs r)
+  \gamma^\mu \fs k \gamma^\nu (\fs k - \fs q) \gamma^\lambda
\right. \nonumber\\
&&\qquad \left.
+  \gamma^\mu \fs k \gamma^\nu  \gamma^\lambda (\fs k - \fs q -\fs r)
+  \gamma^\mu \gamma^\nu (\fs k - \fs q) \gamma^\lambda
                                              (\fs k - \fs q -\fs r)
\right]
(P+K+Q+R)
\nonumber\\
&&
+
\left[\gamma^\mu \gamma^\nu \gamma^\lambda \right]
B (P,K,Q,R)
\ec
\end{eqnarray}
and
\begin{eqnarray}
A_\pm (P,K,Q,R) &=&
P^2(Q+R+K)+Q^2(P+R+K)+R^2(P+Q+K)+K^2(P+Q+R)
\nonumber\\
&&
+2(QRK+PRK+PQK+PQR)
\nonumber\\
&&
\pm M(P+K+Q+R)^2 + M^2(P+K+Q+R)
\ec
\\
B (P,K,Q,R) &=& (QRK+PRK+PQK+PQR) + M^2(P+K+Q+R)
\ee
\end{eqnarray}
$r$ is assumed to be the momentum incoming from the external
gauge boson attached to the vertex $\gamma^\lambda$ and
$R$ is defined by $R=\sqrt{M^2-(k-q-r)^2-i \varepsilon}$.



\begin{thebibliography}{10}
\bibitem{fujikawa}
Fujikawa formulated the generalized Pauli-Villars regularization
so as to regulate both even-parity and odd-parity terms\cite{genpv}.
In this case, the bose symmetry with respect to the external gauge
field cannot be maintained.
It is recovered only in the case of the anomaly-free theory.
\bibitem{Latfs}
Frolov and Slavnov discussed the lattice
version of the SO(10) chiral gauge theory\cite{latfs}.
In this case, both the original fermion and the regulators are
written by the Wilson fermion and are not gauge invariant.
They argued that the breaking of the gauge invariance due to
the Wilson-mass cancelle each other in the summation over the infinite
number of contributions.
\end{thebibliography}
\end{document}